\documentclass[3p,times,12pt]{elsarticle}

\journal{Nuclear Physics B}

\usepackage[colorlinks]{hyperref}
\usepackage{amsmath,amssymb}
\allowdisplaybreaks[2]

\usepackage{epstopdf}
\usepackage{amsmath}
\usepackage{amsfonts}
\usepackage{graphicx}
\usepackage{appendix}
\usepackage{dsfont}
\usepackage{amssymb}
\usepackage{bm}
\usepackage{color}
\usepackage{ulem}
\usepackage{soul}
\usepackage{natbib}
\usepackage{nccmath}
\usepackage{array}

\newcommand{\beq}{\begin{equation}}
\newcommand{\eneq}{\end{equation}}
\newcommand{\be}{\begin{equation}}
\newcommand{\ee}{\end{equation}}
\newcommand{\bea}{\begin{eqnarray}}
\newcommand{\eea}{\end{eqnarray}}

\usepackage{multirow}
\usepackage{wasysym}

\begin{document}

\begin{frontmatter}



\title{Tunable Kondo screening length at a Y-junction of three inhomogeneous spin chains.}


\author[unical,infncos]{Domenico Giuliano\corref{cor1}}
\ead{domenico.giuliano@fis.unical.it}

\author[unical,infncos]{Andrea Nava}
\ead{andrea{\_}nava{\_}@msn.com}

\author[infnpg]{Pasquale Sodano}
\ead{pasquale.sodano01@gmail.com} 

\address[unical]{Dipartimento di Fisica, Universit\`a della Calabria,  
Arcavacata di Rende I-87036, Cosenza, Italy }
\address[infncos]{INFN, Gruppo collegato di Cosenza, 
Arcavacata di Rende I-87036, Cosenza, Italy }  
\address[infnpg]{I.N.F.N., Sezione di Perugia, Via A. Pascoli, I-06123, Perugia, Italy} 
\cortext[cor1]{Corresponding author}

\begin{abstract}
We derive the topological Kondo Hamiltonian describing a Y junction of three 
 XX-spin chains   connected  to outer quantum Ising chains with different tilting angles for the Ising 
axis. We show that the tilting angles in the spin models play the role of   the phases of the superconducting order parameters
at the interfaces between bulk superconductors and one-dimensional conducting normal electronic wires.  As a result,   
different  tilting angles  induce  nonzero  equilibrium spin (super)currents through the junction. 
 Employing the  renormalization group   approach to the  topological Kondo model, 
we   derive the scaling formulas for the equilibrium spin currents.  We argue that,  by 
monitoring the crossover in the currents induced by the 
  Kondo effect, it is possible  to  estimate  the   Kondo screening length.    
  In particular, we prove  how it is possible  to   tune the   Kondo length by acting on 
 the applied phases  only; this  enables us to   map out  the scaling properties  by just 
tuning the tilting angles and the Kondo length accordingly.  

\end{abstract}


\end{frontmatter}
  
\section{Introduction}
\label{intro}

In its original description, the  Kondo effect was evidenced as a low-temperature upturn 
in the  resistance of a metal containing magnetic impurities antiferromagnetically interacting with the spin of 
the itinerant conduction electrons  in the metal (the ``Kondo interaction'')  \cite{kondo64,hewson,kouwenhoven01}. 

The effect is determined by the  
low-energy/low-temperature $T$ proliferation of impurity spin-flip processes. These induce a   nonperturbative, 
strongly correlated, (``Kondo'') state in which the 
electron spins cooperate to ``dynamically screen'' the impurity spin. Letting $s$ be the impurity spin
and $k$ be the number of different electronic spin screening channels, if $k=2s$, 
the impurity spin is perfectly screened and the Nozi\`eres Fermi-liquid state sets in  the $T \to 0$ limit,  
in which the impurity effectively acts as a spinless localized scatterer (the ``Kondo singlet'') \cite{noz_1,noz_2}.
Instead,  when $k > 2s$ (``overscreened Kondo effect''),  
a non Fermi-liquid state with rather peculiar properties sets in  \cite{luda_4,luda_2}. 

Right after  its  explanation \cite{kondo64}, the Kondo effect appeared as a paradigmatic  example of 
a many-body correlated electronic state, eventually becoming  
a testground both for theoretical  many-body techniques  \cite{bulla}, and for designing  
correlated electronic nanodevices \cite{kouwenhoven01}. In particular, the possibility of 
realizing the effect in controlled systems with tunable parameters, such as quantum dots with metallic 
 \cite{alivisatos96,kouwenhoven98,gg98_1,gg98_2,gln}, or superconducting leads \cite{avish,choi,gbab}, 
 allowed for engineering quantum circuits with the maximum   value for the conductance per 
 each single channel \cite{kouwenhoven01}. Moreover,  a recent, remarkable achievement 
has been provided by  the realization of a peculiar, overscreened ``topological'' Kondo effect (TKE), in which the Kondo impurity is determined 
by the   Majorana fermionic modes arising at the endpoints of one-dimensional (1D) topological 
superconductors  \cite{beri_1,topo_2,exact_2,topo_4}.

A key feature of the Kondo effect, that is strictly related to its nonperturbative nature, is the emergence 
of a finite temperature scale $T_K$ (the ``Kondo temperature'') separating the high-$T$ perturbative regime from 
the $T \to 0$ Kondo fixed point. Specifically, $T_K$ emerges within the perturbative   renormalization group (RG)
framework, as a dimensionful scale that is invariant along the RG trajectories \cite{kwilson,noz_1}.

Using the Fermi velocity  $v$ associated to itinerant electrons, it is possible to trade  $T_K$ for a length 
scale $\ell_K \sim v / T_K$. The physical meaning of $\ell_K$ is that the value of any local observable at 
a distance $x$ from the impurity is 
determined by the Kondo fixed point if $x< \ell_K$, while it only takes perturbative 
correction in the Kondo interaction if   $x > \ell_K$. Basically, $\ell_K$ measures the 
size of the spin cloud dynamically screening the impurity spin  
(the ``Kondo cloud''), and is accordingly    dubbed as the ``Kondo screening length'' (KL), an analog of which has also been 
proposed to emerge at a Majorana mode coupled to a 1D quantum wire \cite{gaf_3}. 
The emergence of $\ell_K$ is a direct consequence of the implementation of the scaling 
assumption in the Kondo regime \cite{kwilson}. 

Finding 
an experimental evidence of $\ell_K$  would be a strong confirmation of the validity of the scaling 
assumption. Despite the strong theoretical  background supporting the existence of the KL, so 
far it has never been experimentally detected. Such a failure may be attributed to a number 
of reasons, such as the tiny value of spin correlations over distances of the 
order of $\ell_K$, the finite density of magnetic impurities in a real metal, 
the effects of the interaction between itinerant electrons, etc. \cite{Affleck09}.

A promising route to overcome the difficulties in measuring $\ell_K$  is
  realizing  the Kondo effect in nonconducting systems, such as quantum spin chains (SCs).
Indeed, despite typically being insulating, spin-1/2 quantum SCs have a low-lying
elementary excitation spectrum consisting of spin-1/2 delocalized ``spinons'', collective 
modes carrying spin but no charge, which can effectively screen an isolated magnetic impurity 
antiferromagnetically coupled to the chain
\cite{eggert92,furusaki,sorensen,sirker08}. Lattice spin correlations in real space are typically
more easily measurable than spin density correlations between distant electrons in a metal, which 
makes spin chains a pretty better arena to probe $\ell_K$, compared to metals. In addition,  working with spin chains   
allows for studying Kondo physics by using a series of tools developed for spin systems, such as  entanglement witnesses and negativity 
\cite{bayat10,bayat12}. Remarkably, nowadays technology allows for realizing systems behaving as quantum SCs with tunable parameters 
by using, for instance, cold atoms on an optical lattice  \cite{lukin,grst_13,gsta}, or
pertinently engineered Josephson junction one-dimensional arrays \cite{giuso_1,giuso_x}. 
In this case, the Kondo problem formally emerges by using the Jordan-Wigner (JW)
representation for the spin $1/2$ operators to map the 
lattice spin Hamiltonian   onto a Luttinger liquid model interacting with an isolated
magnetic impurity \cite{grst_13,gsta,grt}.

Working with a  Y-junction of quantum 
spin chains (YSC), allows for the realization of  Kondo Hamiltonians 
even without explicitly introducing a quantum impurity in the chains. Indeed, 
when  implementing the JW transformation for a YSC, in order to preserve 
the correct (anti)commutation relations between spin operators belonging to different chains, 
we have to introduce as many additional real fermionic degrees of freedom (the ''Klein factors'' (KFs)) 
as many chains \cite{crampettoni}. The KFs do now appear in the bulk Hamiltonian of the chains, as 
they have to, but,   when introduced in the boundary interaction Hamiltonian describing the junction, they determine 
an effective, spin-1/2 degree of freedom, interacting with the bulk degrees of freedom of 
the chains through a  topological Kondo Hamiltonian, with the bulk spin density operator being a nonlocal
function of the single chains. By now, a topological Kondo Hamiltonian has been shown to 
describe a junction of three quantum Ising chains \cite{tsve_1,gct}, of three XX chains \cite{crampettoni},
of a pertinently engineered Josephson junction network \cite{giuso_x}, and of three XY chains, 
continuously interpolating between the Ising- and the XX-limit \cite{gsst}. 

As a possible route to estimate  $\ell_K$ at a YSC, it has been proposed to look
at the scaling of a pertinently defined local magnetization at the junction \cite{giuso_x,gsst}.
However,  it would be much more effective to directly extract scaling properties from a (spin, in this case) current 
transport measurement, as  it is typically done with  Kondo effect in a quantum dot with metallic leads.
In fact, measuring the equilibrium (super)current
pattern induced through   similar junctions realized with Josephson junction arrays  connected   to bulk superconductors at fixed phases, 
has provided an effective mean  to monitor the phase diagram of the junction and the 
associated scaling properties \cite{giuso_1,giuliano07,giulianoepl,giuliano09}. An important step in extending 
this approach to a YSC  has recently been provided in  
Ref.\cite{r.1}, where it has been shown how, when applying the JW transformation to the interface 
between an XX-chain and a quantum Ising chain with Ising axis rotated with respect to the $z$-axis (in 
spin space) of the XX-chain, the interface is mapped onto the interface between  a spinless normal 1D conductor and a p-wave superconductor, 
with the phase of the order parameter
equal to twice the tilting angle of the Ising axis.  In the low-energy, long-wavelength limit, once the system
parameters are pertinently chosen, an interface as such stabilizes perfect  Andreev reflection on the normal side, 
which is the same as connecting a ''truly'' fermionic system to a bulk superconductor at fixed phase. As a result, 
this becomes an effective mechanism to induce an equilibrium, nonzero spin current pattern across 
the YSC.

In this paper we first develop an effective field   theory describing the low-energy, long-wavelength limit of 
a junction of $N$ XX-spin chains connected to ''outer'' quantum Ising chains with different tilting angles for the Ising 
axis. Therefore, we use the result to analyze the scaling properties of the TKE arising at 
a three-chain junction. 

Technically, we  argue how, in perfect analogy with the derivation done in 
Refs.\cite{gaf_1,gaf_2} for a normal metal-superconductor 
interface, for a long enough XX-chains, each terminal Ising chain may be traded 
for a pertinent boundary interaction Hamiltonian, only depending 
on an emerging Majorana mode $\gamma$ and on the tilting angle of the corresponding Ising axis. 
In the low-energy, long-wavelength limit, we  prove that the emerging 
Majorana mode stabilizes perfect  Andreev reflection for JW fermions at the interface, with a phase
shift equal to twice the tilting angle of the Ising axis. As a result,  the different tilting angles of the Ising chains work as applied phases 
at the endpoints of the XX-chains, thus inducing   a nonzero equilibrium spin current pattern
across the junction. 

To describe how  the spin currents are affected by   the TKE, we first  map our system  onto  
a Y junction of three quantum Ising chains, with, in general, boundary couplings all different from each other, 
and all explicitly depending on the applied phases. Therefore, combining the RG approach to the 
(anisotropic) TKE, which eventually provides the running couplings as functions of the bare couplings 
and of the running scale, with the functional dependence of the bare couplings on the applied phases, 
we recover the running couplings as a function of the applied phases. This allows us to derive 
scaling formulas for the system  groundstate energy and, by differentiating the energy with respect 
to the applied phases, to derive  scaling formulas for the spin currents across the junction.

Compared to the expected 
scaling of the currents as the first inverse power of the length of the leads $\ell$ \cite{acz,siano,giuso_1,transfer,ngcg}, the TKE induces 
a crossover  in the form of an  upturn in the currens as $\ell \sim \ell_K$. Probing such a crossover would on one hand provide a direct 
evidence of the emergence of the TKE at the YSC, on the other hand it would yield a direct measurement 
of $\ell_K$.

In addition, we prove that $\ell_K$ itself is  a known function of the applied 
phases, whose functional form can be readily inferred from the explicit solution of the RG equations for 
the running couplings. In fact, this is possibly the main advantage of measuring $\ell_K$ in 
our YSC, compared to other Kondo systems. Acting on the applied phases, we  may tune in a controlled 
way the bare couplings and, therefore, we  may in principle tune  $\ell_K$ at will. So, 
$\ell_K$ itself becomes a {\it tunable parameter}, which we may act on by pertinently 
varying the applied phases, that is, the tilting angles of the Ising axes. 

This result eventually leads to  two complementary ways to probe $\ell_K$ in our system. 
Indeed, it is possible to either look at the scaling of the currents with $\ell$ at fixed applied phases, or 
to alternatively fix $\ell$ and vary the phases by, therefore,  tuning $\ell_K$ accordingly. 
In particular, the second method    allows  for recovering the  scaling by tuning $\ell_K$, {\it without changing $\ell$}, 
which is the hardest thing to achieve in a real-life system.

The paper is organized as follows:

\begin{itemize}

 \item In Sec.\ref{modham} we introduce   the model Hamiltonian for a junction of $N$ spin chains connected to each 
other at one of their endpoints. Each chain is modeled as an ``inner'' quantum XX-chain of length $\ell$ connected at an 
``outer'' quantum Ising chain with a tilted Ising axis. Eventually, consistently with the derivation of 
Refs.\cite{gaf_1,gaf_2}  we trade the outer chains for pertinent boundary Hamiltonians localized  at 
the interfaces;

\item In Sec.\ref{n2} we introduce   our method for computing the spin current using the simple example of the $N=2$ junction 
at fixed applied phases. This is equivalent to a single, inhomogeneous spin chain  and, therefore, in principle it does not 
require introducing KFs to resort to JW fermions. For this reason, we   extensively     
use  the $N=2$ chain as a testground of 
our method, showing how it enables us to recover all the known results for a single chain connected to two superconductors at
fixed phase difference \cite{acz,siano,giuso_1};

\item In Sec.\ref{n3} we extend the derivation of Sec.\ref{n2} to the $N=3$ junction, particularly evidencing the emergence of  
 the KFs in the boundary Hamiltonian describing the junction;

\item In Sec.\ref{emek} we derive   the RG equations for the running boundary couplings
in the $N=3$ junction; 

\item In Sec.\ref{sck} we use the results of Sec.\ref{emek} to derive the spin current pattern at the onset 
of the Kondo regime, showing  the explicit dependence of $\ell_K$ on the applied phases and, 
therefore, its tunability, in some 
simple cases in which it can be explicitly derived in a closed-form;

\item In Sec.\ref{concl} we summarize the main results of  our   work;

\item We provide the mathematical details of our derivation in the various Appendices.

 \end{itemize}

\section{Model Hamiltonian for the  junction of $N$ spin chains}
\label{modham}

According to Ref.\cite{r.1} we describe  each quantum spin chain  
by means of a  one-dimensional, inhomogeneous lattice quantum spin Hamiltonian $H_{\rm SC}^\lambda$ over an $L$ site lattice with   
 open boundary conditions at the endpoint at $j=L$. Therefore, denoting with $\lambda$ the chain index, we set

\beq
H_{\rm SC}^\lambda  = -  \sum_{j =1}^{L -1} \: \{ ( t_j + \gamma_j ) \hat{\sigma}^m_{j , \lambda}  \hat{\sigma}^m_{j+1 , \lambda}  
+ ( t_j - \gamma_j ) \hat{\sigma}^n_{j , \lambda} \hat{\sigma}_{j+1 , \lambda}^n \} -  \sum_{ j = 1}^L g_j \sigma_{j , \lambda}^z 
\:\:\:\: , 
\label{e.1}
\eneq
with the parameters chosen as detailed below:

\begin{itemize}

\item {The isotropic contribution to the magnetic exchange $t_j$:}

\beq
t_j = \Biggl\{ \begin{array}{l} J \;\;\; , \;\; (1 \leq j \leq \ell - 1 ) \\
J' \;\;\; , \;\; (j = \ell ) \\
t \;\;\; , \;\; ( \ell + 1 \leq j \leq L-1 ) 
\end{array}
\:\:\:\: ; 
\label{eq.2}
\eneq
\noindent

\item {The anisotropic contribution to the magnetic exchange $\gamma_j$:}

\beq
\gamma_j = \Biggl\{ \begin{array}{l} 0 \;\;\; , \;\; (1 \leq j \leq \ell  ) \\
\gamma \;\;\; , \;\; ( \ell + 1 \leq j \leq L-1 ) 
\end{array}
\:\:\:\: ; 
\label{eq.3}
\eneq
\noindent

\item {The applied transverse field $g_j$:}
 
\beq
g_j = \Biggl\{ \begin{array}{l} H \;\;\; , \;\; (1 \leq j \leq \ell  ) \\
g \;\;\; , \;\; ( \ell + 1 \leq j \leq L  ) 
\end{array}
\:\:\:\: ; 
\label{eq.4}
\eneq
\noindent

\item {The (in-plane) projected spin operators:}

\begin{eqnarray}
\hat{\sigma}_{j , \lambda}^m &=& \hat{m}_{j , \lambda}  \cdot \vec{\sigma}_{j , \lambda}  = \cos ( \phi_{j , \lambda}  ) \sigma^x_{j , \lambda} + \sin ( \phi_{j , \lambda} ) \sigma^y_{j , \lambda} = 
e^{- i \phi_{j , \lambda} } \sigma_{j , \lambda}^+ + e^{ i \phi_{j , \lambda} } \sigma_{j , \lambda}^- \nonumber \\
\hat{\sigma}_{j , \lambda}^n &=& \hat{n}_{j , \lambda}  \cdot \vec{\sigma}_{j , \lambda}  = - \sin ( \phi_{j , \lambda}  ) \sigma^x_{j , \lambda} +
 \cos ( \phi_{j , \lambda}  ) \sigma_{j , \lambda}^y = 
- i e^{ - i \phi_{j , \lambda} } \sigma_{j , \lambda}^+ + i e^{i \phi_{j , \lambda} } \sigma_{j , \lambda}^- 
\:\:\:\: ,
\label{eq.5}
\end{eqnarray}
\noindent
with 

\beq
\phi_{j , \lambda} = \Biggl\{ \begin{array}{l} 0 \;\;\; , \;\; 1 \leq j \leq \ell \\
\phi_\lambda \;\;\; , \;\; \ell + 1 \leq j \leq L 
\end{array}
\label{eq.6}
\eneq
\noindent
(note that, differently from all the other parameters, to induce a nonzero spin current pattern 
through the junction, we  choose the phase $\phi$ to be 
dependent on the  chain index $\lambda$). 
\end{itemize} 

The chains are connected to each other at the  $j=1$ endpoint. This defines the actual 
junction, which is described by the boundary Hamiltonian $H_\Delta$,  given by 

\beq
H_\Delta = - J_\Delta \: \sum_{ \lambda = 1}^N \: \{\sigma_{1,\lambda}^x \sigma_{1 , \lambda  + 1 }^x + 
 \sigma_{1,\lambda}^y \sigma_{1 , \lambda  + 1 }^y \} 
\:\:\:\: , 
\label{eq.7}
\eneq
\noindent
with $N$ being the number of chains and $\lambda + N \equiv \lambda$.

\begin{figure} 
\includegraphics*[width=.8\linewidth]{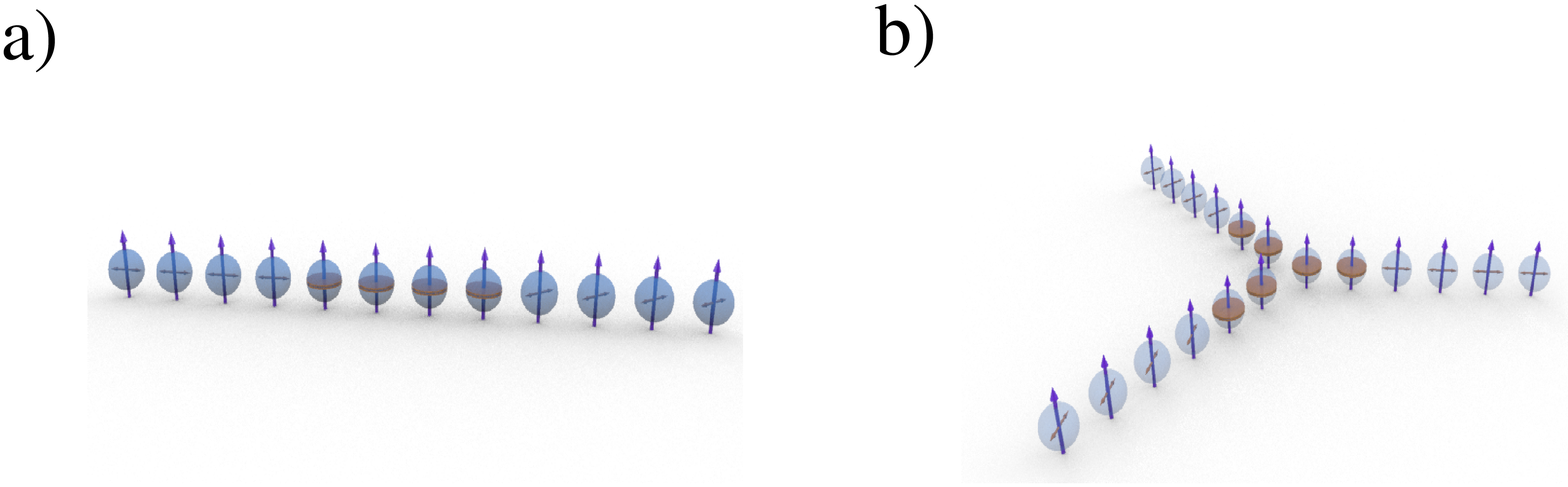}
\caption{ {\bf a)} Sketch of a single chain with inhomogeneous parameters
corresponding (via the Jordan-Wigner transformation) to a spinless, SNS junction. 
Following the drawing code of Ref.\cite{r.1}, each sphere represents a quantum 
spin. Spheres with the equatorial plane colored in red represent spins interacting 
with an isotropic magnetic interaction lying within the XY-plane in spin space (the 
XX-part of the whole chain), while spheres with just one colored segment within the 
equatorial plane correspond to spins with an Ising interaction directed along the segment.
The tilting angle between the Ising interaction axes in the two external leads is 
mapped onto the phase difference between the superconducting leads of the SNS junction;
{\bf b)} Sketch of the $N=3$ junction analyzed in the paper, realized with 
three inhomogeneous spin chains, each one  consisting of an (inner) XX-part joined to 
an (outer) Ising part, with the applied phases to each chain defined by the 
direction of the Ising interaction between the spins.   } 
 \label{cartoon}
\end{figure}

In Fig.\ref{cartoon}, we provide a sketch of a single, inhomogeneous chain, 
and of the junction with $N=3$ chain, to which we devote our attention in 
this paper.

 \noindent
 
 The equilibrium spin current through a chain is obtained as  the average   of the
 $z$-component of the spin current density operator,   $j_{ [j , j+1 ] ; \lambda}^z$. This is 
 a link operator, which can be derived from the continuity equation for the spin 
 density operator at a site $j$ ($1 \leq j < \ell$). Indeed, from the  Heisenberg 
equations of motion for the lattice spin operator, we   obtain

\beq
\partial_t \sigma_{j , \lambda}^z = - i [ \sigma_{j , \lambda}^z , H_{{\rm SC} , \lambda}  ] = j^z_{[ j , j+1 ] ; \lambda } - j^z_{[ j- 1 , j] ; \lambda}
\:\:\:\: , 
\label{e.8}
\eneq
\noindent
with 
\beq
j_{[ j , j+1] ; \lambda }^z = J \: \{ \sigma_{j , \lambda}^x \sigma_{j+1 , \lambda }^y - \sigma_{j , \lambda}^y \sigma_{j+1 , \lambda}^x \}
\:\:\:\: . 
\label{e.9}
\eneq
\noindent
 
To map the  quantum spin-1/2 spin chain onto  
 an equivalent spinless fermion model, we employ the generalized JW  transformation  
introduced  in Ref.\cite{crampettoni}. This requires introducing   as many KFs $\eta_\lambda$  
as many chains, and  setting   \cite{crampettoni,tsve_1,giuso_x,gsst}

\begin{eqnarray}
\sigma_{ j , \lambda}^+  &=& i \eta_\lambda c_{j , \lambda}^\dagger e^{ i \pi \sum_{ t = 1}^{j-1} c_{t , \lambda}^\dagger c_{t , \lambda} } \nonumber \\
\sigma_{ j , \lambda}^-  &=& i \eta_\lambda c_{j , \lambda}  e^{ i \pi \sum_{ t = 1}^{j-1} c_{t , \lambda}^\dagger c_{t , \lambda} } \nonumber \\
\sigma_{j ,\lambda}^z &=& c_{j , \lambda}^\dagger c_{ j , \lambda}- \frac{1}{2}
\:\:\:\: .
\label{3sc.1}
\end{eqnarray}
\noindent

In Eq.(\ref{3sc.1}),  $\{c_{ j , \lambda}  , c_{j , \lambda}^\dagger \}$ ($j=1,\ldots, L ; \lambda= 1, \ldots ,N$) is  a set of $L\times N$ spinless lattice 
fermion operators, while the  Klein factors $\eta_\lambda$ are  fermion operators satisfying the 
anticommutation algebra 

\begin{eqnarray}
&& \{\eta_\lambda , \eta_{\lambda'} \} = 2 \delta_{\lambda, \lambda'} \nonumber \\
&& \{ \eta_\lambda  , c_{j , \lambda'} \} = \{ \eta_\lambda , c_{j , \lambda'}^\dagger \} = 0 
\:\:\:\: . 
\label{qua.1}
\end{eqnarray}
\noindent
Upon inserting the JW  formulas into the (''bulk'') Hamiltonian operators in 
Eq.(\ref{e.1}) the Klein factors cancel. The corresponding   
Hamiltonian for the $\lambda$-chain   is given by

\begin{eqnarray}
H_{\rm SC}^\lambda &=& - J \sum_{j = 1}^{\ell - 1} \: \{ c_{j , \lambda}^\dagger c_{j + 1 , \lambda } + 
c_{j + 1 , \lambda}^\dagger c_{ j , \lambda} \} - H \sum_{ j = 1}^\ell c_{j , \lambda}^\dagger c_{j , \lambda}
- J' \{ c_{ \ell , \lambda} ^\dagger c_{ \ell + 1 , \lambda} + c_{ \ell + 1 , \lambda}^\dagger c_{ \ell , \lambda }\}
\nonumber \\
&-& t \sum_{ j = \ell + 1}^{L-1} \: \{ c_{ j , \lambda}^\dagger c_{ j + 1 , \lambda} + c_{ j + 1 , \lambda}^\dagger c_{ j , \lambda} \}
- \gamma \sum_{ j = \ell + 1}^{L-1} \: \{ c_{ j , \lambda} c_{ j + 1 , \lambda} e^{ - 2 i \phi_\lambda} 
+ c_{j+1, \lambda}^\dagger c_{ j , \lambda}^\dagger e^{ 2 i \phi_\lambda} \} - g \sum_{ j = \ell + 1}^L 
c_{j , \lambda}^\dagger c_{ j , \lambda} 
\:\:\:\: . 
\label{qua.2}
\end{eqnarray}
\noindent
In terms of  JW  fermions, the right-hand side of Eq.(\ref{qua.2}) describes   a junction between a normal wire 
(ranging from $j=1$ to $j=\ell$), and a p-wave 
topological superconductor (ranging from $j= \ell + 1$ to $j=L$). 

To further simplify our derivation,  in the following we employ the ``long-$\ell$'' approximation of Refs.\cite{gaf_1,gaf_2}, 
by trading the lead Hamiltonian in Eq.(\ref{qua.2})  for a simple 
 boundary Hamiltonian depending on the degrees of freedom in the ``normal'' part of the chain \cite{gaf_1,gaf_2}, as 
well as on the emerging, ``Majorana-like'' zero-mode operator at the endpoint of the superconducting 
lead \cite{kita_1}. To better ground such an approximation, in Appendix \ref{simple}, we exactly derive 
the boundary Hamiltonian  in the   limit
  $\gamma = t$ and $g = 0$ on the ''p-wave'' side of the junction (that, for 
the chain $\lambda$, is given by 
$H_{{\rm F} , \lambda}$ in Eq.(\ref{quax.7}) of Appendix \ref{simple}). As a result, the  ``bulk'' Hamiltonian of the system 
in fermionic representation takes the form \cite{kita_1}

\beq
H_{\rm Bulk} = \sum_{ \lambda = 1}^N H_{{\rm F} , \lambda}
= \sum_{ \lambda =1}^N \left\{  - J \sum_{ j = 1}^{\ell - 1 } [  c_{ j , \lambda}^\dagger c_{ j + 1 , \lambda} + 
c_{ j + 1 , \lambda}^\dagger c_{ j , \lambda} ]  - H \sum_{ j = 1}^\ell c_{ j , \lambda}^\dagger c_{ j , \lambda} 
+ i \frac{ \tau}{2}  \gamma_\lambda  \: [  e^{ -  i \phi_\lambda}  c_{ \ell , \lambda} + e^{ i \phi_\lambda} c_{\ell , \lambda}^\dagger  ] \right\}
\:\:\:\: , 
\label{qua.3}
\eneq
\noindent 
with $\tau$ and the real, zero-energy Majorana mode $\gamma_\lambda$ defined   in Appendix \ref{simple}.

Resorting to the JW fermion for $H_\Delta$, we  obtain  
 
\beq
H_\Delta = J_\Delta \: \sum_{ \lambda = 1}^N \: [ i \eta_\lambda \eta_{ \lambda + 1} ] \: \{ i [ c_{ 1 , \lambda}^\dagger c_{1 , \lambda + 1 } - 
c_{1 , \lambda + 1}^\dagger c_{ 1 , \lambda } ] \} 
\:\:\:\: , 
\label{qua.4}
\eneq
\noindent
with, again, $\lambda + N \equiv \lambda$. Eq.(\ref{qua.4}) shows that, differently from what happens with 
$H_{\rm Bulk}$, the KFs do contribute to $H_\Delta$. In particular, for $N=3$ we  obtain  
a special case of  the topological Kondo Hamiltonian at a junction of the three quantum spin  \cite{tsve_1,giuso_x,gsst}. 

Once expressed in terms of JW fermions, the current density $j_{[ j , j+1 ] ; \lambda}^z$ is given by

\beq
j_{[ j , j+1 ] ; \lambda}^z = - 2 i J \: \{ c_{j , \lambda}^\dagger  c_{j+1 , \lambda} - c_{j + 1 , \lambda}^\dagger  c_{j , \lambda} \}
\:\:\:\: . 
\label{qua.5}
\eneq
\noindent
Using the continuity equation over the link $[ \ell , \ell + 1 ]$, we  eventually find  that, under stationary conditions, 
the average value of $j_{[ j , j+1 ] ; \lambda}^z $ is the same as the average value, over the reference state, of the 
operator $I_\lambda$, defined as 

\beq
I_\lambda = \frac{\tau}{2} \gamma_\lambda \: \{ e^{ - i \phi_\lambda} c_{ \ell , \lambda} - 
e^{ i \phi_\lambda} c_{ \ell , \lambda}^\dagger \} = \frac{\partial H_{B , \lambda} }{\partial \phi_\lambda } 
\:\:\:\: . 
\label{qua.6}
\eneq
\noindent
 Eq.(\ref{qua.6}) provides a straightforward  way  to derive  the equilibrium current pattern 
through the junction by just differentiating the groundstate energy with respect to the 
applied phases. Therefore, in the following we systematically use 
Eq.(\ref{qua.6}) to evaluate the currents.

Before concluding this Section, it is worth stressing how, in general, 
we expect  that  the periodicity of the spin equilibrium current through a chain depends  on whether the number of site 
$\ell$ is even, or odd. The analysis of the even-odd chains is detailed  in Appendix \ref{evenodd}. For the sake of simplicity, 
in the following   we will be focusing onto the even-$\ell$ case only.  

\section{Spin supercurrent through the $N=2$-junction}
\label{n2}

Before analyzing  the $N=3$ YSC, in this Section we illustrate our approach to computing  the equilibrium spin current 
using the example of  the $N=2$ junction. Indeed,  the 
 junction between two spin chains is   equivalent to a single, inhomogeneous spin chain, with 
 the   Hamiltonian being  exactly solvable, with no need of introducing the KFs. 
 
In JW fermionic coordinates, the  $N=2$  junction  Hamiltonian, $H_\Delta^{(2)}$,  is given by

\beq
H_\Delta^{(2)}  = J_\Delta \: [ i \eta_1 \eta_{ 2} ] \: \{ i [ c_{ 1 , 1}^\dagger c_{1 , 2 } - 
c_{1 , 2}^\dagger c_{ 1 , 1} ] \} 
\:\:\:\: , 
\label{quak.4}
\eneq
\noindent
with the bulk Hamiltonian $H_{\rm Bulk}^{(2)} = \sum_{\lambda = 1,2}  H_{\rm SC}^\lambda$, 
and $H_{\rm SC}^\lambda$ given in Eq.(\ref{qua.2}).  

  $H_\Delta^{(2)}$ in Eq.(\ref{quak.4}) is the only term, in the junction Hamiltonian, 
containing the KF's in the product $i \eta_1 \eta_2$. Rewriting this operator as $2 \xi^\dagger \xi - 1$, 
with $\xi = \frac{\eta_1 + i \eta_2 }{2}$, we see that it commutes with the whole Hamiltonian and that 
its eigenvalues are $\pm 1$. Accordingly, for all the practical purposes, it can be dropped from 
$H_\Delta^{(2)}$ and substituted with $\pm 1$. As a double check of the conclusion that KFs are 
unessential for $N=2$, we should  verify that the final result for the equilibrium spin supercurrent is 
independent of the sign of $J_\Delta$. 

 After dropping  $[ i \eta_1 \eta_{ 2} ]$,  the  boundary Hamiltonian, 
as well as the  ``bulk'' Hamiltonian describing the chains, 
are both quadratic in the fermion operators; as a result, they   can be exactly diagonalized
and the spin current can be evaluated.

 \begin{figure} 
\includegraphics*[width=.95\linewidth]{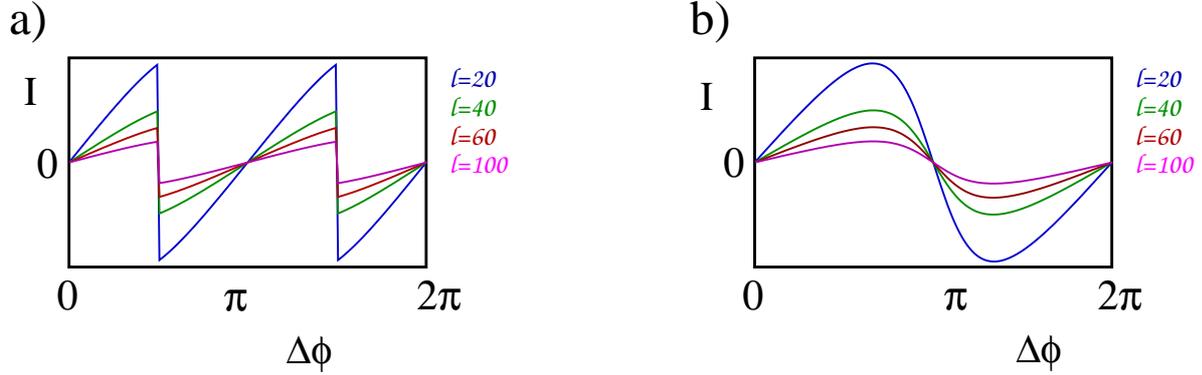}
\caption{  {\bf a)} $I [ \Delta \phi]$ as a function of $\Delta \phi$ through an $N=2$-junction with $J=1$
computed by exact diagonalization with 
$J_\Delta = 0.4$, $\tau =1$, $\mu = 0.05$, and $\ell = 20$ (blue curve), $\ell =40$ (green curve),
$\ell = 60$ (red curve), $\ell = 100$ (magenta curve). At fixed $J_\Delta$ basically   the curve
collapse onto each other, once they are rescaled by $\ell$. Here we are assuming that
fermion parity is not conserved, which yields finite jumps in $ I [ \Delta \phi]$ at the level-crossing 
values of the phase difference, $\Delta \phi = \frac{\pi}{2} + k \pi$, $k$ integer, by accordingly 
setting to $\pi$ the period of $I [ \Delta \phi ]$;  
{\bf b)} Same as in panel {\bf a)}, but now assuming fermion parity conservation. There are accordingly 
no more jumps at  $\Delta \phi = \frac{\pi}{2} + k \pi$ and the full periodicity of $2 \pi$ in
$\Delta \phi$ has been restored.} 
 \label{plot_1}
\end{figure}
 \noindent
 
In Fig.\ref{plot_1} we provide a sample of the results for the equilibrium spin
current through the junction. To derive the current,  we numerically perform the exact diagonalization of
the real-space Hamiltonian.    As a result, we find that $I_1 = - I_2$ and that, as expected, both currents 
only depend on the phase difference $\Delta \phi = \phi_1 - \phi_2$. 

In computing the spin current, an important point to address is whether  the total JW 
fermion parity (that is, the $z$-component of the total spin) is conserved, or not. 
 To account for both possibilities,  in Fig.\ref{plot_1}, we draw $I_1 [ \Delta \phi ] \equiv  I  [ \Delta \phi ]$ 
as a function of $\Delta \phi$, computed, both by assuming that fermion parity is not 
conserved (Fig.\ref{plot_1}{\bf a)}), and by assuming that fermion parity is 
conserved (Fig.\ref{plot_1}{\bf b)}), for the values of the  parameters reported in 
the figure caption.  In drawing all the plots we have set $\tau = 1$. For $\tau=1$ we recover  pure-Andreev 
reflection at both boundaries  as soon as 
$\ell \geq \frac{2 \pi J^2 \sin (k_F ) }{\tau^2} \sim 5$.  Accordingly, to describe the 
results of Fig.\ref{plot_1} we may safely rely on the field-theory approach  developed in 
Appendix \ref{lelw} by approximating the lattice fermion operator $c_{j , \lambda}$ 
at time $t$ ($\lambda = 1,2$) as 

\beq
c_{ j , \lambda } (t ) \approx e^{ - i \phi_\lambda} \: \{ e^{ i k_F j} \psi_\lambda ( x_j - \ell - v t )
- e^{ - i k_F j } \psi_\lambda^\dagger ( \ell - x_j - v t ) \}
\:\:\:\: , 
\label{n2.1}
\eneq
\noindent
with $\psi_\lambda ( x - v t )$ being chiral fermionic fields, 
$- 2 J \cos ( k_F ) - H = 0$, $v_F = 2 J \sin ( k_F )$ and with $x_j = a j$, $a$ being the lattice step (which we set to 1 henceforth). 
Inserting  Eq.(\ref{n2.1}) into Eq.(\ref{quak.4}) and getting rid, of the operator $[ i \eta_1 \eta_{ 2} ]$, we  reexpress  $H_\Delta^{(2)}$
in terms of the continuum field operators as 

\beq
H^{(2)}_\Delta \to   i J_\Delta \: \{ e^{ i \Delta \phi} \: [ e^{ - i k_F} \psi_1^\dagger ( - \ell ) - e^{ i k_F} \psi_1 ( \ell ) ] 
[ e^{  i k_F} \psi_2 ( - \ell ) - e^{ - i k_F} \psi_2^\dagger ( \ell ) ]  - {\rm h.c.} \}  
\:\:\:\: . 
\label{n2.2}
\eneq
\noindent

The operator at the right-hand side of Eq.(\ref{n2.2}) is bilinear in the local fermionic fields 
at $x=\pm \ell $, and  it corresponds to a purely marginal perturbation, not inducing any scaling 
with $\ell$ in the boundary operator itself. Therefore, we expect  no additional scaling in 
$ I [ \Delta \phi ]$, besides the one with $\ell^{-1}$ that characterizes the equilibrium
supercurrent across a noninteracting fermionic system \cite{acz,siano,giuso_1,transfer,ngcg}. 
Apparently, this is fully consistent with  the plots we draw in  Fig.\ref{plot_1} 
at different values of $\ell$. 

 Regarding fermion parity conservation we note that, 
 in  a ``fermionic'' SNS junction, the conservation of the total 
fermion parity is   expected to hold, especially in the absence of gapless,  
Fermi liquid-like, quasiparticle baths and/or in the presence of ``fast'' variations
in time of the system parameters, which do not allow the system to relax toward the 
actual minimum energy state, at the cost of changing its total fermion parity. 
At variance, in a spin system, fermion parity corresponds to the total spin conservation along the $z$-axis, 
which can be readily broken by means of, e.g., impurities, local magnetic field fluctuations, 
etc.

The non conservation of fermion parity leads to   the discontinuity  in 
$ I [ \Delta \phi ]$ at $\Delta \phi = \frac{\pi}{2} + k \pi$.  To discuss this point,  we rely on  
the formalism of Appendix \ref{lelw}. In particular, considering   the weak coupling limit  $J_\Delta / J \ll 1$,
we note that  we may consistently assume that both chains terminate 
at $j=1$ (open boundary conditions). The allowed energy eigenvalues in each chain are therefore
determined by solving Eqs.(\ref{lollolove.z3}) of Appendix \ref{lelw}. These always 
 take a zero-energy solution, with the 
Bogoliubov-de Gennes (BDG) wavefunction in chain-$\lambda$ given by 

\beq
\left[ \begin{array}{c}
u_{ j ; 0 ; \lambda}          \\
v_{ j ; 0 ; \lambda} 
       \end{array} \right] = 
       \frac{1}{\sqrt{2 ( \ell + 1) }}  \: \left[ \begin{array}{c}
i e^{ - i \phi_\lambda} \: \sin ( k_F j ) \\
- i e^{  i \phi_\lambda} \: \sin ( k_F j )
                                 \end{array} \right]
\:\:\:\: .
\label{n2.3}
\eneq
\noindent 
The corresponding zero-mode operators, $\Gamma_{0 ; \lambda}$, are therefore given by 

\beq
\Gamma_{0 ; \lambda} =   \frac{e^{ - i \phi_\lambda} }{\sqrt{ 2 ( \ell + 1) }} \: \sum_{ j = 1}^\ell \: \{ i \sin ( k_F j ) \: [ c_{ j , \lambda} - c_{ j , \lambda}^\dagger ] \} 
- \frac{2 J e^{ - i \phi_\lambda}  \sin [ k_F ( \ell + 1 ) ] }{\tau \sqrt{2 ( \ell + 1 ) }}  \: \gamma_\lambda
\:\:\:\: .
\label{2n.4}
\eneq
\noindent

Aside from  the over-all phase factor $e^{ - i \phi_\lambda}$, $\Gamma_{0 ; \lambda}$ is a real fermion operator. When considering the two 
(still disconnected) chains all together, the two real zero-modes $\Gamma_{0 ; 1}$ and $\Gamma_{0 ; 2}$, can be 
combined into a complex fermionic zero-mode operator $a_0 = \frac{1}{2} \; \{ \Gamma_{ 0 ; 1 } + i \Gamma_{0 ; 2 } \}$. 
In the disconnected limit, the $N=2$-junction spectrum is twofold degenerate, with the two degenerate states (for each 
energy eigenvalue) corresponding to the mode $a_0$ being empty, or full (that is, with different JW fermion parity). 
On turning on the interaction, a finite hybridization between the 
zero-mode operators at the two chains sets in, with a strength proportional to $J_\Delta$ and modulated by
$\Delta \phi$. In fact, this can be readily inferred from Eq.(\ref{n2.2}) by truncating the mode decomposition of the 
fermion field operators to the zero-modes, thus getting  the ''restricted'' Hamiltonian  involving the zero-mode 
operators, given by

\beq
H_{\Delta ; 0}^{(2)} = \frac{4 i \sin^2 ( k_F ) J_\Delta \cos ( \Delta \phi ) }{ \ell + 1 } \: \Gamma_{0 ; 1} \Gamma_{0 ; 2} 
=  \frac{4  \sin^2 ( k_F ) J_\Delta \cos ( \Delta \phi ) }{ \ell + 1 } \: \{ 2 a_0^\dagger a_0 - 1 \}
\:\:\:\: . 
\label{2n.5}
\eneq
\noindent
From Eq.(\ref{2n.5}) we   see  that, for $0 \leq \Delta \phi < \frac{\pi}{2}$, the actual   groundstate
corresponds to having the  $a_0$-mode empty. At variance, for 
$\frac{\pi}{2} < \Delta \phi \leq \pi$, the groundstate corresponds to  the $a_0$-mode filled by one JW fermion, 
with  opposite fermion parity. If fermion parity is not conserved, then 
the level crossing at $\Delta \phi = \frac{\pi}{2}$ implies a finite discontinuity in $I [ \Delta \phi ]$, which is the feature 
evidenced in the plots of Fig.\ref{plot_1}{\bf a)}. At variance, if fermion parity is conserved,  the finite jump is substituted by a smooth, continuous curve, 
determined by the impossibility for the system to undergo the switch toward the ``true'' groundstate at $\Delta \phi = \frac{\pi}{2}$
without changing the total fermion parity, as it appears in Fig.\ref{plot_1}{\bf b)} \cite{pik,gaf_3,transfer}. 

\begin{figure} 
\includegraphics*[width=.8\linewidth]{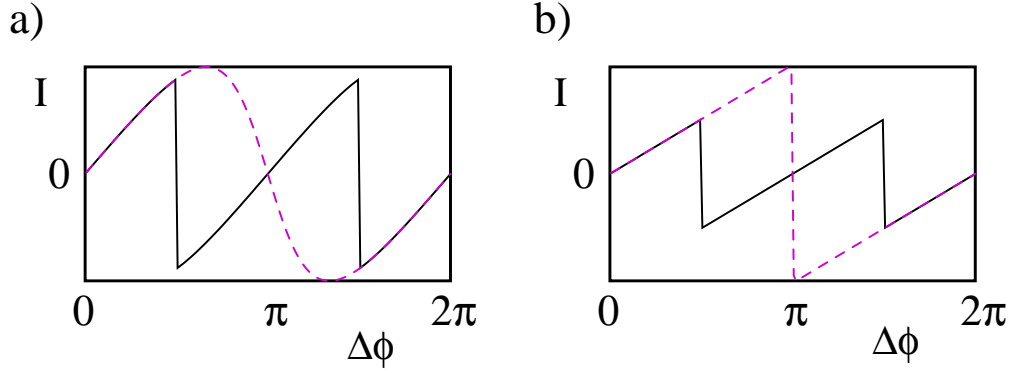}
\caption{  {\bf a):} $I [ \Delta \phi]$ as a function of $\Delta \phi$ through an $N=2$-junction with $J=1$
computed by exact diagonalization with $\tau =1$, $\mu = 0.05$, $\ell = 100$, and 
$J_\Delta = 0.4$ in the case in   fermion parity  is not preserved (black full curve), as well as in the 
case in which it is preserved (dashed magenta curve);  
{\bf b):} Same as in panel {\bf a)}, but with $J_\Delta = 1.0$.   } 
 \label{plot_2}
\end{figure}

An additional comment is in order to deal with the periodicity of $I [ \Delta \phi ]$ as a function of $\Delta \phi$ both in 
the case in which the fermion parity ${\cal P}$ is not conserved, as well as in the case in which it is conserved. In the former case, $I [ \Delta \phi ]$ 
is periodic with period equal to $\pi$, that is, to the minimal interval of values of $\Delta \phi$ separating two consecutive level 
crossings as described by Eq.(\ref{2n.5}) (see also the analysis of Appendix \ref{evenodd} for a comprehensive discussion of this point). In the latter case, 
the periodicity is restored back to $2 \pi$, as we display in Fig.\ref{plot_2}, where we draw a synoptic plot of $I [ \Delta \phi ]$ for 
the same values of the system parameters and of $\ell$, but computed with, and without, assuming that ${\cal P}$ is conserved. 
The two periodicities are halved, with respect to what we expect  to get in the case of a   fermionic quantum wire 
between two  topological superconductors at fixed phase difference, which is  expected,  
 as a consequence of the  JW transformation applied to the quantum spin chain \cite{r.1}.

\begin{figure} 
\includegraphics*[width=1.\linewidth]{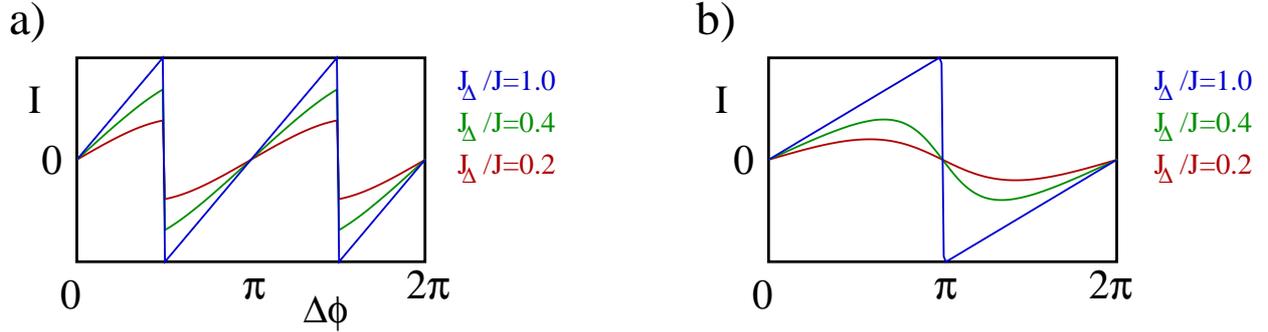}
\caption{ {\bf a):} $I [ \Delta \phi]$ as a function of $\Delta \phi$ through an $N=2$-junction with $J=1$
computed by exact diagonalization by assuming that  fermion parity is not conserved, 
with $\tau =1$, $\mu = 0.05$, $\ell = 100$, and 
$J_\Delta = 1.0$ (blue curve), $J_\Delta=0.4$ (green curve),
$J_\Delta=0.2$ (red curve);   {\bf b):} The same as in panel {\bf a)}, but now by assuming that 
  fermion parity is conserved. The crossover from a sinusoidal to a sawtooth-like dependence of 
$I [ \Delta \phi ]$ on $\Delta \phi$ on increasing $J_\Delta / J$ is apparent.    } 
 \label{plot_3}
\end{figure}

 \noindent
Finally, while, as 
$H_\Delta^{(2)}$ is a truly marginal interaction, with no induced running of the coupling strengths,  tuning ''by hand'' $J_\Delta / J$, 
it is still possible to  trigger a crossover between the sinusoidal dependence of $I [ \Delta \phi ]$ on $\Delta \phi$, which 
is typical of the weakly coupled regime $J_\Delta / J \ll 1$, and the sawtooth one, which takes place when $J_\Delta \sim J$ \cite{glazrak,giuso_1}.  To address 
this (non-dynamically induced) crossover in $I [ \Delta \phi]$, in Fig.\ref{plot_3} we draw $I [ \Delta \Phi ]$ as a 
function of $\Delta \phi$ for the same values of the parameters as we used to draw Fig.\ref{plot_1}; we consider both cases in which   
the fermion parity is not conserved (Fig.\ref{plot_3}{\bf a)}), and  is conserved (Fig.\ref{plot_3}{\bf b)}).
We set $\ell = 100$ and vary $J_\Delta / J$, as discussed in the caption. The crossover from the sinusoidal to the sawtooth
behavior is apparent, whether ${\cal P}$ is conserved, or not.

 \noindent
We now
move to discussing the $N=3$ junction, in which the KFs are expected to 
play a crucial role in determining the emergence of the topological Kondo effect \cite{crampettoni,tsve_1,gsst}.

\section{Effective Hamiltonian and groundstate structure of the  $N=3$-junction}
\label{n3}

Differently from the $N=2$-junction,  the $N=3$-junction is not 
exactly solvable, due to the nontrivial effect of the KFs $ \{ \eta_\lambda \}$ on the boundary interaction. 
  Indeed, in this case the  KFs combine into an effective impurity spin-1/2 degree of freedom,
thus determining  a peculiar realization of the TKE at our junction. The TKE emerges in our system 
just as at   a junction of three quantum Ising chains \cite{tsve_1,gct}, or of  three XX chains \cite{crampettoni},
of three one-dimensional  Josephson junction arrays \cite{giuso_x}, and of three XY chains   \cite{gsst}.  
For  this reason, we attack the problem by means of the standard RG approach to a boundary impurity model, 
within the field theory framework developed in Appendix \ref{lelw}. 

In terms of the continuum fermionic fields, the junction Hamiltonian, $H^{(3)}_\Delta$, is given by 
 
 \beq
H^{(3)}_\Delta  =   J_\Delta \: \sum_{\lambda = 1}^3 \: \{ e^{ i [ \phi_\lambda - \phi_{\lambda + 1} ] } \: [ i \eta_\lambda 
\eta_{\lambda + 1}] 
i [ e^{ - i k_F} \psi_\lambda^\dagger ( - \ell ) - e^{ i k_F} \psi_\lambda ( \ell ) ] 
[ e^{  i k_F} \psi_{\lambda + 1}  ( - \ell ) - e^{ - i k_F} \psi_{\lambda + 1}^\dagger ( \ell ) ]  + {\rm h.c.} \} 
\:\:\:\: , 
\label{n3.1}
\eneq
\noindent
with $\lambda + 3 \equiv \lambda$. 

  The key feature of $H^{(3)}_\Delta $ in Eq.(\ref{n3.1}) is the explicit dependence of the Kondo interaction 
on the phase differences $\phi_\lambda - \phi_{\lambda + 1}$. This induces a   dependence on the applied phases 
in the groundstate energy of our system. Thus, when differentiating the groundstate energy with respect to 
the applied phases, one has a  nonzero equilibrium spin current pattern through the junction. Monitoring the spin current at 
different scales provides an effective tool to map out the phase diagram of the system. 

The idea of probing the phase diagram of junctions of one-dimensional systems by measuring the equilibrium current pattern 
through the system has been largely exploited in the literature regarding junctions of one-dimensional Josephson junction arrays \cite{giuso_1,giulianoepl,giuso_x,giuliano09,cmgs}.
Here, we show how our approach extends this technique  to junctions of quantum spin chains, by means of a pertinent generalization of 
the methods developed in Ref.\cite{r.1} for a single spin chain. 

In the  weak coupling limit, $J_\Delta / J \ll 1$, we assume  open boundary conditions for the lattice fields $c_{j , \lambda}$ at the inner boundary, that is, 
 $c_{j=0,\lambda} = 0$, $\forall \lambda$. As a result,   Eq.(\ref{n3.1}) becomes 

\beq
H^{(3)}_\Delta = 4 \sum_{ \lambda = 1}^3 \:  J_{\lambda , \lambda+1} \sin^2 ( k_F )   \: [ i \eta_\lambda \eta_{\lambda +1} ] 
[ i \xi_\lambda ( 0 ) \xi_{ \lambda + 1} ( 0 ) ] 
\:\:\:\: , 
\label{n3.2}
\eneq
\noindent
with $\xi_\lambda ( x )$ being chiral real fermionic fields and $J_{\lambda , \lambda + 1} = 
J_\Delta \: \cos [ \phi_\lambda - \phi_{\lambda + 1} ] $.  $H^{(3)}_\Delta$ in Eq.(\ref{n3.2}) 
corresponds to the  (in general anisotropic) Kondo interaction arising at a junction of 
three quantum Ising chains \cite{tsve_1,gsst}. The anisotropy is determined by 
the phase differences and, for large enough values of the phase differences, the 
interaction strengths can even take different signs.  

To set up the field theory approach to the interacting boundary problem defined by $H^{(3)}_\Delta$
in Eq.(\ref{n3.2}), we have to first construct the system's groundstate by pertinently taking 
into account the emerging  real-fermion zero-mode operators $\Gamma_{0 ; \lambda}$, as well
as the possible degeneracy associated to different eigenvalues of the total fermion parity operator. 
To do so, we single out the zero-mode contribution to the mode expansion of the 
$\xi_\lambda$ fields at the right-hand side of Eq.(\ref{n3.2}), by writing the 
corresponding contribution to $H_\Delta^{(3)}$,   $H^{(3)}_{ \Delta ; 0 }$, as

\beq
  H^{(3)}_{ \Delta ; 0 } =   4  \sum_{ \lambda = 1}^3 \: \frac{ J_{\lambda , \lambda+1} \sin^2 ( k_F )  }{ \ell + 1 } \: [ i \eta_\lambda \eta_{\lambda +1} ] 
[ i \Gamma_{0 ; \lambda}  \Gamma_{0 ; \lambda + 1} ] 
\:\:\:\: . 
\label{n3.3}
\eneq
\noindent

$ H^{(3)}_{ \Delta ; 0 }$ in Eq.(\ref{n3.3}) describes a dipole interaction between two effective spin-1/2 spin operators. 
A key point is that,  naively rewriting it down as 
 $H^{(3)}_{ \Delta ; 0 }  \to \sum_{\lambda = 1}^3 G_\lambda \: \sigma_\eta^\lambda \sigma_\Gamma^\lambda$,
 with $\sigma_\eta^\lambda$ and $ \sigma_\Gamma^\lambda$ being Pauli matrices acting over orthogonal spaces and 
 $G_\lambda$ being pertinently defined constants, would 
lead to an incorrect state counting (6 independent real Majorana modes would correspond to 3 pairs of 
complex Dirac modes, together with their Hermitean conjugate, which would yield a total of 8 independent states.
At variance, the construction with the Pauli matrices would imply a total of 4 independent states). 
In fact, the correct way of realizing the fermion operators entering $H^{(3)}_{ \Delta ; 0 } $ is 
provided by a straightforward generalization of the Lee-Wilczek 
construction \cite{lwilk}, which we reformulate and adapt to our model  in Appendix \ref{lw}.

In order to properly diagonalize   $H^{(3)}_{ \Delta ; 0 }$ in Eq.(\ref{n3.3}), 
following the derivation of Appendix \ref{lw}, we define the state $ | a_\Gamma , a_\eta \rangle_\gamma$ 
so that 

\begin{eqnarray}
 && \sigma_\Gamma^3 |  a_\Gamma , a_\eta \rangle_\gamma = [ i \Gamma_{0 ; 1} \Gamma_{ 0 ; 2}  ]  |  a_\Gamma , a_\eta \rangle_\gamma = 
 a_\Gamma  |  a_\Gamma , a_\eta \rangle_\gamma \nonumber \\
  && \sigma_\eta^3 |  a_\Gamma , a_\eta \rangle_\gamma = [ i \eta^1 \eta^2  ]  |  a_\Gamma , a_\eta \rangle_\gamma = 
 a_\eta  |  a_\Gamma , a_\eta \rangle_\gamma \nonumber \\
 && {\cal P}_\tau   |  a_\Gamma , a_\eta \rangle_\gamma \ = \gamma  |  a_\Gamma , a_\eta \rangle_\gamma 
 \;\;\;\; , 
 \label{n3.4}
\end{eqnarray}
\noindent
with ${\cal P}_\tau = P_\Gamma \cdot P_\eta$,
$P_\Gamma$  being the fermion parity associated to the triple $ \{ \Gamma_{ 0 ; 1 } , \Gamma_{0 ; 2 } , \Gamma_{0 ; 3 } \} $,  
 $P_\eta$  being the fermion parity associated to the triple $ \{ \eta^1, \eta^2  , \eta^3  \} $, 
and $a_\Gamma , a_\eta , \gamma = \pm 1$. In addition, consistently with the mode expansion of Eq.(\ref{lollolov.extra}) of 
Appendix \ref{lelw}, we set $\xi_{n , \lambda}    |  a_\Gamma , a_\eta \rangle_\gamma = 0$, $\forall n > 0$ and 
$\forall a_\Gamma , a_\eta , \gamma , \lambda$. As we show in Eqs.(\ref{lw.13}), what is the actual groundstate 
of $ H^{(3)}_{ \Delta ; 0 }$ (plus the bulk Hamiltonian in the disconnected junction limit)
depends on the relative values of the coupling strengths $J_{\lambda , \lambda + 1}$ and, in particular, on their sign. 
Expressing $J_{\lambda , \lambda + 1}$ in terms of the   independent phase 
differences $\Delta \phi_a = \phi_1 - \phi_2$ and $\Delta \phi_b = \phi_1 - \phi_3$, from Eqs.(\ref{lw.13}) we 
obtain  that the groundstate  has 
energy ${\cal E}_3^{(0)} [ \Delta \phi_a , \Delta \phi_b ]$, given by 

\begin{eqnarray}
&& {\cal E}_3^{(0)} [ \Delta \phi_a , \Delta \phi_b ] = - \frac{ 4 J_\Delta \sin^2 ( k_F) }{\ell + 1}
\: \times \nonumber \\
&& {\rm max}_{ \lambda_{a} , \lambda_b  = \pm 1  } \: \{ \lambda_a \cos (  \Delta \phi_a ) 
+  \lambda_b \cos ( \Delta \phi_b ) +  \lambda_a \lambda_b  \cos ( \Delta \phi_a - \Delta \phi_b )   \} 
\:\:\:\: . 
\label{n3.5}
\end{eqnarray}
\noindent

From the discussion above, we conclude that there are two states, corresponding to different 
values of $\gamma$, that minimize the energy, for  each choice of $\lambda_a$ and $\lambda_b$. 
Therefore, whether, on varying $\Delta \phi_a$ and/or $\Delta \phi_b$, at a level crossing for 
the groundstate, the system remains within the initial states or ``jumps'' into the 
actual groundstate, is not a matter of whether the fermion parity  is conserved, or not, but rather of whether the 
system is allowed to crossover from, e.g., the singlet state at the first line of 
Eq.(\ref{lw.13}) to the triplet state at the last line of the same equation. 
In the latter case, the equilibrium spin currents within each one of the three chains, $I_1 [ \Delta \phi_a , \Delta \phi_b ] , 
I_2 [ \Delta \phi_a , \Delta \phi_b ] , I_3 [ \Delta \phi_a , \Delta \phi_b ]$ are respectively given by 
(to leading order in $J_\Delta$)

\begin{eqnarray}
 I_1 [ \Delta \phi_a , \Delta \phi_b ] &=& \frac{ \partial {\cal E}_3^{(0)} [ \Delta \phi_a , \Delta  
 \phi_b ]   }{\partial \Delta \phi_a} + 
 \frac{ \partial  {\cal E}_3^{(0)} [ \Delta \phi_a , \Delta \phi_b ]   }{\partial \Delta \phi_b} = 
 \frac{ 4 J_\Delta \sin^2 ( k_F) }{\ell + 1}  [ \lambda_a \sin  (  \Delta \phi_a ) 
+   \lambda_b \sin ( \Delta \phi_b ) ] \nonumber \\
 I_2 [ \Delta \phi_a , \Delta \phi_b ] &=& -  \frac{ \partial {\cal E}_3^{(0)} [ \Delta \phi_a , \Delta  
 \phi_b ]   }{\partial \Delta \phi_a} = -  \frac{ 4 J_\Delta \sin^2 ( k_F) }{\ell + 1}  [  \lambda_a \sin  (  \Delta \phi_a ) 
+  \lambda_a \lambda_b   \sin ( \Delta \phi_a - \Delta \phi_b ) ] \nonumber \\
  I_3 [ \Delta \phi_a , \Delta \phi_b ] &=& -  \frac{ \partial {\cal E}_3^{(0)} [ \Delta \phi_a , \Delta   
 \phi_b ]   }{\partial \Delta \phi_b} = -  \frac{ 4 J_\Delta \sin^2 ( k_F) }{\ell + 1}  [   \lambda_b \sin  (  \Delta \phi_b ) 
-  \lambda_a \lambda_b    \sin ( \Delta \phi_a - \Delta \phi_b ) ]
\:\:\:\: .
\label{n3.6}
\end{eqnarray}
\noindent 
In the former case, instead, the currents are determined by just the initial state of
the system, which sets once, and forever, the values of $\lambda_a$ and $\lambda_b$, regardless
of $\Delta \phi_a$ and $\Delta \phi_b$. In a real life experiment, fluctuations, local fields, 
impurities, as well as Landau-Zener like transitions induced by nonadiabatic changes in 
the applied phases \cite{laze}, are likely to favor the scenario described by Eqs.(\ref{n3.6}). 
Yet, for the sake of completeness, in the following we 
keep discussing both scenarios, when possible. As a main remark, 
it is worth pointing out that, for any choice of $\lambda_a , \lambda_b$ (and, therefore, 
both when the system keeps within its true ground state, or it does not), the 
currents in Eqs.(\ref{n3.6}) are consistent with ``Kirchoff law'' at the junction, 
$\sum_{\lambda = 1}^3  I_\lambda [ \Delta \phi_a , \Delta \phi_b ]  = 0$.

We note that the current pattern in 
Eqs.(\ref{n3.6}) might look like what one  would expect at a junction between three spinless normal conducting 
wires connected to three topological superconductors at fixed 
phases of the superconducting leads. However, in this latter case, changing $\ell$ would 
simply result in a rescaling of   $I_\lambda [ \Delta \phi_a , \Delta \phi_b ]$ with 
$\ell^{-1}$. Eventually, including the effects of the dynamical, finite-energy bulk modes of 
the wires would just provide a slight change in the functional dependence of the currents on 
$\Delta \phi_a , \Delta \phi_b$, without affecting  the scaling with $\ell^{-1}$. Instead, as we 
discuss in the following, in 
a junction between spin chains, TKE does affect the scaling 
properties of the currents, as $\ell$ becomes of the order of $\ell_K$.

\begin{figure} 
\includegraphics*[width=.4\linewidth]{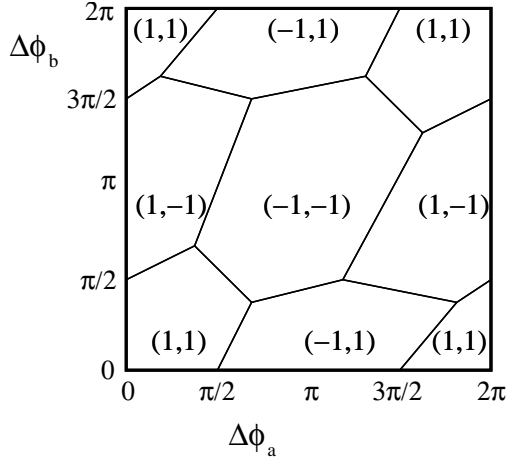}
\caption{Regions in the $\Delta \phi_a -  \Delta \phi_b$ plane in which the 
current pattern through the $N=3$-junction is determined by Eqs.(\ref{n3.6})
with the values of $(\lambda_a , \lambda_b) = (\pm 1 , \pm 1)$ reported in the figure. Due to the periodicity 
of the currents in both $\Delta \phi_a$ and $\Delta \phi_b$, 
we limit the plot to the square $0 \leq \Delta \phi_a , \Delta \phi_b \leq 2 \pi$. 
If the system keeps within its actual groundstate when crossing a borderline between different 
regions, the currents 
are expected to exhibit finite discontinuities at the crossings, similar to what happens to $I [\Delta \phi ]$ in the $N=2$ junction
at $\Delta \phi = \frac{\pi}{2}$.    } 
 \label{regions}
\end{figure}

To provide a synoptic view of the changes in the groundstate of the system as functions of 
the applied phases,   in Fig.\ref{regions} we report the regions in the   $\Delta \phi_a - \Delta \phi_b $-plane 
corresponding to different values of $\lambda_a,  \lambda_b$. Assuming that the current 
pattern through the junction is always determined by the ``actual'' groundstate of the system, 
Fig.\ref{regions} also provides a synoptic view of how the branches of the  currents in Eqs.(\ref{n3.6})
vary depending on the applied phase differences $\Delta \phi_a , \Delta \phi_b$.

 \noindent
We now resort to  the RG approach,  to  
discuss  the   nonperturbative effects that arise when $\ell \sim \ell_K$.

\section{Renormalization group approach to the topological  Kondo effect at the $N=3$ junction}
\label{emek}

To implement the RG approach, we resort to the imaginary time framework and  describe
the boundary interaction in terms of the imaginary time action $S_\Delta^{(3)} = \int_0^\beta \: d \tau \: H_\Delta^{(3)} ( \tau)$, 
with $H_\Delta^{(3)} ( \tau)$ being the boundary action in the interaction representation at imaginary time $\tau$ and 
$\beta = (k_B T)^{-1}$.
To  carefully  take into account the role of 
the zero-mode operators,  we   write the  operator 
$\xi_\lambda ( 0 )$ at imaginary time $\tau$,  $  \xi_\lambda ( \tau  )$, as

\beq
\xi_\lambda ( \tau ) = \frac{\Gamma_{0 ; \lambda}}{\sqrt{\ell + 1}} + \bar{\xi}_\lambda ( \tau ) 
\:\:\:\: . 
\label{ek.1}
\eneq
\noindent
 From Eq.(\ref{ek.1}) we obtain  
$S_\Delta^{(3)} = S_{\Delta ; 1}^{(3)} + 
S_{\Delta ; 2}^{(3)} + S_{\Delta ; 3}^{(3)}$, with 

\begin{eqnarray}
 S_{\Delta ; 1}^{(3)} &=& \frac{4 \sin^2 ( k_F) }{\ell + 1} \: \sum_{ \lambda = 1}^3 J_{\lambda , \lambda + 1} \: \int_0^\beta \: d \tau \: [ i \eta_\lambda ( \tau ) 
 \eta_{\lambda + 1} ( \tau ) ] \: [ i \Gamma_{ 0 ; \lambda} ( \tau ) \Gamma_{ 0 ; \lambda + 1} ( \tau ) ] \nonumber \\
  S_{\Delta ;2}^{(3)} &=& \frac{4 \sin^2 ( k_F) }{\sqrt{\ell + 1}}  \: \sum_{ \lambda = 1}^3 J_{\lambda , \lambda + 1} \: \int_0^\beta \: d \tau \: [ i \eta_\lambda ( \tau ) 
 \eta_{\lambda + 1} ( \tau ) ] \:i \{ \Gamma_{ 0 ; \lambda} ( \tau ) \bar{\xi}_{\lambda + 1} ( \tau ) + 
 \bar{\xi}_\lambda ( \tau ) \Gamma_{ 0 ; \lambda + 1} ( \tau ) \} \nonumber \\
  S_{\Delta ; 3}^{(3)} &=&  4 \sin^2 ( k_F)   \: \sum_{ \lambda = 1}^3 J_{\lambda , \lambda + 1} \: \int_0^\beta \: d \tau \: [ i \eta_\lambda ( \tau ) 
 \eta_{\lambda + 1} ( \tau ) ] \: [ i  \bar{\xi}_\lambda ( \tau )  \bar{\xi}_{\lambda + 1} ( \tau ) ] 
 \:\:\:\: . 
 \label{ek.3}
\end{eqnarray}
\noindent
Out of the three contributions in Eqs.(\ref{ek.3}),    $S_{\Delta ; 1}^{(3)}$ is exactly 
accounted for by   diagonalizing $H^{(3)}_{ \Delta ; 0 }$ and by determining the groundstate accordingly. 
 The interaction between the zero-modes and the dynamical modes of the $\xi_\lambda$ fields, as well as the interaction between
the dynamical modes themselves, provides a nontrivial renormalization to the $J_{\lambda , \lambda + 1}$ and, therefore, 
to the groundstate energy. 

To explicitly  derive the corresponding RG  equations, we   introduce  a high-energy cutoff $D_0 \sim 2J$ and 
 then we progressively reduce $D_0$  to $D = D_0 - \delta D$, by integrating over the modes lying in the energy windows between
$-D_0$ and $-D_0 + \delta D$ and $D_0 - \delta D$ and $D_0$. To do so, we 
write down   the 
partition function ${\cal Z}$ as 
\beq
{\cal Z} = {\cal Z}_0 \: \langle {\bf T}_\tau \: e^{ - S_{\Delta ; I}^{(3)} } \rangle_0
\:\:\:\: , 
\label{ek.4}
\eneq
\noindent
with $\langle \ldots \rangle_0$ denoting averaging over the bulk action at disconnected junction 
plus  $S_{\Delta ; 1}^{(3)}$, $S_{\Delta ; I}^{(3)} =   S_{\Delta ;2}^{(3)} + S_{\Delta ;3}^{(3)}$ and 
${\cal Z}_0$ being the partition function of the ``unperturbed'' system (with only 
$ S_{\Delta ;1}^{(3)}$ as a nonzero boundary action).

Expanding ${\cal Z}$ up to second-order in the $J_{\lambda , \lambda + 1}$, we have to  perform the contractions
leading to the terms that renormalize  $S_{\Delta ; 1}^{(3)}$. In doing so, we have to 
pay particular attention to the correlation function of the Klein factors, 
$~_\gamma \langle \psi_1 |  \eta^\lambda ( \tau_1 ) \eta^\lambda ( \tau_2 ) | \psi_1 \rangle_\gamma$. Specifically, 
using Lehman's representation for the correlation function in combination with 
the results of Appendix \ref{lw} and assuming that $ | \psi_1 \rangle_\gamma$ is the groundstate of 
the junction, we obtain  

\beq
~_\gamma \langle \psi_1 |  \eta^\lambda ( \tau_1 ) \eta^\lambda ( \tau_2 ) | \psi_1 \rangle_\gamma
= e^{ - \left[  \frac{ 8 \sin^2 ( k_F )  J_{\lambda + 1 , \lambda + 2}}{\ell + 1 }  \right] \: ( \tau_1 - \tau_2 ) } 
\:\:\:\: . 
\label{ekk.00}
\eneq
\noindent
In addition, we need  the finite-temperature, imaginary time ordered correlation function of the dynamical modes,
which is  given by

\beq
\langle {\bf T}_\tau \bar{\xi}_\lambda ( \tau )  \bar{\xi}_{\lambda'} ( \tau' ) \rangle = - \frac{2 \delta_{\lambda , \lambda'}}{\frac{\beta}{\pi} \sin \left[ \frac{\pi }{\beta}
( \tau - \tau' ) \right] } 
\:\:\:\: . 
\label{ek.2}
\eneq
\noindent
Using the result of Eq.(\ref{ekk.00},\ref{ek.2}), introducing 
the short-imaginary-time distance cutoff $\tau_c$ and rescaling $\tau_c$ to $\tau_c + \delta \tau_c$, we
find that  $S_{\Delta ;1}^{(3)}$ is corrected by a term  $\delta S_{\Delta ;1}^{(3)}$ given by 

\beq
\delta S_{\Delta ;1}^{(3)} = - \frac{4  [ 4 \sin^2 ( k_F ) ]^2 \: \delta \tau_c}{\tau_c ( \ell + 1 ) } \: 
\sum_{ \lambda = 1}^3 \: J_{\lambda , \lambda + 1} J_{\lambda + 1 , \lambda + 2} \: \int_0^\beta \: d \tau \: 
e^{ - \left[  \frac{ 8 \sin^2 ( k_F )  J_{\lambda + 1 , \lambda + 2} \tau_c }{\ell + 1 }  \right]  } \: [ i \eta_{\lambda + 2 } ( \tau ) \eta_\lambda ( \tau ) ]
\: [ i \Gamma_{ 0 , \lambda + 2} ( \tau ) \Gamma_{ 0 , \lambda} ( \tau ) ] 
\:\:\:\: . 
\label{ekk.01}
\eneq
\noindent
 From Eq.(\ref{ekk.01}), we eventually  infer  the RG equations for 
the running couplings by looking at how the cutoff-dependent corrections vary as a function of the cutoff itself.
As a result, defining the dimensionless running couplings $G_{\lambda , \lambda + 1}$ as 

\beq
G_{\lambda , \lambda + 1} = \frac{ 4 \sin^2 ( k_F ) J_{\lambda , \lambda +1} }{ v}
= 2 \sin ( k_F ) \frac{J_{\lambda , \lambda + 1}}{J}
\;\;\;\; , 
\label{intermezzo}
\eneq
\noindent
we  obtain  the RG equations for the running  couplings, given by 
 
\begin{eqnarray}
\frac{d G_{1 , 2}}{d l} &=& e^{ - \frac{  2   | G_{1,2} | }{\ell + 1}   } \: G_{2 , 3} G_{3 , 1}  \equiv \hat{\beta}_{1,2} [ \{ G_{\lambda , \lambda +1} \} ]  \nonumber \\
\frac{d G_{2 , 3}}{d l} &=&  e^{ -  \frac{  2  | G_{2,3} |  }{\ell + 1}   } \:G_{3 , 1} G_{1 , 2}  \equiv \hat{\beta}_{2,3} [ \{ G_{\lambda , \lambda +1} \} ]  \nonumber \\
\frac{d G_{3 , 1}}{d l} &=&  e^{ -   \frac{   2   | G_{3,1 } |  }{\ell + 1}    } \:G_{1 , 2} G_{2 , 3} \equiv \hat{\beta}_{3,1} [ \{ G_{\lambda , \lambda +1} \} ] 
\:\:\:\: , 
\label{ek.5a}
\end{eqnarray}
\noindent
with    $l = \ln \left( \frac{D_0}{D} \right)$,  and 
$D \sim \pi v / \ell$ being the running energy scale. Note that, in writing Eqs.(\ref{ek.5a}), we have 
set $\tau_c = a / v$, with $a$ being the lattice step. Importantly, we note that the same equations arise when 
deriving the renormalization of the coupling strengths in  $S_{\Delta ;3}^{(3)}$ to second order in 
the $J_{\lambda , \lambda + 1}$. Also, as we have introduced the absolute values of the running couplings 
at the exponents of the right-hand side of Eqs.(\ref{ek.5a}), they hold regardless the system groundstate
corresponds to $ | \psi_1 \rangle_\gamma$, or to any other of the states listed in Eqs.(\ref{lw.13}).

An important observation is that,   as long as $ | G_{\lambda , \lambda + 1 } ( D )  | / (\ell + 1 )  \ll 1$, 
we may neglect the exponential factors at the right-hand side of Eqs.(\ref{ek.5a}), so that they reduce to 

\begin{eqnarray}
\frac{d G_{1 , 2}}{d l} &=&   G_{2 , 3} G_{3 , 1}  \equiv \beta_{1,2} [ \{ G_{\lambda , \lambda +1} \} ]  \nonumber \\
\frac{d G_{2 , 3}}{d l} &=&  G_{3 , 1} G_{1 , 2}  \equiv \beta_{2,3} [ \{ G_{\lambda , \lambda +1} \} ]  \nonumber \\
\frac{d G_{3 , 1}}{d l} &=&  G_{1 , 2} G_{2 , 3} \equiv \beta_{3,1} [ \{ G_{\lambda , \lambda +1} \} ] 
\:\:\:\: . 
\label{ek.5}
\end{eqnarray}
\noindent
Eqs.(\ref{ek.5}) are the standard RG equations for the topological Kondo effect \cite{tsve_1}. In appendix \ref{rgsol} we discuss in 
detail the general features of the solutions of Eqs.(\ref{ek.5}) for the various possible sign 
assignment of the bare couplings. Here, we focus onto the specific consequences of Eqs.(\ref{ek.5})  for our junction.

As a first observation, we note that, except for some specific lines 
in parameter space (see the next Section and  Appendix \ref{rgsol} for details),
Eqs.(\ref{ek.5}) always imply a flow toward the Kondo fixed point. In particular, 
to double-check the validity of the approximation leading to Eqs.(\ref{ek.5}), we note that
  the energy splitting between the  groundstate of $H_{\Delta ; 0}^{(3)}$ and 
its first excited state is of order of $\bar{\epsilon}_G = 8 | J_{\lambda , \lambda + 1} | \sin^2 (k_F ) / (\ell + 1 )$, 
while the energy required to excite a ``dynamical'' mode of $\bar{\xi}_\lambda ( \tau )$ is, 
instead, as large as $\delta \epsilon = \pi v / ( \ell + 1 ) = 2 \pi J \sin ( k_F ) / ( \ell + 1 )$. As 
a result, we find  that $ \bar{\epsilon}_G / \delta \epsilon \sim | G_{\lambda , \lambda + 1} |/ \pi$.
Accordingly, as long as  $| G_{\lambda , \lambda + 1} | \ll 1$ (that is, within the perturbative regime),
finite-energy, dynamical modes of $\xi_\lambda$ lie pretty higher in energy than the excited states of 
$H_{\Delta ; 0}^{(3)}$. This enables us to derive the spin currents just as we have done in 
Sec.\ref{n3}, by simply substituting the bare Kondo couplings with the renormalized (running) ones. 
 
The running couplings depend on $\Delta \phi_a , \Delta \phi_b$ via their initial values $G_{\lambda , \lambda + 1}^{(0)}$. Thus, it is in principle 
straightforward to derive the spin current pattern from the integral curves of Eqs.(\ref{ek.5a}) by 
just differentiating with respect to the phase differences.  
This picture breaks down   at the scale $\tilde{D}$ at which 
$ | G_{\lambda , \lambda + 1} ( \tilde{D} )  | \sim 1$. This condition is a signal of the onset 
of the nonperturbative regime and, accordingly, we identify  $\tilde{D}$ with $D_K$. 
As a result, we conclude that the (improved) formula expressing the spin current pattern across the 
junction in terms of derivatives of the running couplings with respect to the phase differences
holds all the way down to  $D \sim  D_K$.

To infer the behavior of the junction near the strongly coupled Kondo fixed point, we 
note that, as we point out in 
Appendix \ref{rgsol}, the anisotropy between the (absolute values of the) boundary 
couplings is in general suppressed along the RG  trajectories. 
For this reason, we temptatively construct the effective boundary Hamiltonian at 
the strongly coupled Kondo fixed point by pertinently adapting the derivation 
done in   Ref.\cite{gsst} in the isotropic case. Specifically, our Kondo Hamiltonian corresponds to the 
 realization of the two-channel spin-1/2 Kondo model discussed in 
Refs.\cite{tci,giuta}. At the fixed point, this exhibits a 
remarkable ``fractional degeneracy'' \cite{adestri,luda_1,luda_2,luda_3,luda_4},
which is encoded in the emergence of two energy degenerate total spin 
singlet groundstates at the strongly coupled fixed point, $ | {\bf \Sigma } \rangle_1 , | {\bf \Sigma } \rangle_2$
\cite{tci,giuta,gsst}.

As discussed above, the isotropic fixed point is expected to faithfully describe also the 
strongly coupled regime  for  boundary couplings different from 
each other. Since the differences in the boundary 
couplings are directly related to their dependence on the applied phases, we readily conclude that all the spin currents 
through the junction must be equal to zero at the Kondo fixed point. 
In order to build the leading boundary 
operator at the strongly coupled fixed point, 
  we assume that
close to, but not exactly at, the  Kondo fixed point, the coupling strengths 
keep (slightly) different from each other. Therefore,  we repeat the construction of  Appendix A of Ref.\cite{gsst}, 
getting, as final result, the boundary perturbation that, in terms of the lattice fields $\{ c_{ j , \lambda} \}$, 
is given by 

\beq
H_{\Delta ; {\rm Sc} }^{(3)} = i {\bf V}^y \: \left\{ \sum_{\lambda = 1}^3 \: \frac{3 J^3}{  [ J_{\lambda +1 , \lambda + 2} + J_{\lambda + 2 , \lambda } ]^2 } \right\}
\: \prod_{\lambda  = 1}^3 \: [ c_{2,\lambda}^\dagger + c_{2, \lambda} ] 
\:\:\:\: , 
\label{kfp.1}
\eneq
\noindent
with  ${\bf V}^y$ acting on the degenerate singlets as $ {\bf V}^y | {\bf \Sigma } \rangle_{1,2} = \mp i | {\bf \Sigma } \rangle_{2,1}$, and 
the operators $ \{ c_{1 , \lambda } , c_{1 , \lambda}^\dagger \}$ hybridized with the topological spin determined 
by the Klein factors into either one of the degenerate singlets \cite{tci,gsst}. Since the lattice field operators at 
$j=1$ are hybridized with the topological spin operator, in order to resort to the analog of the 
low-energy, long-wavelength expansion in Eq.(\ref{lollolov.ter}), we have to impose open boundary conditions on 
the lattice fields at $j=2$. Once the boundary conditions corresponding to perfect Andreev reflection at the outer 
boundaries are accounted for, as well, Eq.(\ref{kfp.1}) yields, in the continuum field theory framework

\beq
H_{\Delta ; {\rm Sc} }^{(3)} \to  i {\bf V}^y \: \left\{ \sum_{\lambda = 1}^3 \: \frac{3 [ \sin ( k_F ) J]^3}{  [ J_{\lambda +1 , \lambda + 2} + J_{\lambda + 2 , \lambda } ]^2 } \right\}
\: \prod_{\lambda  = 1}^3 \: \xi_\lambda ( 0 )   
\:\:\:\: . 
\label{kfp.2}
\eneq
\noindent
The operator at the right-hand side of Eq.(\ref{kfp.2}) has scaling dimension $d=\frac{3}{2}$. It is, therefore, a strongly irrelevant operator
in the infrared. Thus, its effects, including a possible nonzero contribution to the spin currents, are expected to vanish 
as we let the system flow to the Kondo fixed point. In fact, in order to evaluate such a contribution, we should know the specific 
dependence of the $J_{\lambda , \lambda + 1}$ on $ \Delta \phi_a , \Delta \phi_b$ in the strongly coupled regime. In principle, this 
could be derived by, e.g., employing techniques such as the ones developed in Refs.\cite{exact_1,exact_2}. However, this lies outside of 
the scope of this work, as we eventually show how the peculiar scaling properties of the spin current pattern through the junction
at the onset of the nonperturbative Kondo  regime provide an effective way of monitoring the emergence of the topological Kondo 
effect at our $N=3$ junction of quantum spin chains.

In the following, we  provide a guideline about how to do so by discussing
a few, simple,  paradigmatic cases of interest.

\section{Spin current pattern at the onset of the Kondo regime}
\label{sck}
 
We explicitly solve  Eqs.(\ref{ek.5})  in Appendix \ref{rgsol} where we show that, for 
generic values of the $G_{\lambda , \lambda + 1}^{(0)}$,   the solution is  
expressed in terms of  the incomplete elliptic integral in Eq.(\ref{eko.3}). 
At the same time, we show how the solution is   remarkably  simplified if two of the three bare couplings are equal to each other, 
say $G_{\lambda + 1 , \lambda + 2}^{(0)} = G_{\lambda + 2 , \lambda}^{(0)}$. In this case, 
since $G_{\lambda + 1 , \lambda +2}^2 ( l ) - G_{\lambda + 2  , \lambda  }^2 ( l )$ is 
constant along the RG  trajectories, we find  that 
$G_{\lambda + 1 , \lambda +2}  ( l ) =  G_{\lambda + 2  , \lambda  } ( l )$ at any 
scale $l$. This extra constraint allows for providing explicit, closed-form formulas for 
the solution of Eqs.(\ref{ek.5}), which we discuss in detail in Appendix \ref{rgsol}. Using those 
solutions with appropriate values for the initial boundary couplings  $G_{\lambda , \lambda +1}^{(0)}$,
in the following we explicitly derive the scaling of the spin currents for $\ell \leq \ell_K$ in two paradigmatic  cases.

The first case corresponds to setting  $\phi_1 = \phi_2 \neq \phi_3$, which implies $\Delta \phi_a = 0 , 
\Delta \phi_b = \phi_1 - \phi_3 \neq 0$. In this case, we obtain  

\beq
G_{1,2}^{(0)} = G_\Delta \;\;\; , \;\; G_{2,3}^{(0)} = G_{3,1}^{(0)} = G_\Delta \cos ( \Delta \phi_b ) 
\:\:\:\: , 
\label{skc.1}
\eneq
\noindent
with $G_\Delta = \frac{4 \sin^2 ( k_F ) J_\Delta}{v}$. Accordingly, 
Eqs.(\ref{ek.5}) reduce to a set of two differential equations, 
given by 

\begin{eqnarray}
 \frac{d G_{1,2} }{d l} &=& G_{2,3}^2 \nonumber \\
 \frac{ d G_{2,3} }{d l } &=& G_{1,2} G_{2,3} 
 \:\:\:\: .
 \label{skc.2}
\end{eqnarray}
\noindent
Solving Eqs.(\ref{skc.2}) by using,  as a running parameter, $l =   \ln \left( \frac{\ell}{\ell_0} \right)$, with 
$\ell$ being the chain length and $\ell_0$ a reference scale, we obtain  

\begin{eqnarray}
 G_{1,2} \left[ l = \ln \left( \frac{\ell}{\ell_0} \right) \right] &=& G_\Delta \sin ( \Delta \phi_b ) 
 \: \left\{ \frac{1 + \sin ( \Delta \phi_b ) + [ 1 - \sin ( \Delta \phi_b ) ] \left( \frac{\ell}{\ell_0} \right)^{2 G_\Delta \sin ( \Delta \phi_b )}  }{
 1 + \sin ( \Delta \phi_b ) - [ 1 - \sin ( \Delta \phi_b ) ] \left( \frac{\ell}{\ell_0} \right)^{2 G_\Delta \sin ( \Delta \phi_b )} }  \right\} \nonumber \\
  G_{2,3} \left[ l = \ln \left( \frac{\ell}{\ell_0} \right) \right] &=& G_\Delta \sin ( \Delta \phi_b ) 
 \:  \left\{ \frac{ 2 \cos ( \Delta \phi_b ) \left( \frac{\ell}{\ell_0} \right)^{  G_\Delta \sin ( \Delta \phi_b )}  }{
 1 + \sin ( \Delta \phi_b ) - [ 1 - \sin ( \Delta \phi_b ) ] \left( \frac{\ell}{\ell_0} \right)^{2 G_\Delta \sin ( \Delta \phi_b )} }  \right\}  
 \:\:\:\: . 
 \label{skc.3}
\end{eqnarray}
\noindent
Apparently, Eqs.(\ref{skc.3}) imply that, at any value of $\Delta \phi_b \neq \frac{\pi}{2} + k \pi$, with $k$ integer, either 
all three the running couplings are $>0$, or two of them are $<0$, the third being $>0$. As we discuss in Appendix 
\ref{rgsol}, this implies a flow towards the Kondo fixed point in both cases. This is evidenced by the explicit solutions
at the right-hand side of Eqs.(\ref{skc.3}), which let us identify the $\Delta \phi_b$-dependent Kondo length $\ell_K [ \Delta \phi_b ]$ given by 

 \beq
\ell_K [ \Delta \phi_b ] = \ell_0 \: \left\{ \frac{1 + | \sin ( \Delta \phi_b ) |   }{1 - | \sin ( \Delta \phi_b ) | }  \right\}^{\frac{1}{2 G_\Delta | \sin ( \Delta \phi_b ) | } }
\:\:\:\: . 
\label{skc.4}
\eneq
\noindent
Inserting Eqs.(\ref{skc.4}) into Eqs.(\ref{skc.3}), we get  the expected scaling of the running couplings
with $\ell / \ell_K [ \Delta \phi_b ]$ \cite{hewson}, that is

\begin{eqnarray}
 G_{1,2} \left[ l = \ln \left( \frac{\ell}{\ell_0} \right) \right] &=& G_\Delta | \sin ( \Delta \phi_b ) |  
 \: \left\{ \frac{1  +  \left( \frac{\ell}{\ell_K [ \Delta \phi_b ] } \right)^{2 G_\Delta | \sin ( \Delta \phi_b ) | }  }{
 1    -  \left( \frac{\ell}{\ell_K [ \Delta \phi_b ] } \right)^{2 G_\Delta | \sin ( \Delta \phi_b ) | } }  \right\} \nonumber \\
  G_{2,3} \left[ l = \ln \left( \frac{\ell}{\ell_0} \right) \right] &=& G_\Delta | \sin ( \Delta \phi_b ) |   \: \left\{ \frac{ 2 \cos ( \Delta \phi_b ) 
  \left( \frac{\ell}{\ell_K [ \Delta \phi_b ] } \right)^{ G_\Delta | \sin ( \Delta \phi_b ) | }   }{
 1    -  \left( \frac{\ell}{\ell_K [ \Delta \phi_b ] } \right)^{2 G_\Delta | \sin ( \Delta \phi_b ) | } }  \right\}
 \:\:\:\: . 
 \label{skc.5}
\end{eqnarray}
\noindent

In the specific case   discussed  here, none of the running couplings changes sign along the RG
trajectories. Therefore, rescaling $\ell$ at fixed $\Delta \phi_b$ does not induce switches in the ``actual'' groundstate
of the system: this either corresponds to the singlet state $ | \psi_1 \rangle_\gamma$, or to the $M=0$ component of 
the triplet, $ | \psi_2 \rangle_\gamma$, of Eqs.(\ref{lw.13}), depending on whether $\cos ( \Delta \phi_b ) > 0$, or 
$\cos ( \Delta \phi_b ) < 0$. On rescaling  $\ell$ at a given $\Delta \phi_b$, Eqs.(\ref{skc.3}) imply that 
$G_{1,2}  \left[ l = \ln \left( \frac{\ell}{\ell_0} \right) \right]$ and $ G_{2,3} \left[ l = \ln \left( \frac{\ell}{\ell_0} \right) \right]$
are scaling functions of $\ell / \ell_K [ \Delta \phi_b ]$, but also that the explicit form of 
the scaling function parametrically depends on the  RG  invariant
$\kappa [ \Delta \phi_b ] = [ G_{1,2}^{(0)}]^2 - [ G_{2,3}^{(0)}]^2 = G_\Delta^2 \sin^2 ( \Delta \phi_b )$. 
The key point is that, by simply acting on $\Delta \phi_b$ and/or on $G_\Delta$ (that is, on 
$J_\Delta$), we may change $\ell_K [ \Delta \phi_b ]$, by leaving the parametric function 
unchanged (that is, by simultaneously varying the boundary exchange strength $J_\Delta$ so that 
$J_\Delta \sin ( \Delta \phi_b )$ does not change). We can vary at will $\ell_K [ \Delta \phi_b ]$ and therefore 
recover the pertinent setup to directly probe  the (Kondo) scaling by directly tuning 
$\ell_K [ \Delta \phi_b ]$. As a probe of the emergence of $\ell_K  [\Delta \phi_b ]$,   we can measure the equilibrium spin current 
 through the junction. 

\begin{figure} 
\includegraphics*[width=.5\linewidth]{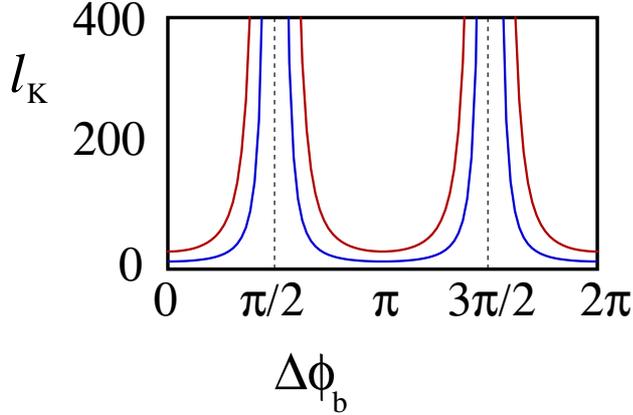}
\caption{$\ell_K [ \Delta \phi_b]$ as a function of $\Delta \phi_b$ for $0 \leq \Delta \phi_b \leq 2 \pi$ and for 
$G_\Delta = 0.3$ (red curve), and for $G_\Delta = 0.4$ (blue curve). Due to the divergence
of the right-hand side of Eq.(\ref{skc.4}) at $\Delta \phi_b = \frac{\pi}{2} , \frac{3 \pi}{2}$, 
the plots have been cutoff around these values of the applied phase difference.  } 
 \label{klength}
\end{figure}

From Eq.(\ref{skc.4}) we see that $\ell_K [ \Delta \phi_b ]$ is minimum when $\Delta \phi_b = k \pi$, with 
integer $k$. At these values of $\Delta \phi_b$ (and, in general, within small intervals centered on 
these values), the system rapidly evolves toward the Kondo regime, already for $\ell$ as large as 
30 sites (for $G_\Delta = 0.3$) or even 10 sites (for $G_\Delta = 0.4$). Moving from $0$ to 
larger values of $\Delta \phi_b$, $\ell_K [ \Delta \phi_b ]$ increases, implying that longer chains are 
required (larger $\ell$), in order for the junction to reach the Kondo regime. Eventually, 
$\ell_K [ \Delta \phi_b ]$ diverges at $\Delta \phi_b = \frac{\pi}{2}$ and, by periodicity, at 
any $\Delta \phi_b = \frac{\pi}{2} + k \pi$, with $k$ integer. This means that, for $\Delta \phi_b$ close 
to $\frac{\pi}{2} + k \pi$, in practice the junction never reaches the Kondo regime. As a result, 
we conclude that the same system does, or does not, exhibit Kondo effect depending on just a single parameter,
in principle tunable from the outside, such as the value of the angle $\Delta \phi_b$ between the Ising axis 
in the external leads of chains 1 and 2 and the axis in the external lead of chain 3. To evidence this behavior, 
in Fig.\ref{klength} we plot $\ell_K [ \Delta \phi_b ]$ as a function of $\Delta \phi_b$ for 
$0 \leq \Delta \phi_b \leq 2 \pi$ and for $G_\Delta = 0.3$ (red curve), and for $G_\Delta = 0.4$ (blue curve).
Aside from  the features above, the plot evidences the $\Delta \phi_b = \pi$ periodicity of 
$\ell_K [ \Delta \phi_b ]$, which is implied by Eq.(\ref{skc.4}), and the over-all decrease of the 
curves as $G_\Delta$ is increased. Due to the divergences at $\Delta \phi_b = \frac{\pi}{2} , \frac{3 \pi}{2}$, 
the plots have been cutoff around these values of the applied phase difference.

 \noindent
To determine the spin currents through the three chains, we differentiate the groundstate energy with 
respect to the phases $ \{ \phi_\mu \}$. By substituting, in Eq.(\ref{n3.5}), the bare boundary coupling 
strengths with the renormalized ones, we obtain, at generic values of $\Delta \phi_a , 
\Delta \phi_b$, the $\ell$-dependent energy 

\beq
{\cal E}^{(0)} [ \ell ; \Delta \phi_a ,  \Delta \phi_b ] = - \frac{v }{\ell + 1} \: {\rm max}_{\lambda_a , \lambda_b = \pm 1} \: 
\{ \lambda_a G_{1,2} [ \ell ; \Delta \phi_a , \Delta \phi_b ] + \lambda_b G_{2,3} [ \ell ; \Delta \phi_a , \Delta \phi_b ]
+   \lambda_a \lambda_b G_{3,1} [ \ell ; \Delta \phi_a , \Delta \phi_b ] \}
\:\:\:\: , 
\label{skc.6}
\eneq
\noindent
with the dependence of the running couplings on the scale $\ell$ and on $\Delta \phi_a , 
\Delta \phi_b$ explicitly evidenced. Taking into account that $\Delta \phi_a , 
\Delta \phi_b$ enter the explicit formulas for the running couplings only 
through the  $G_{\lambda , \lambda + 1}^{(0)}$'s, 
we  readily recover  the formulas for the currents through the three chains, which 
are given by 

\beq
I_\mu [ \ell ; \Delta \phi_a  , \Delta \phi_b ] = \frac{v G_\Delta \: \sin [ \phi_\mu - \phi_{ \mu + 1} ]  }{\ell + 1} \sum_{\lambda = 1}^3 \frac{\partial 
\hat{G}_{\lambda , \lambda + 1}   }{\partial 
G_{\mu , \mu + 1  }^{(0)}  } - \frac{v G_\Delta \: \sin [ \phi_{\mu - 1}  - \phi_{ \mu } ]  }{\ell + 1} \sum_{\lambda = 1}^3 \frac{\partial 
\hat{G}_{\lambda , \lambda + 1}   }{\partial 
G_{\mu - 1 , \mu }^{(0)}  } 
\:\:\:\: , 
\label{skc.7}
\eneq
\noindent
with $\hat{G}_{3 ,  1}  = \lambda_a \lambda_b G_{3,1}$, $\hat{G}_{1 ,  2}  = \lambda_a  G_{1,2} $, $\hat{G}_{2 ,  3}  = \lambda_b
G_{2,3}$.  In our specific case, 
taking into account the system symmetries, we obtain 

\begin{eqnarray}
 I_1 [ \ell ; \Delta \phi_b ] &=&  I_2 [ \ell ; \Delta \phi_b ]  = \frac{v G_\Delta \sin ( \Delta \phi_b ) }{\ell + 1} \: \sum_{\lambda = 1}^3 \frac{ \partial 
 \hat{G}_{\lambda , \lambda + 1} }{ \partial G_{3,1}^{(0)}}
 \nonumber \\
 I_3  [ \ell ; \Delta \phi_b ] &=& - 2  I_1 [ \ell ; \Delta \phi_b ]  = 
 - \frac{v G_\Delta \sin ( \Delta \phi_b ) }{\ell + 1} \: \sum_{\lambda = 1}^3 \left\{ \frac{\partial \hat{G}_{\lambda , \lambda + 1} }{ \partial G_{3,1}^{(0)}} +  
 \frac{\partial \hat{G}_{\lambda , \lambda + 1} }{\partial G_{2,3}^{(0)}} \right\} =  \frac{v  }{\ell + 1} \: \sum_{\lambda = 1}^3   \frac{\partial \hat{G}_{\lambda , \lambda + 1} }{
 \partial \Delta \phi_b}  
 \:\:\:\: . 
 \label{skc.8}
\end{eqnarray}
\noindent
Clearly, the onset of the nonperturbative regime in the running couplings implies, via Eqs.(\ref{skc.8}), 
an analogous feature in the equilibrium spin currents. This can be detected by two alternative means, that is, 
by either looking at the scaling of  $I_\mu [ \ell ; \Delta \phi_b ]$ as a function of $\ell$ at a given $\Delta \phi_b$, 
or by  looking at the current pattern throughout the whole interval of periodicity in $\Delta \phi_b$ at different (and 
increasing) values of $\Delta \phi_b$.

\begin{figure} 
\includegraphics*[width=1.\linewidth]{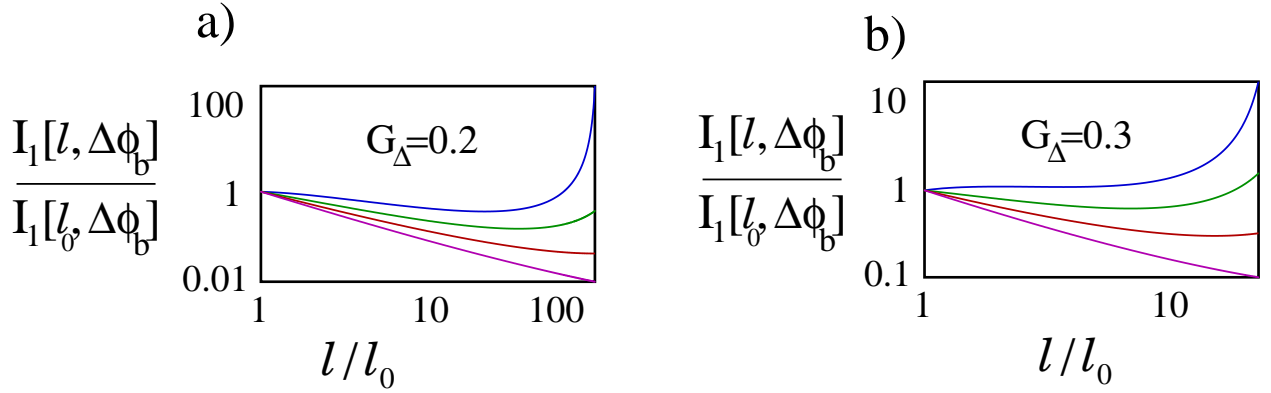}
\caption{ {\bf a)}  $I_1 [ \ell ; \Delta \phi_b ] / I_1 [ \ell_0 ; \Delta \phi_b ]$ as a function of $\ell / \ell_0$ 
for $G_\Delta = 0.2$ and for (from top to bottom) $\Delta \phi_b = 0.1 \pi$ (blue curve), 
$\Delta \phi_b = 0.2 \pi$ (green curve), $\Delta \phi_b = 0.3 \pi$ (red curve), $\Delta \phi_b = 0.4\pi$ (magenta curve).
The increase of $\ell_K [ \Delta \phi_b ]$ with $\Delta \phi_b$ as the phase difference evolves from 0 
to $\frac{\pi}{2}$ is apparent in the switch of the upturn of the curve as $\ell \sim 100 \ell_0$ for $\Delta \phi_b = 0.1\pi$ 
to a (roughly) $\ell^{-1}$ scaling at $\Delta \phi_b = 0.4 \pi$, which is what we would expect in the absence of 
a Kondo-like boundary interaction;  
{\bf b)} Same as in panel {\bf a)}, but with $G_\Delta = 0.3$.} 
 \label{scall}
\end{figure}

Within the former approach, we expect to see 
the onset of the nonperturbative regime in the spin current that takes place at different scales $\ell$ for 
different values of $\Delta \phi_b$, reflecting the dependence of $\ell_K$ on $\Delta \phi_b$. To verify such a prediction, 
in Fig.\ref{scall} we present logarithmic plots of $I_1 [ \ell ; \Delta \phi_b ] / I_1 [ \ell_0 ; \Delta \phi_b ]$ as a function of 
$\ell$ at fixed $\Delta \phi_b (=0.1\pi, 0.2\pi,0.3\pi,0.4\pi)$, and for two different values of $G_\Delta$, as detailed in the figure caption. 
The smallest value of $\Delta \phi_b$ we use to draw  Fig.\ref{scall}{\bf a)} and   Fig.\ref{scall}{\bf b)}
is $\Delta \phi_b = 0.1 \pi$. To obtain readable plots, we therefore draw diagrams up to 
a maximum value of $\ell$ slightly lower than $\ell_K [ \Delta \phi_b = 0.1 \pi]$, which is 
$\sim 170 \ell_0$ for $G_\Delta = 0.2$ (Fig.\ref{scall}{\bf a)}) and $\sim 27  \ell_0$ for $G_\Delta = 0.3$ (Fig.\ref{scall}{\bf b)}).
As expected, we see that, on increasing $\Delta \phi_b$ from values close to 0 to values close to $\frac{\pi}{2}$, 
the current plots evolve from diagrams exhibiting a clear upturn for $\Delta \phi_b = 0.1 \pi$ 
at a scale $\ell \sim \ell_K [ 0.1 \pi ]$, to a simple decrease with $\ell$ roughly $\propto \ell^{-1}$ times 
corrections from  higher-order contributions in the boundary couplings at $\Delta \phi_b = 0.4 \pi$. 
Given our result for $\ell_K [ \Delta \phi_b ]$, we  may therefore readily interpret
Fig.\ref{scall}{\bf a)} and   Fig.\ref{scall}{\bf b)} as an evidence for $\ell_K [ \Delta \phi_b ]$ to 
increase at $\Delta \phi_b$ increasing from $0$ to $\frac{\pi}{2}$. This is, in fact, a striking 
feature of our system: by just acting on $\Delta \phi_b$ keeping all the other system parameters 
fixed, we  may tune, or not, the onset of the Kondo regime at a given scale,  given the large window of variation 
of $\ell_K [ \Delta \phi_b ]$ evidenced in Fig.\ref{klength}.

 \noindent
 
\begin{figure} 
\includegraphics*[width=1.\linewidth]{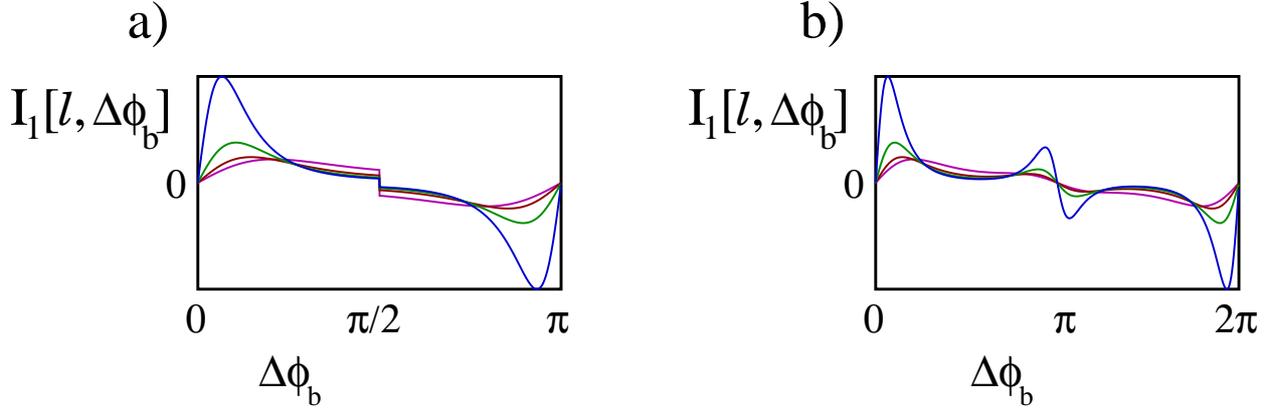}
\caption{ {\bf a)}  $I_1 [ \ell ; \Delta \phi_b ]  $ as a function of $\Delta \phi_b $ 
for $G_\Delta = 0.2$ and for (from top to bottom) $\frac{\ell}{\ell_0} = 120$ (blue curve), 
$\frac{\ell}{\ell_0} = 90$ (green curve), $\frac{\ell}{\ell_0} = 60$  (red curve), $\frac{\ell}{\ell_0} = 30$ (magenta curve).
The plots have been drawn by assuming that the system always occupies its true groundstate, which 
determines the finite jump in the current at $\Delta \phi_b = \frac{\pi}{2}$ and the halving of the period to 
$\pi$; 
{\bf b)} Same as in panel {\bf a)}, but  in this case it is assumed that the system does not relax to its actual 
groundstate when $\Delta \phi_b$ crosses $\frac{\pi}{2}$ and $\frac{3 \pi}{2}$. } 
 \label{spincur}
\end{figure}

To complement the results reported in Fig.\ref{scall}, we may alternatively analyze   
$I_1 [ \ell ; \Delta \phi_b ] $ as a function of $\Delta \phi_b$ at fixed chain length, for different 
values of $\ell$. Since the scaling of $I_1 [ \ell ; \Delta \phi_b ] $ with $\ell$ is different for 
different values of $\Delta \phi_b$, as we discuss above, we expect that monitoring the spin current 
across a full periodicity interval at increasing values of $\ell$, the 
growth of the current with $\ell$ is faster in the regions of values of $\Delta \phi_b$ where 
$\ell_K [ \Delta \phi_b]$ is lower. An important point here is that, differently from the previous analysis, 
now the plots are drawn by   varying $\Delta \phi_b$ at fixed $\ell$. Thus, the question 
arises whether, at a groundstate level crossing of the junction triggered by the change in 
$\Delta \phi_b$, the system ``adiabatically'' keeps  within the same state, or whether, at 
the level crossing, it ``jumps'' back into its actual groundstate. Apparently, this issue 
bears a close resemblance with the   fermion parity conservation which we 
discuss in Sec.\ref{n2} in the context of the $N=2$ junction. However, as we evidence
in Appendix \ref{lw}, it is possible to realize singlet, as well as
triplet, groundstates at either value of the total   fermion parity.  
In our specific case, starting from $\Delta \phi_b = 0$ and increasing the phase 
difference, from Eqs.(\ref{skc.3}),  we  see  that all three the $G_{\mu , \mu+1}$'s 
keep $>0$ as long as $0 \leq \Delta \phi_b < \frac{\pi}{2}$. Therefore, in this 
range of values of $\Delta \phi_b$, the junction groundstate corresponds to the 
singlet $ | \psi_1 \rangle_\gamma$ of Eq.(\ref{lw.13}), with $\gamma = \pm 1$. Accordingly, the spin currents are 
given by Eqs.(\ref{skc.8}) with $\hat{G}_{\mu , \mu+1} = G_{\mu , \mu+1}$. At 
$\Delta \phi_b = \frac{\pi}{2}$ a level crossing takes place in the junction 
groundstate between $ | \psi_1 \rangle_\gamma$ and the  $| \psi_2  \rangle_\gamma$ component
of the triplet. Correspondingly, the   spin currents are still 
given by Eqs.(\ref{skc.8}), with $\hat{G}_{2 , 3} = - G_{2 , 3}$ and 
$\hat{G}_{3 , 1} = - G_{3 , 1}$. Whether, when going across the level crossing,
the system keeps within   $ | \psi_1 \rangle_\gamma$, or it switches to 
 $| \psi_2  \rangle_\gamma$, may depend on a number of factors, such as, for 
 instance, how ``adiabatically'' we vary  $\Delta \phi_b$. For what concerns 
the spin current pattern, just as it happens for the $N=2$ junction, a switch 
in the groundstate at $\Delta \phi_b = \frac{\pi}{2}$ determines a finite discontinuity
in the currents and a corresponding halving of the period in $\Delta \phi_b$.
To evidence the main features of the spin current in both cases, in Fig.\ref{spincur}
we plot the current $I_1$ as a function of $\Delta \phi_b$ for $G_\Delta = 0.2$ by both 
assuming that the system is always able to relax into the actual groundstate (Fig.\ref{spincur}{\bf a)} - 
note that the period in this case is halved and $=\pi$-) and by assuming that the system does 
not relax and keeps within the same state when we go  across $\Delta \phi_b = \frac{\pi}{2}, \frac{3 \pi}{2}$
(Fig.\ref{spincur}{\bf b)}). Aside from the differences in the discontinuity at $\frac{\pi}{2}$ and in the over-all 
period, the two plots share the same feature. Specifically, in both cases we see that, on increasing $\ell$, the
current is enhanced around the values of $\Delta \phi_b$ at which $\ell_K [ \Delta \phi_b ]$ is minimum, that 
is, $\Delta \phi_b = 0 , \pi , 2 \pi$, due to the onset of the Kondo regime. At variance, around $\Delta \phi_b = \frac{\pi}{2} , \frac{3 \pi}{2}$, where 
$\ell_K [ \Delta \phi_b ]$ is maximum,  the lead length is consistently smaller than the corresponding 
value of $\ell_K$, the Kondo effect does not set in and, as a result, the current decreases with $\ell$ roughly as 
$\ell^{-1}$, as it would be appropriate in the absence of Kondo effect.

 \noindent
  For the sake of completeness, we now 
briefly discuss a different situation,  still easily tractable 
analytically, corresponding to  $\phi_1 = - \phi_2 = \frac{\Delta \phi_a}{2} $, $\phi_3 = 0$.
In this case, we obtain

\beq
G_{1,2}^{(0)} = G_\Delta \cos ( \Delta \phi_a ) \;\;\; , \;\;
G_{2,3}^{(0)} = G_{3,1}^{(0)} = G_\Delta \cos \left( \frac{\Delta \phi_a}{2} \right)
\:\:\:\: .
\label{skc.9}
\eneq
\noindent
Pointing out that now, on letting the phase $\phi_1 (\phi_2 )$ go through a full period, we 
get that the result must be periodic in $\Delta\phi_a$ with period equal to $4 \pi$, 
we note  that, regardless of the specific sign of the boundary couplings, to analytically 
solve the problem it is useful to separately treat the case 
$ | \cos ( \Delta \phi_a ) | < \left| \cos \left( \frac{\Delta \phi_a}{2} \right) \right|$,
which corresponds to $0< \Delta \phi_a < \frac{2 \pi}{3}$,  to 
$\frac{4 \pi}{3} < \Delta \phi_a < 8 \pi /3$, and to $10 \pi / 3 < \Delta \phi_a < 4 \pi$, and the case 
$ | \cos ( \Delta \phi_a ) | > \left| \cos \left( \frac{\Delta \phi_a}{2} \right) \right|$,
which corresponds to $\frac{2 \pi}{3}< \Delta \phi_a < \frac{4 \pi}{3}$ and  $8 \pi /3 < \Delta \phi_a < 10 \pi / 3$.
In the former case, the running couplings are given by 

\begin{eqnarray}
  G_{1,2} \left[ l = \ln \left( \frac{\ell}{\ell_0} \right) \right] &=& G_\Delta  \Omega ( \Delta \phi_a ) 
  \: \tan \Biggl\{ {\rm arctan} \left[ 
  \frac{\cos ( \Delta \phi_a ) }{ \Omega ( \Delta \phi_a ) }
 \right]+   G_\Delta  \Omega ( \Delta \phi_a )  \ln \left( \frac{\ell}{\ell_0} \right)\Biggr\} \nonumber \\
   G_{2,3} \left[ l = \ln \left( \frac{\ell}{\ell_0} \right) \right] &=& \frac{ G_\Delta  \Omega ( \Delta \phi_a )  \: \epsilon \left[ \cos 
   \left( \frac{ \Delta \phi_a }{2} \right) \right]   }{
   \cos\Biggl\{ {\rm arctan} \left[ 
  \frac{\cos ( \Delta \phi_a ) }{ \Omega ( \Delta \phi_a ) }
 \right] +   G_\Delta \Omega ( \Delta \phi_a )  \ln \left( \frac{\ell}{\ell_0}\right) \Biggr\}  }
 \:\:\:\: , 
 \label{skc.10}
\end{eqnarray}
\noindent
with   $\Omega ( \Delta \phi_a ) = \sqrt{\cos^2 \left( \frac{\Delta \phi_a}{2} \right) - \cos^2 ( \Delta \phi_a ) }$ and 
$\epsilon ( \phi )$ being the sign function, 
and, clearly, $  G_{2,3} \left[ l = \ln \left( \frac{\ell}{\ell_0} \right) \right]  = 
G_{3,1} \left[ l = \ln \left( \frac{\ell}{\ell_0} \right) \right] $.

Eqs.(\ref{skc.10}) imply 
that the running couplings diverge (either by positive, or negative values), at a scale 
$\ell_K [ \Delta \phi_a ]$ given by \cite{gsst,grt}

\beq
\ell_K  [\Delta \phi_a ] =  \ell_0  
\: \exp \left\{\left[ \frac{1 }{  G_\Delta  \Omega ( \Delta \phi_a )  }  \right] \left[  \frac{\pi}{2} -  {\rm arctan} \left[ 
  \frac{\cos ( \Delta \phi_a ) }{ \Omega ( \Delta \phi_a ) } \right] 
 \right] \right\}
\:\:\:\: . 
\label{skc.11}
\eneq
\noindent
As stated in Appendix \ref{rgsol}, we expect that, for $0 \leq \Delta \phi_a \leq \frac{2 \pi}{3}$, the junction flows towards 
the Kondo fixed point with all the three running coupling flowing to $+ \infty$ (after a change in sign of $G_{1,2}$ along 
the renormalization group trajectories, if $G_{1,2}^{(0)} < 0$), and that the same thing happens for  $\frac{10 \pi}{3} \leq \Delta \phi_a \leq 2 \pi$.
Eq.(\ref{skc.11}) implies that $\ell_K [ \Delta \phi_a ] \to \infty$ for $\Delta \phi_a \to \frac{2 \pi}{3}^-$, as well as 
for $\Delta \phi_a \to \frac{10 \pi}{3}^-$. Therefore, we  conclude  that no crossover to Kondo regime can in practice 
take place close to those boundaries of the intervals of validity of Eqs.(\ref{skc.10}).

In the complementary case, $\frac{2 \pi}{3}< \Delta \phi_a < \frac{4 \pi}{3}$ and  $8 \pi /3 < \Delta \phi_a < 10 \pi / 3$, we  obtain  

\begin{eqnarray}
   &&  G_{1,2} \left[ l = \ln \left( \frac{\ell}{\ell_0} \right) \right]  =  G_\Delta\: \tilde{\Omega} ( \Delta \phi_a )  \left\{ 
  \frac{ \cos ( \Delta \phi_a ) +  \tilde{\Omega} ( \Delta \phi_a ) + 
   [ \cos ( \Delta \phi_a ) - \tilde{\Omega} ( \Delta \phi_a ) ] \left( \frac{\ell}{\ell_0} \right)^{
   2 G_\Delta \tilde{\Omega} ( \Delta \phi_a ) }}{\cos ( \Delta \phi_a ) + \tilde{\Omega} ( \Delta \phi_a )  - 
   [ \cos ( \Delta \phi_a ) - \tilde{\Omega} ( \Delta \phi_a ) ] \left( \frac{\ell}{\ell_0} \right)^{
   2 G_\Delta \tilde{\Omega} ( \Delta \phi_a ) }} \right\} \nonumber \\
     &&  G_{2,3} \left[ l = \ln \left( \frac{\ell}{\ell_0} \right) \right]  =  G_\Delta\: \tilde{\Omega} ( \Delta \phi_a ) \: \left\{
     \frac{ 2\cos  \left( \frac{\Delta \phi_a}{2}  \right)  \left( \frac{\ell}{\ell_0} \right)^{
    G_\Delta \tilde{\Omega} ( \Delta \phi_a )}}{\cos ( \Delta \phi_a ) + \tilde{\Omega} ( \Delta \phi_a )- 
   [ \cos ( \Delta \phi_a ) - \tilde{\Omega} ( \Delta \phi_a )] \left( \frac{\ell}{\ell_0} \right)^{
   2 G_\Delta \tilde{\Omega} ( \Delta \phi_a )}} \right\}
 \:\:\:\: , 
 \label{skc.12}
\end{eqnarray}
\noindent
with $\tilde{\Omega} ( \Delta \phi_a ) = \sqrt{ \cos^2 ( \Delta \phi_a ) - \cos^2 \left( \frac{\Delta \phi_a}{2} \right)  }$.

From the right-hand side of Eqs.(\ref{skc.12}), we  readily see  that, whenever $\cos ( \Delta \phi_a ) < 0$, there is no 
onset of the Kondo regime at the junction. Indeed, since $\cos \left( \frac{\Delta \phi_a}{2} \right)< 0 $
for $\frac{2 \pi}{3} < \Delta \phi_a < \frac{10 \pi}{3}$, having $\cos ( \Delta \phi_a ) < 0$ corresponds to the 
case $G_{2,3}^{(0)} = G_{3,1}^{(0)}$ and $G_{\lambda , \lambda + 1}^{(0)} < 0$, $\forall \lambda$.  
As we discuss in detail in Appendix \ref{rgsol}, 
no Kondo effect is expected to set in this case, which is ultimately consistent with Eqs.(\ref{skc.12}). 
At variance, the crossover to the Kondo regime takes place when $\cos ( \Delta \phi_a ) > 0$, with an associated 
Kondo length $\ell_K [ \Delta \phi_a ]$ given by 

\beq
\ell_K [ \Delta \phi_a ] = \left\{ \frac{\cos ( \Delta \phi_a ) + \tilde{\Omega} ( \Delta \phi_a )}{
\cos ( \Delta \phi_a )-  \tilde{\Omega} ( \Delta \phi_a )} \right\}^{\frac{1}{2 
G_\Delta \tilde{\Omega} ( \Delta \phi_a )}}
\:\:\:\: . 
\label{skc.13}
\eneq
\noindent
We therefore conclude that both Eqs.(\ref{skc.3}) and Eqs.(\ref{skc.10},\ref{skc.12}) are consistent with the general RG
analysis of Appendix \ref{rgsol}, of which they constitute a special case. In both cases,  analyzing the 
scaling properties of the equilibrium spin currents through the junction provides an effective tool to map out the 
phase diagram associated to the corresponding RG  trajectories. For the sake of simplicity, here  we do not discuss 
further our second example, as the corresponding analysis would be exactly analogous to what we have done in 
the first example.  
 
 As a general comment on the emerging TKE at our YSC, it is worth stressing that, differently
from what happens with Y junctions of fermionic quantum wires \cite{oca_1,oca_2} and of Josephson junction 
chains \cite{frustr,giuliano09,cmgs}, here we recover a nontrivial phase diagram for the junction even in the absence of a
bulk interaction in the chain. This is a remarkable effect of the Kondo interaction, which is marginally
relevant and is able to take the system out of the trivial, weakly coupled regime, even with effectively 
(in terms of JW fermions) noninteracting leads.

\section{Conclusions}
\label{concl}

In this paper we have derived the topological Kondo Hamiltonian describing a Y junction of three 
inhomogeneous spin chains  in which the inner  XX-spin chains are connected to each other at their inner boundary,
while, at the outer boundary,  they are connected to   quantum Ising chains with different tilting angles for the Ising 
axis. Mapping the system Hamiltonian onto a pertinent boundary model, we have shown that the tilting angles 
effectively act as phases applied to the XX chains, thus triggering a nontrivial equilibrium spin current 
pattern through the junction. 

Employing  the renormalization group   approach to this topological Kondo model, 
we have been able to express the running couplings as functions of the bare couplings and of 
the running scale. Substituting the corresponding formulas in the expression of the system
groundstate energy, we have eventually derived the energy itself at a generic value 
of the running scale $l$ as a function of $l$ and of the applied phases. This allowed us 
to derive scaling formulas for the spin currents, by simply differentiating the running 
groundstate energy with respect to the applied phases. We have therefore argued how 
it is possible to directly measure the Kondo screening length $\ell_K$ by 
monitoring the crossover in the currents induced by the 
onset of the Kondo regime. 

Along our derivation,  as evidenced by the examples we provide in Sec.\ref{sck}, we have shown 
that   $\ell_K$ is a known function of the applied 
phases. This has  provided us with  two complementary ways to probe the Kondo length, by  
  either looking at the scaling of the currents with $\ell$ at fixed applied phases, or 
by   fixing  $\ell$ and tuning $\ell_K$ by varying the applied phases.

Incidentally, it is worth stressing how  our proposed YSC is likely to be within the reach of 
nowadays technology, both for what concerns
the practical realization of the system we propose, as well as regarding the experimental 
probe of the spin currents.  In principle, it could be realized by means of, e.g.,   Josephson junction arrays, which 
are well-known to effectively behave as quantum spin chains with the properties required to 
realize our YSC \cite{glazrak,giuso_1,pino}.  Also, several effective methods to 
efficiently detect the spin currents through the junction are already potententially available to 
experimentalists as extensively discussed in, e.g.,  Ref.\cite{r.1}. 

To summarize our results, we have shown how a  $N=3$ YSC
provides a rather unique Kondo setting  in which we may easily tune the  Kondo length by acting on 
 the phase differences  only. Tuning the Kondo length allows for mapping out the scaling properties 
of the system without, e.g., changing the length of the chains and/or varying the energy/temperature  scale(s) associated to 
the measurement, which should not be easy to do in a realistic system, thus paving the way to 
the possibility of a clear-cut experimental measurement of the so far pretty elusive Kondo
scaling length \cite{Affleck09}.

\vspace{0.5cm}

{\bf Acknowledgements --}
A. N. was financially supported  by POR Calabria FESR-FSE 2014/2020 - Linea B) Azione 10.5.12,  grant no.~A.5.1.
D. G. acknowledges  financial support  from Italy's MIUR  PRIN project  TOP-SPIN (Grant No. PRIN 20177SL7HC).

\appendix

\section{Derivation of the effective boundary Hamiltonian for the topological superconductor-normal wire junction }
\label{simple}
 
In this Appendix  we recover, in terms of JW fermions,  the effective boundary Hamiltonian 
corresponding to $H_{\rm SC}^\lambda$ in Eq.(\ref{e.1}). 
 
In particular,  the boundary Hamiltonian exactly describes the interface between the XX-chain and the 
outer Ising chain in the   limit  $\gamma = t $ and $g=0$ \cite{kita_1}.  As a result, we obtain

\begin{eqnarray}
&-& t \sum_{ j = \ell + 1}^{L-1} \: \{c_{ j , \lambda}^\dagger c_{ j + 1 , \lambda}+ c_{ j + 1 , \lambda}^\dagger c_{ j , \lambda} \}
- \gamma \sum_{ j = \ell + 1}^{L-1} \: \{ c_{ j , \lambda}c_{ j + 1 , \lambda} e^{- 2 i \phi_\lambda} 
+ c_{j+1, \lambda}^\dagger c_{ j , \lambda}^\dagger e^{ 2 i \phi_\lambda} \} - g \sum_{ j = \ell + 1}^L 
c_{j , \lambda}^\dagger c_{j , \lambda} \nonumber \\
&\to& - t \sum_{ j = \ell + 1}^{L-1} \: \{ [c_{ j , \lambda}^\dagger e^{   i \phi_\lambda} + c_{ j , \lambda} e^{ - i \phi_\lambda} ]
 [ c_{ j + 1 , \lambda}e^{ - i \phi_\lambda} -  c_{ j + 1 , 
\lambda}^\dagger e^{ i \phi_\lambda} ] \} \equiv - i t  \sum_{ j = \ell + 1}^{L-1} \: 
\xi_{ j , \lambda}\eta_{ j + 1 , \lambda} 
\:\:\:\: , 
\label{quax.2}
\end{eqnarray}
\noindent
with the real lattice fermions $\xi_{ j , \lambda}, \eta_{ j , \lambda}$ respectively given by 

\begin{eqnarray}
\xi_{ j , \lambda} &=& c_{ j , \lambda}e^{  -  i \phi_\lambda} + c_{ j , \lambda}^\dagger   e^{  i \phi_\lambda} \nonumber \\
\eta_{j , \lambda} &=& - i \: \{ c_{ j , \lambda} e^{  -  i \phi_\lambda} - c_{ j , \lambda}^\dagger   e^{  i \phi_\lambda} \}
\:\:\:\: . 
\label{quax.3}
\end{eqnarray}
\noindent
Defining new, ``nonlocal'' Dirac fermions $d_{j , \lambda}$ ($j=\ell + 1 , \ldots , L-1$) as 
$d_{ j , \lambda} = \frac{1}{2} \: \{ \xi_{ j , \lambda} - i \eta_{ j + 1 , \lambda} \}$, we 
find that 

\beq
- i t  \sum_{ j = \ell + 1}^{L-1} \: 
\xi_{ j , \lambda}\eta_{ j + 1 , \lambda}  = 
2 t \sum_{ j = \ell + 1 }^{L-1} \: \left[ d_{j, \lambda}^\dagger d_{ j , \lambda} - \frac{1}{2} \right] 
\:\:\:\: , 
\label{quax.4}
\eneq
\noindent
which evidences the emergence of    the zero-mode operators 
at the two endpoints,  respectively given by 
 
\begin{eqnarray}
 \eta_{ \ell + 1 , \lambda } &=& - i \: \{ c_{ \ell + 1  , \lambda} e^{  -  i \phi_\lambda} - c_{ \ell + 1 , \lambda}^\dagger   e^{  i \phi_\lambda} \} \nonumber \\
 \xi_{L , \lambda}  &=& c_{ L , \lambda}^\dagger e^{   i \phi_\lambda} + c_{ L , \lambda} e^{ - i \phi_\lambda} 
 \;\;\;\; . 
 \label{quax.5}
\end{eqnarray}
\noindent
Finally,  we project the term in the model Hamiltonian that is $\propto$ to  $J'$ in Eq.(\ref{eq.2}) onto the  
subspace spanned by the zero-modes in Eq.(\ref{quax.5}), thus obtaining   the boundary Hamiltonian 
$H_{B , \lambda}$, given by 

\beq
H_{B , \lambda} = i \frac{ \tau}{2}  \gamma_\lambda  \: \{ e^{ -  i \phi_\lambda}  c_{ \ell , \lambda} + e^{ i \phi_\lambda} c_{\ell , \lambda}^\dagger \} 
\:\:\:\: , 
\label{quax.6}
\eneq
\noindent
with $\gamma_\lambda \equiv \eta_{ \ell + 1 , \lambda}$ and $\tau \propto J'$. Based on the derivation illustrated in this 
Appendix, throughout all the paper we used as effective fermionic realization of the 
model Hamiltonian for each spin-chain the Hamiltonian $H_{{\rm F} , \lambda}$, given by 

\beq
H_{{\rm F} , \lambda} = - J \sum_{ j = 1}^{\ell - 1 } \{ c_{ j , \lambda}^\dagger c_{ j + 1 , \lambda} + 
c_{ j + 1 , \lambda}^\dagger c_{ j , \lambda} \} - H \sum_{ j = 1}^\ell c_{ j , \lambda}^\dagger c_{ j , \lambda} 
+ H_{B , \lambda}
\:\;\;\;\; , 
\label{quax.7}
\eneq
\noindent
with $H_{B , \lambda}$ given in Eq.(\ref{quax.6}). 
 
At generic values of the system parameters,   the boundary model   provides a reliable approximation at energies $\leq \Delta_{\rm Eff}$, with 
the effective gap  $\Delta_{\rm Eff} \sim | 2 t - g | $, in which case we effectively describe  the interface  
by retaining the low-energy emerging Majorana mode as the only effective degree of freedom on the gapped side \cite{gaf_1,gaf_2}.

\section{Periodicity in the spin equilibrium current through a single chain with an even/odd number of sites $\ell$}
\label{evenodd}

In Sec.\ref{modham} we mentioned how, for a single inhomogeneous chain, corresponding to an $N=2$
junction,   the periodicity of the spin equilibrium current is expected to depend on whether the number 
of sites in the chain,  $\ell$, is even, or odd \cite{r.1}. Since, throughout all the paper, we focus onto 
symmetric junctions only, which behave as the even-$\ell$ chain, in the following we consider also 
chains with odd $\ell$.  

For the sake of completeness and also to allow for a detailed comparison of our results with 
the ones obtained in Ref.\cite{r.1}, we devote this Appendix to carefully 
investigate how the periodicity  in a single chain depends on whether $\ell$ is even, or odd.
In doing so, we   relate the current periodicity to the structure of the low-lying energy eigenmodes of 
the chain Hamiltonian and to their dependence on the applied phase difference.
 
 To simplify  our discussion, here we consider the simple  model for the $N=2$ junction, that is, a single, homogeneous chain, connected 
to two Ising chains at its endpoints, with tilting angles corresponding to phases   $\phi_1$ and  
$\phi_2$. 

According to the derivation of Appendix \ref{simple}, we describe the chain in terms of  the lattice boundary 
Hamiltonian $H_2 = H_{\rm Bulk}^{(2)} +  H_{B}$, with 

\begin{eqnarray}
 H_{\rm Bulk}^{(2)} &=& - J \sum_{j = 1}^{\ell - 1} \{ c_j^\dagger c_{j + 1} + c_{ j + 1}^\dagger c_j \} - \mu \sum_{j = 1}^\ell c_j^\dagger c_j \nonumber \\
 H_B &=&   i \frac{ \tau}{2}  \gamma_1  \: \{ e^{ -  i \phi_1}  c_{ 1} + e^{ i \phi_1} c_{ 1}^\dagger \} 
 + i \frac{ \tau}{2}  \gamma_2  \: \{ e^{ -  i \phi_2}  c_{ \ell  } + e^{ i \phi_2} c_{\ell  }^\dagger \} 
 \:\:\:\: . 
 \label{x.1}
\end{eqnarray}
\noindent
By exactly diagonalizing $H_2$ at fixed phase difference $\phi_1 - \phi_2$, we have 
computed the spin current $ I [   \phi ] = I [ \phi_1 - \phi_2]$ when  $\mu = 0$ and 
$\tau / J = 0.25$, for $\ell=40$ and for  $\ell=41$.

\begin{figure} 
\includegraphics*[width=.8\linewidth]{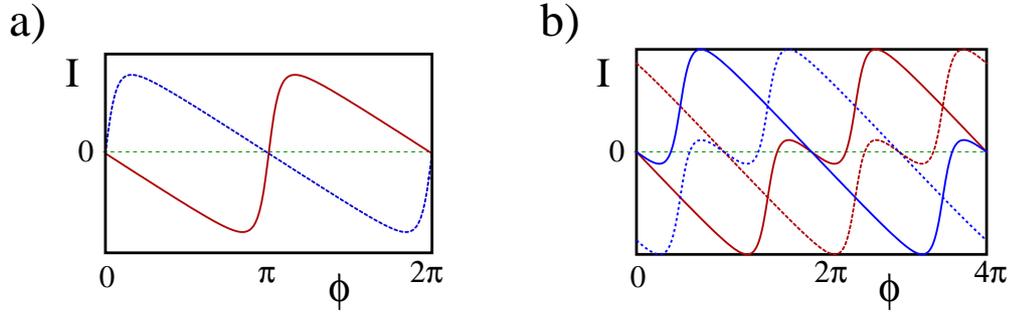}
\caption{{\bf a)} Current $ I [ \phi] $ {\it vs.} $\phi$ in the boundary model with Hamiltonian $H_2$ for 
$\tau / J = 0.25$ and $\ell = 40$; 
{\bf b)} Same as in {\bf a)}, but drawn for $\ell =41$.
 }  \label{pl_01}
\end{figure}

We draw  the relevant plots in  Fig.\ref{pl_01}, which 
we have constructed assuming that fermion parity is  always preserved. For $\ell=40$ (Fig.\ref{pl_01}{\bf a)}), the system 
realizes the so-called ${\cal Z}_2$-periodicity, with the current  periodic, with period equal to $2 \pi$. Correspondingly, there are 
two branches for the spin supercurrent $ I  [  \phi ] $. For $\ell=41$ (Fig.\ref{pl_01}{\bf b)}), 
the system realizes the ${\cal Z}_4$-periodicity, with the current   periodic   with period 
equal to $4 \pi$, and four different branches. 

\begin{figure} 
\includegraphics*[width=.8\linewidth]{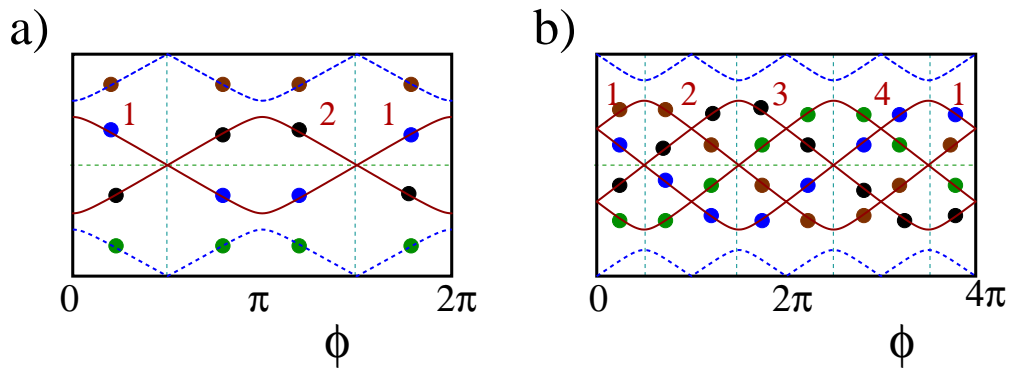}
\caption{{\bf a)} Plot of the single-particle levels (solid red lines) closest to the Fermi level (dashed green line) 
as a function of $\phi$ in the boundary model  with Hamiltonian $H_2$ for $\tau / J = 0.25$ and $\ell = 40$ as
a function of $\phi$. Varying $\phi$ from $0$ to $2 \pi$, there are two crossing between 
many-body groundstates with different total fermion parity, marked by the vertical, dashed cyan lines.
The regions with different fermion parities are labelled by 1 and 2 in the plot;  
{\bf b)} Same as in {\bf a)}, but drawn for $\ell =41$. As $\phi$ varies from 0 to $4 \pi$ here are now four different regions (see 
discussion in the main text), corresponding to a doubled (${\cal Z}_4$) periodicity in $ I [  \phi ]$ and to the emergence of 
four different branches in the current.
 }  \label{pl_02}
\end{figure}

To provide a  physical interpretation of the  current plots in Fig.\ref{pl_01},  in  Fig.\ref{pl_02} we show the 
sigle-quasiparticle energy levels crossing  the Fermi level  as $\phi$ varies. Fig.\ref{pl_02}{\bf a)} and 
Fig.\ref{pl_02}{\bf b)} are drawn for systems with the same parameters as  the ones 
corresponding to Fig.\ref{pl_01}{\bf a)} and to Fig.\ref{pl_01}{\bf b)}. 

Let us focus on   Fig.\ref{pl_02}{\bf a)} first. With the green and the red dots we mark the levels that 
 are neirest neighbors to the ones that cross as $\phi$ varies. In this case, they play no
role in determining the behavior of $ I [ \phi ]$. At variance, what matters is the position 
of the levels that we mark with respectively a blue and a black dot with respect to the Fermi level, 
which we mark with a dashed green line. We see that, as long as $\phi < \frac{\pi}{2}$, the groundstate is 
determined by a pair of a black and a green dot. This corresponds to a given fermion parity, say $+1$. As $\phi$
crosses $\frac{\pi}{2}$, the new groundstate is 
determined by a pair of a blue  and a green dot, which corresponds to the filled (with one additional fermion) level
close to the Fermi energy becoming lower in energy than the corresponding empty one (the black dot), with a 
net change in the fermion parity of the system, that now has become $-1$. The fermion parity keeps $-1$ till
$\phi = \frac{3 \pi}{2}$, which corresponds to the region we mark with 2 in Fig.\ref{pl_02}{\bf a)}. Then, it 
becomes again $+1$. Clearly, requiring fermion parity to be conserved means that the system groundstate, at 
any value of $\phi$, always corresponds to either a   black and a green dot, or to a blue and a green dot, which 
implies two branches in the total current and a total periodicity of $2 \pi$. As highlighted by our discussion, 
this behavior is strictly related to the dynamics of the two low-lying states, related to each other by a ${\cal Z}_2$-transformation, 
and is accordingly dubbed ${\cal Z}_2$-periodicity of the current.

 Let us now consider   Fig.\ref{pl_02}{\bf b)}. In this case, which corresponds to $\ell=41$, we  have four single-quasiparticle 
 energy level that cross, at various values of $\phi$, the Fermi level. In different intervals of values of $\phi$
(this time ranging from $\phi = 0$ to $\phi = 4 \pi$), there are four possible branches, corresponding to 
the different pairs of colored dot that characterize the system state, which imply the four 
different branches for $ I ( \phi )$ in Fig.\ref{pl_01}{\bf b)}. Differently from the 
even-$\ell$ case, now the system behavior is related to the dynamics of the four   low-lying states,
which are mapped onto each other by means of pertinent  ${\cal Z}_4$-transformations, 
and the ($4 \pi$) periodicity of $ I ( \phi )$  is  accordingly dubbed ${\cal Z}_4$-periodicity of the current.

To understand the behavior of 
the system for $\ell$ odd,  we consider as a reference limit the one in which the boundary 
Majorana modes   are fully decoupled from the rest of the chain (that is, the 
$\tau  \to 0$-limit). Here, when $\ell$ is odd and the chemical potential $\mu = 0$, we find 
two different Dirac zero-mode operators, the former being determined by a linear combination of 
$\gamma_1$ and $\gamma_2$ as $a = \frac{1}{2} \: ( \gamma_1 + i \gamma_2 )$, the latter being 
given by $b$, with

\beq
b = \sqrt{\frac{2}{\ell +1}} \: \sum_{ j = 1}^\ell \: (-1)^j c_j 
\:\:\:\: . 
\label{mp.9}
\eneq
\noindent
On turning on $\tau$ and on varying $\phi$, $a$ and $b$, together with their Hermitean conjugate, $a$ and $b$ combine 
together to determine the fermion low-lying states that cross with  each other in Fig.\ref{pl_01}{\bf b)},
which explain why, for $\ell$ odd, one obtains four different branches for the spin current, rather than 
two \cite{r.1}. Incidentally, before concluding this Appendix, it is worth pointing out the striking 
similarity between our Figs.\ref{pl_01},\ref{pl_02} and the  plots derived in Ref.\cite{r.1} under 
similar conditions, but using  the ``full'' model Hamiltonian (including the leads). Apparently, 
this is another piece of evidence of the reliability of our simplified boundary model to correctly 
recover the spin supercurrent in the large-$\ell$ limit. Incidentally, we also note that, given the 
system parameters we are considering, the boundary model already describes well the spin current 
dynamics at $\ell$ as large as 10, which evidences the high level of reliability of our   boundary model Hamiltonian
to describe the current pattern through the junction.

 \noindent
 
 \noindent

\section{Low-energy, long-wavelength effective field theory for the Jordan-Wigner fermion operators}
\label{lelw}

In this Appendix we describe  the  low-energy, long-wavelength field theory description of the JW fermion 
operators which we used throughout our paper to discuss the boundary interaction at the junction between the
XX-chains  and the outer Ising spin chains.
To do so, we start by decomposing the lattice fermion operators in the basis of the 
eigenmodes of  $H= H_{\rm Bulk} + H_\Delta$, $\Gamma_{\epsilon ; A}$. Using the additional 
label $_A$ to discriminate between independent eigenmodes corresponding to the same 
energy $\epsilon$, we set

\beq
\Gamma_{\epsilon ; A } = \sum_{\lambda = 1}^N \: \sum_{ j = 1}^\ell \: \{ u_{ j , \epsilon , ( A  ; \lambda) }^* c_{ j , \lambda} + v_{ j , \epsilon , ( A ; \lambda) }^* c_{ j , \lambda}^\dagger \}
+ \sum_{ \lambda = 1}^N \: w_{ \epsilon ,  ( A ; \lambda )} \gamma_\lambda 
\:\:\:\: , 
\label{lollolov.1}
\eneq
\noindent
with $ \{ u_{ j , \epsilon , ( A ; \lambda) } , v_{ j , \epsilon , ( A  ; \lambda ) } , w_{ \epsilon ,  ( A  ; \lambda ) } \}$ being the BDG eigenfunctions
for the state $_A$ with energy $\epsilon$. 

On imposing the canonical commutation relation $ [ \Gamma_{\epsilon ; A} , H ] = \epsilon \Gamma_{\epsilon ; A}$, we  find  the 
BDG that, for   $1<j<\ell$, are given by 

\begin{eqnarray}
 \epsilon u_{ j , \epsilon , (A  ; \lambda )  } &=& - J \: \{ u_{ j + 1 , \epsilon, ( A ; \lambda ) } + u_{ j - 1 , \epsilon, ( A  ; \lambda ) } \} - 
 H u_{ j , \epsilon ,  (A ; \lambda ) } \nonumber \\
  \epsilon v_{ j , \epsilon , ( A ; \lambda )  } &=&   J \: \{ v_{ j + 1 , \epsilon, ( A ; \lambda ) } + v_{ j - 1 , \epsilon, ( A ; \lambda) } \} +  H v_{ j , \epsilon , ( A ;  \lambda ) }
  \:\:\:\: . 
  \label{lollolov.2}
\end{eqnarray}
\noindent
For  $j=\ell$, we get  

\begin{eqnarray}
 \epsilon u_{ \ell , \epsilon ,  ( A; \lambda )  } &=& - J \: u_{ \ell - 1 , \epsilon ,  ( A ; \lambda ) } - H u_{\ell , \epsilon ,  ( A ; \lambda ) } +
 \frac{i \tau}{2} e^{ - i \phi_\lambda} w_{\epsilon ,  ( A ; \lambda ) }   \nonumber \\
  \epsilon v_{ \ell , \epsilon , ( A ; \lambda ) } &=&  J \: v_{ \ell - 1 , \epsilon ,  ( A ; \lambda ) } +  H v_{\ell , \epsilon ,  ( A ; \lambda ) } + 
  \frac{i \tau}{2} e^{  i \phi_\lambda} w_{ \epsilon ,  ( A ; \lambda ) }  \nonumber \\
  \epsilon  w_{ \epsilon,  (A; \lambda ) }  &=& - i \tau \: \{ e^{  i \phi_\lambda}u_{ \ell , \epsilon ,  ( A ; \lambda )}  +  e^{ - i \phi_\lambda}v_{ \ell , \epsilon ,  ( A ; \lambda )} \}
  \:\:\:\: . 
\label{lollolov.3}
\end{eqnarray}
\noindent
We look for solutions of  Eqs.(\ref{lollolov.2}) of the form 

\beq
\left[ \begin{array}{c}
u_{ j , \epsilon ,  ( A; \lambda )} \\ 
v_{ j , \epsilon ,  ( A ; \lambda )} 
       \end{array} \right] = \left[ \begin{array}{c}
\alpha_{\epsilon ,  ( A ; \lambda )} e^{ i k j } + \beta_{\epsilon ,  ( A ; \lambda )} e^{ - i k j } \\
\gamma_{\epsilon ,  ( A ; \lambda )} e^{ - i k' j } + \delta_{\epsilon ,  ( A ; \lambda )} e^{  i k' j } 
                                    \end{array} \right]
\:\:\:\: , 
\label{lollolov.4}
\eneq
\noindent
with $\epsilon = - 2 J \cos ( k ) - H = 2 J \cos ( k' ) + H$ and 
$\alpha_{\epsilon ,  ( A ; \lambda )} , \beta_{ \epsilon ,  ( A ; \lambda )} , \gamma_{\epsilon ,  ( A; \lambda )} , \delta_{\epsilon ,  ( A ; \lambda )}$ 
amplitudes independent of $j$. On inserting Eqs.(\ref{lollolov.4}) into Eqs.(\ref{lollolov.3}) and on getting 
rid of $ w_{\epsilon ,  ( A ; \lambda ) } $, we eventually obtain 

\begin{eqnarray}
 0 &=& J u_{\ell + 1 , \epsilon ,(A ; \lambda )} + \frac{\tau^2}{2 \epsilon} \: \{ u_{\ell , \epsilon , (A; \lambda )} + e^{ - 2 i \phi_\lambda} v_{\ell , \epsilon , ( A ; \lambda )} \} \nonumber \\
  0 &=& - J v_{\ell + 1 , \epsilon , ( A ; \lambda )} + \frac{\tau^2}{2 \epsilon} \: \{ e^{ 2 i \phi_\lambda} u_{\ell , \epsilon , ( A ; \lambda )} +  v_{\ell , \epsilon , ( A ; \lambda )} \} 
 \;\;\;\; . 
 \label{lollolov.5}
\end{eqnarray}
\noindent
In the low-energy, long-wavelength limit, Eqs.(\ref{lollolov.5}) imply $ u_{\ell , \epsilon , (A; \lambda )} + e^{ - 2 i \phi_\lambda} v_{\ell , \epsilon , ( A ; \lambda )} =0$.
To recover the low-energy description about the Fermi point $k_F$ defined by $- 2 J \cos ( k_F ) - H = 0$, we set $k \approx k_F + \frac{\epsilon}{v}$, and 
$k' \approx k_F - \frac{\epsilon}{v}$, with $v = 2 J \sin ( k_F )$. Accordingly, Eq.(\ref{lollolov.4}) becomes 

\beq
\left[ \begin{array}{c}
u_{ j , \epsilon ,  ( A ; \lambda )} \\ 
v_{ j , \epsilon ,  ( A ; \lambda )} 
       \end{array} \right]  \approx 
\left[ \begin{array}{c}
\alpha_{ ( A ; \lambda) ; \epsilon } e^{ i k_F j } e^{ i \frac{\epsilon x_j}{v} } +   \beta_{ ( A ; \lambda) ; \epsilon } e^{-  i k_F j } e^{ - i \frac{\epsilon x_j}{v} }     \\
\gamma_{ (A; \lambda) ; \epsilon} e^{ - i k_F j } e^{ i \frac{\epsilon x_j}{v} } +   \delta_{ ( A ; \lambda) ; \epsilon } e^{  i k_F j } e^{ - i \frac{\epsilon x_j}{v} }   
       \end{array} \right] 
\;\;\;\; , 
\label{lollolov.6}
\eneq
\noindent
with $x_j = a j$, $a$ being the lattice step, while Eqs.(\ref{lollolov.5}) are equivalent to the condition

\beq
e^{i \phi_\lambda}\: \{e^{ i k_F \ell } \alpha_{ ( A ; \lambda ); \epsilon  }e^{ i \frac{\epsilon \ell}{v} }
+ e^{ - i k_F \ell } \beta_{ ( A ; \lambda ) ; \epsilon }e^{ - i \frac{ \epsilon \ell}{v} } \} + 
e^{- i \phi_\lambda}\: \{e^{ - i k_F \ell } \gamma_{ ( A ; \lambda ); \epsilon  }e^{ i \frac{ \epsilon \ell}{v} }
+ e^{  i k_F \ell } \delta_{ ( A ; \lambda ) ; \epsilon }e^{ - i \frac{\epsilon \ell}{v} }  \}  = 0 
\:\:\:\: . 
\label{lollolov.bis}
\eneq
\noindent
Inverting Eq.(\ref{lollolov.1}) and using Eqs.(\ref{lollolov.6}), we obtain  the low-energy, long-wavelength 
mode expansion for the lattice field operator 
in the Heisenberg representation at time $t$, $c_{j , \lambda} (t)$, which is given by 

\beq
c_{j , \lambda} ( t ) \approx e^{i k_F j }\psi_{ R , \lambda} ( x_j - v t ) + e^{ - i k_F j } \psi_{L , \lambda} ( - x_j - v t ) 
\:\:\:\: , 
\label{lollolov.ter}
\eneq
\noindent
with 

\begin{eqnarray}
\psi_{ R , \lambda} ( x - v t ) &=& \sum_A  \: \sum_\epsilon \: \{ \alpha_{( A ; \lambda ); \epsilon } \Gamma_{ \epsilon ;A}
+ [ \gamma_{ ( A; \lambda); \epsilon } ]^* \Gamma_{ - \epsilon ; A}^\dagger  \} \: e^{ i \frac{\epsilon}{v} ( x - v t ) } \nonumber \\
\psi_{ L , \lambda} ( - x - v t ) &=& \sum_A \: \sum_\epsilon \: \{ \beta_{( A ; \lambda ); \epsilon } \Gamma_{ \epsilon ; A}
+ [ \delta_{ ( A ; \lambda); \epsilon } ]^* \Gamma_{ - \epsilon ; A}^\dagger  \} \: e^{ - i \frac{\epsilon}{v} ( x + v t ) }  
\:\:\:\: . 
\label{lollolov.quater}
\end{eqnarray}
\noindent
Furthermore, from Eqs.(\ref{lollolov.bis}) we  find  that the chiral fields $\psi_{ R , \lambda} , \psi_{L , \lambda}$  
can be expressed in terms of a single, chiral field $\psi_\lambda$, such that

\begin{eqnarray}
\psi_{ R , \lambda} ( x - v t ) &=&    e^{ - i \phi_\lambda} \psi_\lambda ( x_j - \ell - v t )
\nonumber \\
\psi_{ L  , \lambda} ( x +  v t ) &=&    
  - e^{ - i \phi_\lambda} \psi_\lambda^\dagger ( - x_j + \ell - v t )
\:\:\:\: . 
\label{lollolove.x1}
\end{eqnarray}
\noindent

Eq.(\ref{lollolove.x1}) is independent of the boundary conditions at the inner boundaries 
of the chains. Once these are defined, as well, they induce further constraints among 
the fields $  \psi_\lambda , \psi_\lambda^\dagger $. For instance, choosing 
open boundary conditions at the inner boundary, that is, setting  $c_{ j  = 0 , \lambda}  = 0$, 
$\forall \lambda = 1 , \ldots , N$ (which corresponds to the  disconnected junction limit, $J_\Delta = 0$), we  
recover  the corresponding boundary conditions on the BDG wavefunctions, given by   $u_{ j = 0 , \epsilon ,  ( A ; \lambda )}  = 
v_{ j = 0 , \epsilon ,  (A ; \lambda )} = 0$. Combining these conditions with the general result of 
Eqs.(\ref{lollolov.bis}), we  readily find  as many independent solutions of the BDG equations at each allowed value of 
$\epsilon$, as many chains, each solution being nonzero over a single chain only. 
In particular, the solution being nonzero over chain-$\lambda$ only is given by 

\beq
\left[ \begin{array}{c}
u_{j ; \epsilon ; \lambda} \\
v_{j ; \epsilon ; \lambda } 
       \end{array} \right] = 
\left[ \begin{array}{c} e^{ - i \phi_\lambda} \: \{ \alpha_{  \lambda   ; \epsilon  }  e^{ i k_F j } e^{ i \frac{\epsilon}{v} ( x_j - \ell ) } + 
 \beta_{   \lambda  ; \epsilon }  e^{ -  i k_F j } e^{ - i \frac{\epsilon}{v} ( x_j - \ell ) }  \} \\ - e^{  i \phi_\lambda} \: \{ 
 \alpha_{   \lambda   ; \epsilon  }  e^{ i k_F j } e^{ - i \frac{\epsilon}{v} ( x_j - \ell ) } + 
 \beta_{   \lambda  ; \epsilon }  e^{ -  i k_F j } e^{   i \frac{\epsilon}{v} ( x_j - \ell ) }  \}
 \end{array}
 \right]
\;\;\;\;, 
\label{lollolove.z2}
\eneq
\noindent
with $\alpha_{  \lambda   ; \epsilon  } , \beta_{   \lambda  ; \epsilon }$ constants. 

The boundary conditions at the inner boundary imply 

\begin{eqnarray}
 &&   \alpha_{   \lambda  ; \epsilon  } e^{- i \frac{\epsilon}{v} \ell  } + 
\beta_{  \lambda  ; \epsilon }  e^{  i \frac{\epsilon}{v} \ell   }  = 0 \nonumber \\
  &&   \alpha_{   \lambda   ; \epsilon  }  e^{  i \frac{\epsilon}{v} \ell  } + 
\beta_{   \lambda   ; \epsilon }  e^{ - i \frac{\epsilon}{v} \ell   }  = 0 
 \:\:\:\: , 
 \label{lollolove.z3}
\end{eqnarray}
\noindent
which yields energy levels independent of $\phi_\lambda$ and the constants  $( \alpha_{  \lambda   ; \epsilon  } , \beta_{   \lambda  ; \epsilon } )
\equiv ( \alpha_\lambda , \beta_\lambda)$ independent of $\epsilon$  (as it must be). Accordingly, the 
mode expansion in Eq.(\ref{lollolov.ter}) reduces to 

\begin{eqnarray}
 c_{ j , \lambda } ( t )  &\approx & e^{ i k_F j } \: e^{ - i \phi_\lambda} \: \sum_\epsilon \:   \{ \alpha_{   \lambda   } 
 \Gamma_{ \epsilon  }  + \alpha_{   \lambda   }^*  \Gamma_{ -  \epsilon  }^\dagger    \}  \: e^{ i \frac{\epsilon}{v} ( x_j   - v t ) }
 - e^{ - i k_F j } \:e^{ - i \phi_\lambda}  \: \sum_\epsilon \:   \{ \alpha_{   \lambda   } 
 \Gamma_{ \epsilon  }  + \alpha_{   \lambda   }^*  \Gamma_{ -  \epsilon  }^\dagger    \}  \: e^{- i \frac{\epsilon}{v} ( x_j   + v t ) }  \nonumber \\
 & \equiv & 
e^{ i k_F j } \: e^{ - i \phi_\lambda} \: \xi_\lambda ( x_j - v t ) - e^{-  i k_F j } \: e^{ - i \phi_\lambda} \: \xi_\lambda ( - x_j - v t ) 
 \:\:\:\: , 
 \label{lollolov.24}
\end{eqnarray}
\noindent
with $\xi_\lambda ( x - v t )$ being a chiral, real fermionic field, 
given by 

\beq
\xi_\lambda ( x - v t ) = \frac{1}{\sqrt{\ell + 1}} \: \sum_{n = -\infty}^\infty \: \xi_{n , \lambda}  \: e^{ i \frac{\pi n}{\ell} ( x - v t ) } 
\:\:\:\: , 
\label{lollolov.extra}
\eneq
\noindent
with $\xi_{n , \lambda} = \xi_{-n , \lambda }^\dagger$ and $\{ \xi_{n , \lambda} , \xi_{n' , \lambda'} \} = 2 \delta_{n +n' , 0} \delta_{ \lambda , \lambda'}$. 
Eq.(\ref{lollolov.extra}) is what we have used in the main text in the disconnected junction limit.

\section{Fermion parity and state counting in  real fermion Hamiltonians}
\label{lw}

In this Appendix we show  how to   count the eigenstates of the projected boundary 
Hamiltonian $H^{(3)}_{ \Delta ; 0 }$ in Eq.(\ref{n3.3}) of the main text. To do so, we review  
and pertinently adapt to our system the approach originally developed in Ref.\cite{lwilk}
to account for the total fermion parity conservation in a system described by a three real fermion
Hamiltonian. 

Following  Ref.\cite{lwilk}, we start by considering an
Hamiltonian   $H_{\rm Real}$  given by 

\beq
H_{\rm Real} = - i b_1 \gamma_2 \gamma_3 - i b_2 \gamma_3 \gamma_1 - i b_3 \gamma_1 \gamma_2 
\;\;\;\; , 
\label{lw.1}
\eneq
\noindent
with $b_1 , b_2 , b_3$ real parameters. In order to pertinently take into account the fermion
parity conservation, in Ref.\cite{lwilk}, it has been proposed to realize the Majorana 
fermion operators as

\begin{eqnarray}
 \gamma_1 &\to& \sigma^x \otimes {\bf I} = \left[ \begin{array}{cc}
{\bf \sigma}^x & {\bf 0 } \\
{\bf 0 } & {\bf \sigma}^x 
                                                  \end{array} \right] \nonumber \\
  \gamma_2 &\to& \sigma^z \otimes {\bf I} = \left[ \begin{array}{cc}
{\bf \sigma}^z & {\bf 0 } \\
{\bf 0 } & {\bf \sigma}^z 
                                                  \end{array} \right] \nonumber \\   
\gamma_3    &\to& \sigma^y \otimes {\bf \tau}^x = \left[ \begin{array}{cc}
{\bf 0 } & {\bf \sigma}^y   \\
{\bf \sigma}^y & {\bf 0 }   
                                                  \end{array} \right]                                                 
\:\:\:\: , 
\label{lw.2}
\end{eqnarray}
\noindent
with the bilinears that realize the spin-1/2 $su(2)$-algebra given by 

\begin{eqnarray}
 {\cal S}^x &=& \frac{i}{2} \gamma_2 \gamma_3 = \frac{1}{2} \left[ \begin{array}{cc}
{\bf 0 } & {\bf \sigma}^x   \\
{\bf \sigma}^x & {\bf 0 }   
                                                  \end{array} \right]  \nonumber \\
     {\cal S}^y &=& \frac{i}{2} \gamma_3  \gamma_1  = \frac{1}{2} \left[ \begin{array}{cc}
{\bf 0 } & {\bf \sigma}^z   \\
{\bf \sigma}^z & {\bf 0 }   
                                                  \end{array} \right]  \nonumber \\    
      {\cal S}^z &=& \frac{i}{2} \gamma_1  \gamma_2  = \frac{1}{2} \left[ \begin{array}{cc}
 {\bf \sigma}^y  & {\bf 0 } \\
 {\bf 0 }   &{\bf \sigma}^y 
                                                  \end{array} \right]                                                  
\:\:\:\: . 
\label{lw.3}
\end{eqnarray}
\noindent

The fermion parity operator $P$, that anticommutes with all the real fermion operators 
and commutes with all the bilinears, is given by 

\beq
P = \left[ \begin{array}{cc}
 {\bf \sigma}^y  & {\bf 0 } \\
 {\bf 0 }   & - {\bf \sigma}^y 
                                                  \end{array} \right]          
\:\:\:\: . 
\label{lw.4}
\eneq
\noindent
with  $[P ,  H_{\rm Real} ] = 0$. Therefore, it is possible to  diagonalize, 
$H_{\rm Real}$ over subspaces of a  given total fermion parity. As a well-suited parity operator, 
$P$ has eigenvalues $\lambda_P = \pm 1$. The projectors on the two corresponding eigenspaces, 
$P_\pm$, are given by 

\beq
P_\pm = \frac{1}{2} \: \left\{ {\bf I}_4  \pm P \right\} = \left[ \begin{array}{cc}
\frac{1}{2} \: \{ {\bf I} \pm  {\bf \sigma}^y \}  & { \bf 0 } \\
{\bf 0 } & \frac{1}{2} \: \{ {\bf I} \mp  {\bf \sigma}^y \}
                                                                  \end{array} \right] 
\:\:\:\: . 
\label{lw.5}
\eneq
\noindent
In general, a (4-component) vector belonging to the eigenvalue $\lambda_P = \pm 1$  takes the 
form 

\beq
\left[ \begin{array}{c}
w_1 \\ w_2 \\ w_3 \\ w_4         
       \end{array} \right]_+ = \frac{1}{\sqrt{2}} \: \left[ \begin{array}{c}
z_1 \\ i z_1 \\ i z_2 \\ z_2                                                            
                                                          \end{array} \right] 
\;\; , \; 
\left[ \begin{array}{c}
w_1 \\ w_2 \\ w_3 \\ w_4         
       \end{array} \right]_- = \frac{1}{\sqrt{2}} \: \left[ \begin{array}{c}
z_1 \\ - i z_1 \\ -i z_2 \\ z_2                                                            
                                                          \end{array} \right] 
 \:\:\:\: , 
\label{lw.6}
\eneq
\noindent
with $ | z_1 |^2 + | z_2 |^2 = 1$. 

Next, within each fermion parity eigenspace, we  
diagonalize ${\cal S}^z$. In particular, we obtain the following states 

\begin{itemize}
 \item In the $\lambda_P=+1$-sector
 
\beq
| + \rangle_1 = \frac{1}{\sqrt{2}} \: \left[ \begin{array}{c}
1 \\ i \\ 0 \\ 0                                               
                                             \end{array} \right] \;\;\; , \;\;
 | - \rangle_1 = \frac{1}{\sqrt{2}} \: \left[ \begin{array}{c}
0 \\ 0\\ i \\ 1                                              
                                             \end{array} \right] 
\:\:\:\: ; 
\label{lw.7}
\eneq
\noindent
\item In the $\lambda_P = -1$-sector

\beq
| + \rangle_{-1} = \frac{1}{\sqrt{2}} \: \left[ \begin{array}{c}
0 \\ 0 \\ -i  \\ 1                                               
                                             \end{array} \right] \;\;\; , \;\;
 | - \rangle_{-1} = \frac{1}{\sqrt{2}} \: \left[ \begin{array}{c}
1 \\ -i \\ 0 \\ 0                                              
                                             \end{array} \right] 
\:\:\:\: . 
\label{lw.8}
\eneq
\noindent 
\end{itemize}
Clearly, in either one of the two sets of states listed above, $H_{\rm Real}$ acts as 
$- 2 \vec{b} \cdot \vec{\cal S}$.

 It is  important to note how the two sectors 
with different fermion parity are mixed with each other under the action of the $\gamma_a$-operators. 
We obtain 

\beq
~_1 \langle \pm | \gamma_a | \pm \rangle_1 = ~_{-1} \langle \pm | \gamma_a | \pm \rangle_{-1 } = 0 
\:\:\:\: , 
\label{lw.9}
\eneq
\noindent
as well as (listing only the nonzero matrix elements) 

\begin{eqnarray}
 && ~_{-1} \langle \pm   | \gamma_1 | \mp  \rangle_{1 } =- ~_{1} \langle \pm   | \gamma_1 | \mp  \rangle_{-1 } =  i    \nonumber \\
  && ~_{-1} \langle \pm   | \gamma_2 | \mp  \rangle_{1 } =- ~_{1} \langle \pm   | \gamma_2 | \mp  \rangle_{-1 } = 1   \nonumber \\
   && ~_{-1} \langle \pm   | \gamma_3 | \pm  \rangle_{1 } =- ~_{1} \langle \pm   | \gamma_3 | \pm  \rangle_{-1 } = \pm i
   \:\:\:\: . 
   \label{lw.10}
\end{eqnarray}
\noindent
Eqs.(\ref{lw.10}) are the key results   used in deriving the RG equations in Sec.\ref{emek}.

 Naively rewriting $H^{(3)}_{ \Delta ; 0 }$ as  $H^{(3)}_{ \Delta ; 0 }  \to \sum_{\lambda = 1}^3 G_\lambda \: \sigma_\eta^\lambda \sigma_\Gamma^\lambda$,
with $\{ \sigma_\eta^\lambda , \sigma_\Gamma^\lambda \}$ Pauli matrices acting onto orthogonal spaces, would  yield a 
total of 4 independent states. However, since   $H^{(3)}_{ \Delta ; 0 }$ depends on 6 independent real 
fermionic modes, its Hilbert space should contain 8 states in total.To fix this flaw, we resort to the construction discussed above and employ it 
to build two (energy degenerate) copies of each eigenstate of  $H^{(3)}_{ \Delta ; 0 }$, with different eigenvalues 
of a properly defined fermion parity operator.
Since additional contributions to  $H^{(3)}_{ \Delta   }$ including 
nonzero degrees of freedom of the chains commute with the   operator $P_\eta$ measuring the total 
fermion parity associated to the triple $ \{ \eta^1 , \eta^2 , \eta^3 \}$, as well as with the operator  $P_\Gamma$
measuring the total fermion parity associated to the triple $ \{ \Gamma_{ 0 ; 1} , \Gamma_{0 , 2} , \Gamma_{0 , 3 } \}$,
we choose to label the degenerate  eigenstates of  $H^{(3)}_{ \Delta ; 0 }$ with the total 
fermion parity  ${\cal P}_\tau = P_\Gamma \cdot P_\eta$. 
Accordingly, we choose as basis set of the space of states of   $H^{(3)}_{ \Delta ; 0 }$ the 
8 states $ | a_\Gamma , a_\eta \rangle_{\gamma}$, such that 

\begin{eqnarray}
 && \sigma_\Gamma^3 | a_\Gamma , a_\eta \rangle_{\gamma} = i \Gamma_{ 0 , 1} \Gamma_{0 , 2} | a_\Gamma , a_\eta \rangle_{\gamma}
 = a_\Gamma | a_\Gamma , a_\eta \rangle_{\gamma} \nonumber \\
  && \sigma_\eta^3 | a_\Gamma , a_\eta \rangle_{\gamma} = i \eta^1 \eta^2 | a_\Gamma , a_\eta \rangle_{\gamma}
 = a_\eta | a_\Gamma , a_\eta \rangle_{\gamma}\nonumber \\
 && {\cal P}_\tau  | a_\Gamma , a_\eta \rangle_{\gamma} = \gamma  | a_\Gamma , a_\eta \rangle_{\gamma} 
 \:\:\:\: , 
 \label{lw.11}
\end{eqnarray}
\noindent
and $a_\Gamma , a_\eta , \gamma = \pm 1$. 

 $H^{(3)}_{ \Delta ; 0 }$ commutes with $P_\tau$. Therefore, 
at a fixed value of $\gamma$, within the subspace spanned by the states  $ \{ | a_\Gamma , a_\eta \rangle_{\gamma} \}$,
it is represented by the 4$\times$4 matrix $h_{3 ; 0}$, given by 

\beq
h_{3 ; 0 } = \left[ \begin{array}{cccc}
2 J_{1,2} & 0 & 0 & 2 J_{2,3} - 2 J_{3,1} \\
0 & - 2 J_{1,2} &  2 J_{2,3} + 2 J_{3,1}  & 0 \\
0 &   2 J_{2,3} + 2 J_{3,1}  &- 2 J_{1,2} & 0 \\
2 J_{2,3} - 2 J_{3,1} & 0 & 0 &   2 J_{1,2} 
                    \end{array} \right]
\:\:\:\: . 
\label{lw.12}
\eneq
\noindent
$h_{3,0}$ can be readily diagonalized. Below we report the list of 
the eigenvalues ($\{ \epsilon_j \} $), together with the corresponding eigenvectors ($\{ | \psi_\gamma \rangle_j \}$), which 
was the starting point for the derivation of Sec.\ref{n3}:

\begin{eqnarray}
&& \epsilon_1 = - 2 \: \{J_{1,2} + J_{2,3} + J_{3,1} \} \;\;\; , \;\; | \psi_\gamma   \rangle_1 = \frac{1}{\sqrt{2}}
\: \{ | + , - \rangle_\gamma - | - , + \rangle_\gamma \} \nonumber \\
&& \epsilon_2 =  - 2 \: \{J_{1,2} - J_{2,3} - J_{3,1} \} \;\;\; , \;\; | \psi_\gamma  \rangle_2 = \frac{1}{\sqrt{2}}
\: \{ | + , - \rangle_\gamma + | - , + \rangle_\gamma \} \nonumber \\
&& \epsilon_3 =   2 \: \{J_{1,2} +  J_{2,3} - J_{3,1} \} \;\;\; , \;\; | \psi_\gamma  \rangle_3 = \frac{1}{\sqrt{2}}
\: \{ | + , + \rangle_\gamma + | - , - \rangle_\gamma \} \nonumber \\
&& \epsilon_4 =  2 \: \{J_{1,2}-   J_{2,3} + J_{3,1} \} \;\;\; , \;\; | \psi_\gamma   \rangle_4= \frac{1}{\sqrt{2}}
\: \{ | + , + \rangle_\gamma - | - , - \rangle_\gamma \}
\:\:\: . 
\label{lw.13}
\end{eqnarray}
\noindent

\section{Explicit solution of Eqs.(\ref{ek.5}) and renormalization group trajectories}
\label{rgsol}

 In Sec.\ref{emek} we have derived the RG equations for the running couplings
$G_{\lambda , \lambda + 1}$. In particular,  in Eqs.(\ref{ek.5a}), we get  the exact
equations by pertinently taking into account the breaking of the system groundstate degeneracy due to the 
hybridization between the zero-modes of the chains. At the same time, we stressed that, for 
all the practical purposes, Eqs.(\ref{ek.5a}) may be substituted with the simplified 
Eqs.(\ref{ek.5}), which are the ``standard'' RG equations for the anisotropic TKE. 

In this Appendix, we discuss in detail how to recover the Kondo length $\ell_K$ as a function 
of the bare couplings. This is a crucial step of all our derivation,  as the dependence of $\ell_K$ 
on the $G_{\lambda , \lambda + 1}^{(0)}$s, which are known functions of the applied phases, determines
how, and to what extent,  the Kondo length is tuned by acting on $\Delta \phi_a , \Delta \phi_b$. For this reason, we first 
discuss the general case in which the bare couplings are all different from each other and in which the formula 
for $\ell_K$ as a function of the $G_{\lambda , \lambda + 1}^{(0)}$s can only approximately be recovered, and 
then focus onto the cases in which two of the the $G_{\lambda , \lambda + 1}^{(0)}$s are equal to each other, 
when an exact, explicit formula for $\ell_K$ can be provided within the approach of Refs.\cite{gsst,gsta,grt}. 

We note that,  
acting on $\Delta \phi_a , \Delta \phi_b$, we  can not only change the relative magnitudes of the 
running couplings, but also their sign. Accordingly, we have to consider all the possible sign 
assignments for the running couplings. To begin with, we assume that  all three the $G_{\lambda , \lambda + 1}$'s 
have positive sign. In this case, since Eqs.(\ref{ek.5}) 
yield 

\beq
\frac{d [ G_{\lambda , \lambda + 1} - G_{\lambda + 1 , \lambda + 2} ]}{d l} = 
- G_{\lambda + 2 , \lambda } \: \{ G_{\lambda , \lambda + 1} - G_{\lambda + 1, \lambda + 2} \} 
\:\:\:\: , 
\label{eko.1}
\eneq
\noindent
we  conclude  that the difference in the initial values of the running couplings is 
washed out along the renormalization group trajectories and that the boundary Kondo interaction
  flows toward  an isotropic fixed point, which we  identify  
with the one of a Y junction of  three quantum Ising chains \cite{tsve_1,tsve_2,gsst}. 

To   estimate   $\ell_K$, we 
note that Eqs.(\ref{ek.5}) imply that $G_{\lambda , \lambda +1}^2 ( l ) - 
G_{\lambda + 1 , \lambda + 2}^2 ( l)$ are constant along the renormalization group
trajectories for any $\lambda$. Therefore, assuming,for instance,
that $G_{1,2}^{(0)} \leq G_{2,3}^{(0)} \leq G_{3,1}^{(0)}$, we use  the conservation 
laws to trade Eqs.(\ref{ek.5}) for an equation involving $G_{1,2} ( l )$ only, that is 
given by

\beq
\frac{d G_{1,2} ( l ) }{d l} = \sqrt{ [ G_{2,3}^2 ( 0 ) - G_{1,2}^2 ( 0 ) + G_{1,2}^2 ( l ) ] \: 
[ G_{3,1}^2 ( 0 ) - G_{1,2}^2 ( 0 ) + G_{1,2}^2 ( l ) ] } 
\:\:\:\: . 
\label{ek.7}
\eneq
\noindent
Once we  have  solved Eq.(\ref{ek.7}) for $G_{1,2} ( l )$, we  obtain  
$G_{2,3} ( l ) = \sqrt{G_{1,2}^2 ( l ) + [ G_{2,3}^{(0)}]^2 -  [ G_{1,2}^{(0)}]^2}$, and 
$G_{3,1} ( l ) = \sqrt{G_{1,2}^2 ( l ) + [ G_{3,1}^{(0)}]^2 -  [ G_{1,2}^{(0)}]^2}$.
Therefore, the scale at which the perturbation theory breaks down can be uniquely identified 
as the scale at which $G_{1,2} (l)$ diverges. 

Determining $\ell_K$ from  Eq.(\ref{ek.7}) requires  introducing 
the incomplete elliptic integral, so that we  eventually find  

\beq
{\cal F} \left( \frac{\pi}{2} \Biggr| 1 - \frac{[ G_{3,1}^{(0)}]^2 - [ G_{1,2}^{(0)}]^2 }{
[ G_{2,3}^{(0)}]^2 - [ G_{1,2}^{(0)}]^2 } \right) - {\rm arctan} \left[ \frac{G_{1,2}^{(0)} }{ \sqrt{ [ G_{2,3}^{(0)}]^2 - [ G_{1,2}^{(0)}]^2 } } \right]
\approx \sqrt{  [ G_{3,1}^{(0)}]^2 - [ G_{1,2}^{(0)}]^2 } \: \ln \left( \frac{\ell_K}{\ell_0} \right) 
\:\:\:\: , 
\label{eko.3}
\eneq
\noindent
with 

\beq
{\cal F} ( \omega | z ) = \int_0^\omega \: \frac{ d t}{\sqrt{ 1 - z t^2} }
\:\:\:\: . 
\label{eko.4}
\eneq
\noindent 

To investigate the other possibilities, let us first of all assume, without loss of generality, 
that the initial couplings $G_{\lambda , \lambda + 1 }^{(0)}$ are such that 
$G_{1,2}^{(0)} \leq 0 <  G_{2,3}^{(0)} \leq G_{3,1}^{(0)}$. Nothing changes  with respect to the case $G_{1,2}^{(0)} > 0$,
if  $ | G_{1,2}^{(0)} | < G_{2,3}^{(0)}$. In this 
case, while $G_{1,2} ( l )$ start growing  toward 0 as $l$ increases,  $G_{2,3} ( l )$ and  $G_{3,1} ( l )$
decrease toward 0.  $G_{1,2} ( l )$ has to become 0 before  $G_{2,3} ( l )$ and  $G_{3,1} ( l )$ do so. This 
arises from the observation that Eqs.(\ref{ek.5}) imply that the three functions
$G_{\lambda , \lambda +1}^2 ( l ) - G_{\lambda + 1 , \lambda + 2}^2 ( l )$ 
are all constant along the RG  trajectories. If there was a scale $\hat{l}$ at which 
$G_{2,3} ( \hat{l} ) = 0$ and, at the same time,  $G_{1,2} ( \hat{l} ) \neq 0$, then one would get 
$G_{2,3}^2 ( \hat{l} ) - G_{1,2}^2 ( \hat{l} ) = - G_{1,2}^2 ( \hat{l} ) < 0$, which apparently 
contradicts the fact that $G_{2,3}^2 ( \hat{l} ) - G_{1,2}^2 (l) =  [ G_{2,3}^{(0)} ]^2 -  [ G_{1,2}^{(0)} ]^2  > 0$
at any scale $l$. Beyond $\hat{l}$, the RG  trajectories are the same 
as in the case $G_{1,2}^{(0)} > 0$ and $\ell_K$ can be again estimated exactly as in Eq.(\ref{eko.3}), by 
just substituting $G_{\lambda , \lambda + 1}^{(0)}$ with $G_{\lambda , \lambda + 1} ( \hat{l})$ and $\ell_0$ with $\ln ( \hat{l} )$. 

At variance, when $|G_{1,2}^{(0)} | > G_{2,3}^{(0)}$, 
$G_{2,3} ( l)$ becomes 0 at a scale $\bar{l}$ at which we  still have  $G_{1,2} ( \bar{l} ) < 0$, which 
is a consequence of the fact that now the  constant of motion is $[ G_{2,3}^{(0)} ]^2 -   [ G_{1,2}^{(0)} ]^2 <0$.
At scales $l > \bar{l}$, both $G_{1,2} ( l ) $ and $G_{2,3} ( l )$ grow large and negative, while 
$G_{3,1} ( l )$ grows large and positive. In addition, from Eqs.(\ref{ek.5}) we  readily derive  
that, at scales $l > $, $| G_{\lambda , \lambda +1 } ( l ) | - | G_{\lambda + 1,\lambda + 2} ( l ) |$ renormalizes to 0, according to 

\beq
\frac{d [ | G_{\lambda, \lambda + 1} ( l ) | - | G_{\lambda +1 , \lambda + 2} ( l ) | ]}{d l} = 
- |G_{\lambda +2 , \lambda } ( l ) | \:  \{ | G_{\lambda, \lambda + 1} ( l ) | - | G_{\lambda +1 , \lambda + 2} ( l ) | \}
\:\:\:\: . 
\label{eko.2}
\eneq
\noindent
Thus, we infer that the strongly coupled fixed point, in this case, corresponds to 
$ G_{1,2} ( l ) , G_{2,3} ( l ) \to - \infty$ ; $G_{3,1} ( l ) \to  + \infty$, as $l \to \infty$, 
with $ ( | G_{\lambda , \lambda + 1} ( l ) | / | G_{\lambda' , \lambda' + 1} ( l ) | ) \to 1$. 
In fact, this is equivalent to the isotropic strongly coupled fixed point to which the system flows when all 
the boundary couplings are $> 0$, up to the replacement $\eta_2 \to - \eta_2$. Finally, we note that 
the same argument applies equally well to the case in which $| G_{1,2}^{(0)} | > G_{3,1}^{(0)} \geq G_{2,3}^{(0)}$,
so, the same comclusions hold in this latter case, as well. In this case, $\ell_K$ is estimated as above, by
just using $\ell_0 = \ln ( \bar{l} )$ as the reference scale.

The above conclusions apply, as  
well,   to the case $G_{1,2}^{(0)} < G_{2,3}^{(0)} < 0 < G_{3,1}^{(0)}$, except that now 
we  have  to set  $\bar{l} = 0 $. Finally, in the case  $G_{1,2}^{(0)} \leq G_{2,3}^{(0)} \leq G_{3,1}^{(0)} < 0$, all the $\beta$-functions 
at the right-hand side of Eqs.(\ref{ek.5}) are $>0$ and, accordingly, all the running couplings start 
their flow by increasing toward 0, that is, by decreasing their absolute values. At a scale $\bar{l}$,
$G_{3,1} ( \bar{l} ) = 0$. Similar arguments to the ones used above imply $G_{1,2} ( \bar{l} ) < 0 , 
G_{2,3} ( \bar{l} ) < 0$. Thus, being $\beta_{3,1} [ G_{1,2} ( \bar{l} ) , G_{2,3} ( \bar{l} ) , G_{3,1} ( \bar{l} ) ] > 0$, 
$G_{3,1} ( l )$ becomes positive at scales $l> \bar{l}$. Accordingly, we  again obtain  that the strongly coupled fixed point  corresponds to 
$ G_{1,2} ( l ) , G_{2,3} ( l ) \to - \infty$ ; $G_{3,1} ( l ) \to  + \infty$, as $l \to \infty$, 
with $ ( | G_{\lambda , \lambda + 1} ( l ) | / | G_{\lambda' , \lambda' + 1} ( l ) | ) \to 1$. 
$\ell_K$ in this case can be estimated accordingly. 
To summarize the results obtained above,  we  conclude that, if the bare couplings are all different from each 
other, regardless of their relative sign, the junction always flows toward the Kondo fixed point.

Note that from the above discussion we left aside the ``critical lines'' $ |G_{1,2}^{(0)} | = G_{2,3}^{(0)}$ and 
 $ |G_{1,2}^{(0)} | = G_{3,1}^{(0)}$, as well as the partially isotropic cases (when all the three couplings 
 are positive) $ G_{1,2}^{(0)}  = G_{2,3}^{(0)}$ and 
 $ G_{1,2}^{(0)}  = G_{3,1}^{(0)}$. In this special cases it is possible to provide simple, closed-form 
analytical formulas for the running couplings, which we discuss in the following.

To begin with, let us consider the region in which all three the couplings are $>0$ and let us assume that,
without loss of generality, $G_{1,2}^{(0)} = G_{2,3}^{(0)}$. The corresponding RG 
equations are therefore a simplified version of the one discussed in, e.g., Ref.\cite{grt} for 
an impurity embedded within a quantum XXZ spin chain. Indeed, being $G_{1,2}^2 ( l ) - G_{2,3}^2 ( l )$ 
constant along the RG  trajectories, we find that $G_{1,2} ( l  )= G_{2,3} ( l )$ at 
any scale $l$, which allows for simplifying Eqs.(\ref{ek.5}) to 

\begin{eqnarray}
\frac{ d G_{1,2} }{ d l } &=& G_{1,2} G_{3,1} \nonumber \\
\frac{ d G_{3,1} }{d l } &=& G_{1,2}^2 
\:\:\:\: . 
\label{rgs.1}
\end{eqnarray}
\noindent
Depending on the relative values of $G_{1,2}^{(0)}$ and of $G_{3,1}^{(0)}$, we  therefore
obtain  the following explicit solutions (listed together with the corresponding estimate 
of $\ell_K$)

\begin{itemize}

\item {$G_{1,2}^{(0)} = G_{2,3}^{(0)} > G_{3,1}^{(0)}$}

In this case we  obtain  \cite{grt} 

\begin{eqnarray}
G_{3,1} \left[ l = \left( \frac{\ell}{ \ell_0 } \right) \right] &=& 
\sqrt{ [ G_{1,2}^{(0)} ]^2 - [ G_{3,1}^{(0)} ]^2 } \: \tan \left\{
{\rm arctan} \left[ \frac{ G_{3,1}^{(0)} }{ \sqrt{ [ G_{1,2}^{(0)} ]^2 - [ G_{3,1}^{(0)} ]^2 } }
\right] + \sqrt{ [ G_{1,2}^{(0)} ]^2 - [ G_{3,1}^{(0)} ]^2 } \ln \left( \frac{\ell}{\ell_0} \right) \right\} \nonumber \\
G_{1,2}  \left[ l = \left( \frac{\ell}{ \ell_0 } \right) \right] &=& 
\sqrt{ [ G_{1,2}^{(0)} ]^2 - [G_{3,1}^{(0)} ]^2 + G_{3,1}^2  \left[ l = \left( \frac{\ell}{ \ell_0 } \right) \right]  }
\:\:\:\: ,
\label{rgs.2}
\end{eqnarray}
\noindent
which implies 

\beq
 \sqrt{ [ G_{1,2}^{(0)} ]^2 - [ G_{3,1}^{(0)} ]^2 } \ln \left( \frac{\ell_K}{\ell_0} \right)  = 
 \frac{\pi}{2} - {\rm arctan} \left[ \frac{ G_{3,1}^{(0)} }{ \sqrt{ [ G_{1,2}^{(0)} ]^2 - [ G_{3,1}^{(0)} ]^2 } }
\right] 
\:\:\:\: , 
\label{rgs.3}
\eneq
\noindent
in perfect agreement with Eq.(\ref{eko.3}). 

\item {$G_{1,2}^{(0)} = G_{2,3}^{(0)} < G_{3,1}^{(0)}$}

In this case we  obtain

\begin{eqnarray}
 G_{3,1}  \left[ l = \left( \frac{\ell}{ \ell_0 } \right) \right] &=& \sqrt{ [ G_{3,1}^{(0)}]^2 -  [ G_{1,2}^{(0)}]^2} \times \nonumber \\
 && \Biggl\{ 
 \frac{G_{3,1}^{(0)} + \sqrt{ [ G_{3,1}^{(0)}]^2 -  [ G_{1,2}^{(0)}]^2} + [G_{3,1}^{(0)} - \sqrt{ [ G_{3,1}^{(0)}]^2 -  [ G_{1,2}^{(0)}]^2} ] \left( \frac{\ell}{\ell_0} \right)^{2
 \sqrt{ [ G_{3,1}^{(0)}]^2 -  [ G_{1,2}^{(0)}]^2} } }{G_{3,1}^{(0)} + \sqrt{ [ G_{3,1}^{(0)}]^2 -  [ G_{1,2}^{(0)}]^2} -
 [G_{3,1}^{(0)} - \sqrt{ [ G_{3,1}^{(0)}]^2 -  [ G_{1,2}^{(0)}]^2} ] \left( \frac{\ell}{\ell_0} \right)^{2
 \sqrt{ [ G_{3,1}^{(0)}]^2 -  [ G_{1,2}^{(0)}]^2} } } \Biggr\} \nonumber \\
 G_{1,2}  \left[ l = \left( \frac{\ell}{ \ell_0 } \right) \right] &=& 
\sqrt{ [ G_{1,2}^{(0)} ]^2 - [G_{3,1}^{(0)} ]^2 + G_{3,1}^2  \left[ l = \left( \frac{\ell}{ \ell_0 } \right) \right]  }
\:\:\:\: ,
\label{rgs.4}
\end{eqnarray}
\noindent
which yields \cite{gsst}

\beq
\ell_K = \ell_0 \: \left\{ \frac{G_{3,1}^{(0)} + \sqrt{ [ G_{3,1}^{(0)}]^2 -  [ G_{1,2}^{(0)}]^2} }{G_{3,1}^{(0)} - \sqrt{ [ G_{3,1}^{(0)}]^2 -  [ G_{1,2}^{(0)}]^2} }
\right\}^\frac{1}{ 2
 \sqrt{ [ G_{3,1}^{(0)}]^2 -  [ G_{1,2}^{(0)}]^2} }  
 \:\:\:\: , 
 \label{rgs.5}
\eneq
\noindent
that can again be recovered from Eq.(\ref{eko.3}) by going through an appropriate analytical continuation of 
the functions involved. 

\item {$G_{1,2}^{(0)} = G_{2,3}^{(0)} = G_{3,1}^{(0)}$}

Finally, in the fully isotropic case, we  obtain  back the renormalization group equations for 
the standard, isotropic Kondo effect, which is solved by 

\beq
G_{1,2}  \left[ l = \left( \frac{\ell}{ \ell_0 } \right) \right]  = \frac{G_{1,2}^{(0)} }{1 - G_{1,2}^{(0)} \ln \left( \frac{\ell}{\ell_0} \right) }
\:\:\:\: , 
\label{rgs.6}
\eneq
\noindent
implying 

\beq
\ell_K = \ell_0 \: \exp \left[ \frac{1}{G_{1,2}^{(0)} } \right]
\:\:\:\: , 
\label{rgs.7}
\eneq
\noindent
that is, the well-celebrated formula for the Kondo length in the isotropic case
\cite{hewson}. 

\item  {$G_{3,1}^{(0)} <0 <  G_{1,2}^{(0)} = G_{2,3}^{(0)}$, $| G_{3,1}^{(0)} | < G_{1,2}^{(0)}$ }

In this case the solution for the running couplings takes the same form as in Eq.(\ref{rgs.2}), while 
$\ell_K$ is again given by Eq.(\ref{rgs.3}).

\item  {$G_{3,1}^{(0)} <0 <  G_{1,2}^{(0)} = G_{2,3}^{(0)}$, $| G_{3,1}^{(0)} | >  G_{1,2}^{(0)}$ }

For this specific set of values of the ``bare'' parameters the solution of the RG 
equations is again given by Eqs.(\ref{rgs.4}), except that now $G_{3,1}^{(0)}$ is negative, which implies

\begin{eqnarray}
 G_{3,1}  \left[ l = \left( \frac{\ell}{ \ell_0 } \right) \right] &=& \sqrt{ [ G_{3,1}^{(0)}]^2 -  [ G_{1,2}^{(0)}]^2} \times \nonumber \\
 && \Biggl\{ 
 \frac{- | G_{3,1}^{(0)} |  + \sqrt{ [ G_{3,1}^{(0)}]^2 -  [ G_{1,2}^{(0)}]^2} - [| G_{3,1}^{(0)} | +  \sqrt{ [ G_{3,1}^{(0)}]^2 -  [ G_{1,2}^{(0)}]^2} ] \left( \frac{\ell}{\ell_0} \right)^{2
 \sqrt{ [ G_{3,1}^{(0)}]^2 -  [ G_{1,2}^{(0)}]^2} } }{- | G_{3,1}^{(0)} | + \sqrt{ [ G_{3,1}^{(0)}]^2 -  [ G_{1,2}^{(0)}]^2} +
 [| G_{3,1}^{(0)} | +  \sqrt{ [ G_{3,1}^{(0)}]^2 -  [ G_{1,2}^{(0)}]^2} ] \left( \frac{\ell}{\ell_0} \right)^{2
 \sqrt{ [ G_{3,1}^{(0)}]^2 -  [ G_{1,2}^{(0)}]^2} } } \Biggr\} \nonumber \\
 G_{1,2}  \left[ l = \left( \frac{\ell}{ \ell_0 } \right) \right] &=& 
\sqrt{ [ G_{1,2}^{(0)} ]^2 - [G_{3,1}^{(0)} ]^2 + G_{3,1}^2  \left[ l = \left( \frac{\ell}{ \ell_0 } \right) \right]  }
\:\:\:\: .
\label{rgs.8}
\end{eqnarray}
\noindent
Apparently, the solutions at the right-hand side of Eqs.(\ref{rgs.8}) exhibit no divergences anymore. Therefore, 
the interaction keeps perturbative at any scale and, as $\ell \to \infty$, we  eventually get   

\begin{eqnarray}
 && \lim_{\ell \to \infty} \:  G_{3,1}  \left[ l = \left( \frac{\ell}{ \ell_0 } \right) \right]  = -  \sqrt{ [ G_{3,1}^{(0)}]^2 -  [ G_{1,2}^{(0)}]^2} \nonumber \\
 &&  \lim_{\ell \to \infty} \:  G_{1,2}  \left[ l = \left( \frac{\ell}{ \ell_0 } \right) \right]  = 0 
 \;\;\;\; .
 \label{rgs.9}
\end{eqnarray}
\noindent
The RG flow in Eqs.(\ref{rgs.8},\ref{rgs.9}) corresponds to what happens in the region 
of irrelevance of the boundary interaction describing a spin impurity embedded within a quantum XXZ spin
chain \cite{gsta,grt}. The crucial point is that, in order for us to have an effectively irrelevant 
boundary interaction, we have to fine-tune  the coupling strength so that $G_{1,2}^{(0)} = G_{2,3}^{(0)}$.
Would the fine-tuning condition not be satisfied, we  would get back to the flow toward the strongly 
interacting Kondo fixed point, as discussed above.

\item  {$G_{1,2}^{(0)} <0<  G_{2,3}^{(0)},  |G_{1,2}^{(0)} | =  G_{2,3}^{(0)},  G_{2,3}^{(0)} >  G_{3,1}^{(0)} > 0$ }

In this case, Eqs.(\ref{ek.5}) reduce to 

\begin{eqnarray}
 \frac{d   G_{1,2}    }{d l} &=& - G_{1,2}  G_{3,1} \nonumber \\
 \frac{d G_{3,1} }{d l } &=& -  G_{1,2}^2 
 \:\:\:\: . 
 \label{rgs.x1}
\end{eqnarray}
\noindent
The solution now takes the form

\begin{eqnarray}
G_{3,1} \left[ l = \left( \frac{\ell}{ \ell_0 } \right) \right] &=& 
\sqrt{ [ G_{1,2}^{(0)} ]^2 - [ G_{3,1}^{(0)} ]^2 } \: \tan \left\{
{\rm arctan} \left[ \frac{ G_{3,1}^{(0)} }{ \sqrt{ [ G_{1,2}^{(0)} ]^2 - [ G_{3,1}^{(0)} ]^2 } }
\right] - \sqrt{ [ G_{1,2}^{(0)} ]^2 -  [ G_{3,1}^{(0)} ]^2 } \ln \left( \frac{\ell}{\ell_0} \right) \right\} \nonumber \\
G_{1,2}  \left[ l = \left( \frac{\ell}{ \ell_0 } \right) \right] &=& 
- \sqrt{ [ G_{1,2}^{(0)} ]^2 - [G_{3,1}^{(0)} ]^2 + G_{3,1}^2  \left[ l = \left( \frac{\ell}{ \ell_0 } \right) \right]  }
\:\:\:\: .
\label{rgs.x2}
\end{eqnarray}
\noindent
At the scale $\hat{\ell}$ such that  ${\rm arctan} \left[ \frac{ G_{3,1}^{(0)} }{ \sqrt{ [ G_{1,2}^{(0)} ]^2 - [ G_{3,1}^{(0)} ]^2 } }
\right] - \sqrt{ [ G_{1,2}^{(0)} ]^2 -  [ G_{3,1}^{(0)} ]^2 } \ln \left( \frac{\hat{\ell}}{\ell_0} \right)  = 0$,  
$G_{3,1} \left[ l = \left( \frac{\ell}{ \ell_0 } \right) \right] $ becomes negative. At larger scales, the  perturbative
approach breaks down at $\ell = \ell_K$, with

\beq
 \sqrt{ [ G_{1,2}^{(0)} ]^2 - [ G_{3,1}^{(0)} ]^2 } \ln \left( \frac{\ell_K}{\ell_0} \right)  = 
 \frac{\pi}{2} + {\rm arctan} \left[ \frac{ G_{3,1}^{(0)} }{ \sqrt{ [ G_{1,2}^{(0)} ]^2 - [ G_{3,1}^{(0)} ]^2 } }
\right] 
\:\:\:\: . 
\label{rgs.x3}
\eneq
\noindent

\item  {$G_{1,2}^{(0)} <0<  G_{2,3}^{(0)},  |G_{1,2}^{(0)} | =  G_{2,3}^{(0)},  0 < G_{2,3}^{(0)} <  G_{3,1}^{(0)}$ }

Pertinently modifying the result of Eq.(\ref{rgs.4}), we  obtain

\begin{eqnarray}
 G_{3,1}  \left[ l = \left( \frac{\ell}{ \ell_0 } \right) \right] &=& \sqrt{ [ G_{3,1}^{(0)}]^2 -  [ G_{1,2}^{(0)}]^2} \times \nonumber \\
 && \Biggl\{ 
 \frac{G_{3,1}^{(0)} + \sqrt{ [ G_{3,1}^{(0)}]^2 -  [ G_{1,2}^{(0)}]^2} + [G_{3,1}^{(0)} - \sqrt{ [ G_{3,1}^{(0)}]^2 -  [ G_{1,2}^{(0)}]^2} ]
 \left( \frac{\ell}{\ell_0} \right)^{-2
 \sqrt{ [ G_{3,1}^{(0)}]^2 -  [ G_{1,2}^{(0)}]^2} } }{G_{3,1}^{(0)} + \sqrt{ [ G_{3,1}^{(0)}]^2 -  [ G_{1,2}^{(0)}]^2} -
 [G_{3,1}^{(0)} - \sqrt{ [ G_{3,1}^{(0)}]^2 -  [ G_{1,2}^{(0)}]^2} ] \left( \frac{\ell}{\ell_0} \right)^{-2
 \sqrt{ [ G_{3,1}^{(0)}]^2 -  [ G_{1,2}^{(0)}]^2} } } \Biggr\} \nonumber \\
 G_{1,2}  \left[ l = \left( \frac{\ell}{ \ell_0 } \right) \right] &=& 
\sqrt{ [ G_{1,2}^{(0)} ]^2 - [G_{3,1}^{(0)} ]^2 + G_{3,1}^2  \left[ l = \left( \frac{\ell}{ \ell_0 } \right) \right]  }
\:\:\:\: ,
\label{rgs.y4}
\end{eqnarray}
\noindent
which yields 

\begin{eqnarray}
 \lim_{\ell \to \infty}  G_{3,1}  \left[ l = \left( \frac{\ell}{ \ell_0 } \right) \right] &=& \sqrt{ [ G_{3,1}^{(0)}]^2 -  [ G_{1,2}^{(0)}]^2} 
 \nonumber \\
  \lim_{\ell \to \infty}  G_{1,2}  \left[ l = \left( \frac{\ell}{ \ell_0 } \right) \right] &=& 0
  \:\:\:\: , 
  \label{rgs.y5}
\end{eqnarray}
\noindent
that is, again the system flows toward the  region 
of irrelevance of the boundary interaction describing a spin impurity embedded within a quantum XXZ spin
chain \cite{gsta,grt}. 

\item  {$G_{3,1}^{(0)} =  G_{1,2}^{(0)} = G_{2,3}^{(0)} < 0$ }

In this case, the RG flow corresponds to the (irrelevant) ferromagnetic Kondo interaction. Indeed, 
solving the renormalization group equation, we  find  $G_{1,2} ( l ) = G_{2,3} ( l ) = G_{3,1} ( l )$, 
with 

\beq
G_{1,2}  \left[ l = \left( \frac{\ell}{ \ell_0 } \right) \right] = - 
\frac{ | G_{1,2}^{(0)} | }{1 + | G_{1,2}^{(0)} | \ln \left( \frac{\ell}{\ell_0} \right) }
\;\;\;\; , 
\label{rgs.10}
\eneq
\noindent
that implies 

\beq
\lim_{\ell \to \infty} G_{1,2}  \left[ l = \left( \frac{\ell}{ \ell_0 } \right) \right]  = 0 
\:\:\:\: . 
\label{rgs.11}
\eneq
\noindent
\end{itemize}

To  synoptically summarize the results we derived in this Appendix, below we list all the 
fixed points to which the $N=3$ junction flows, for any possible choice of the initial 
values of the couplings (up to trivial exchanges in the indices)
 
\begin{itemize}
 \item {\bf Case 1:} $G_{\lambda , \lambda + 1}^{(0)} > 0$ $\forall \lambda$ (including the 
 case in which one of the $G_{\lambda , \lambda + 1}^{(0)} = 0$).
 
 In this case the boundary interaction flows towards the fixed point 
 describing the anisotropic TKE. All the running couplings flow 
 to $+ \infty$ as $\ell > \ell_K$. Nothing substantially changes
 if one of the bare couplings is = 0. 
 
  \item {\bf Case 2:} $G_{1 , 2}^{(0)} < 0 < G_{2,3}^{(0)} < G_{3,1}^{(0)} $.
  
If $ | G_{1,2}^{(0)} | \leq    G_{2,3}^{(0)}$, then $G_{1,2} ( l )$ crosses 0 
at a scale $\bar{l}$ at which $0 < G_{2,3} (\bar{l} )  < G_{3,1} ( \bar{l} )$. 
For $l > \bar{l}$ the flow is the same as in the case 
$G_{1,2}^{(0)}  = 0 <  G_{2,3}^{(0)} < G_{3,1}^{(0)} $, with all the running couplings flowing 
 to $+ \infty$.

At variance, if   $ | G_{1,2}^{(0)} | >    G_{2,3}^{(0)}$, the system flows toward 
a strongly coupled fixed point with $G_{1,2} ( l ) , G_{2,3} ( l ) \to - \infty$ and 
$G_{3,2} ( l ) \to  + \infty$, which is equivalent to the one discussed at Case 1, provided
$\eta_2 \to - \eta_2$ in the boundary Hamiltonian. 

 Case 2 is trivially equivalent to the case  $G_{1 , 2}^{(0)} < 0 < G_{3,1}^{(0)} < G_{2,3}^{(0)} $,
provided $G_{2,3} ( l )$ and $G_{3,1} ( l )$ are exchanged with each other in 
the discussion.

  \item {\bf Case 3:} $G_{1 , 2}^{(0)} < 0 < G_{2,3}^{(0)} = G_{3,1}^{(0)} $.
  
Also in this case, for  $ | G_{1,2}^{(0)} | \leq    G_{2,3}^{(0)}$,
all the running couplings flow 
 to $+ \infty$ as $\ell > \ell_K$. At variance, for $ | G_{1,2}^{(0)} | >   G_{2,3}^{(0)}$,
 the system flows towards the ``trivial'' fixed point corresponding to  
 Eqs.(\ref{rgs.y5}).
 
   \item {\bf Case 4:} $G_{1 , 2}^{(0)} \leq  G_{2,3}^{(0)} < 0 <  G_{3,1}^{(0)} $. 
   
Employing the substitutions $\eta_2 \to - \eta_2$ and $G_{1,2} \to - G_{1,2}$,
$G_{2,3} \to - G_{2,3}$, this case is mapped onto Case 1, with all three the
$G_{\lambda ,\lambda + 1}^{(0)} > 0$. Thus, we conclude that, in this case, 
$G_{1,2} ( l ) , G_{2,3} ( l ) \to - \infty$ and  $G_{3,1} ( l ) \to + \infty$.
Exactly as in the case 1, nothing changes if either  $G_{2,3}^{(0)} = 0$, or 
 $G_{2,3}^{(0)} = 0 $.
 
    \item {\bf Case 5:} $G_{1 , 2}^{(0)} <  G_{2,3}^{(0)} <  G_{3,1}^{(0)}   \leq   0 $. 
    
Using again the substitutions $\eta_2 \to - \eta_2$ and $G_{1,2} \to - G_{1,2}$,
$G_{2,3} \to - G_{2,3}$, this case becomes equivalent to   case 2. Thus, we conclude that, also in this case, 
$G_{1,2} ( l ) , G_{2,3} ( l ) \to - \infty$ and  $G_{3,1} ( l ) \to + \infty$.
 
    \item {\bf Case 6:} $G_{1 , 2}^{(0)} =   G_{2,3}^{(0)}  < G_{3,1}^{(0)}   \leq   0 $. 
    
Employing again the equivalence with  case 3, we   find that, also in this case, 
$G_{1,2} ( l ) =  G_{2,3} ( l ) \to - \infty$ and  $G_{3,1} ( l ) \to + \infty$.   

   \item {\bf Case 7:} $G_{1 , 2}^{(0)} <  G_{2,3}^{(0)} = G_{3,1}^{(0)}   <  0 $. 
   
In this case 
 the system flows towards the ``trivial'' fixed point corresponding to 
 Eqs.(\ref{rgs.y5}).   
 
   \item {\bf Case 8:} $G_{1 , 2}^{(0)} =  G_{2,3}^{(0)} = G_{3,1}^{(0)}   <  0 $. 
   
In this case the fixed point is a limiting case of the one corresponding to 
 Eqs.(\ref{rgs.y5}) in which all the running couplings flow to 0.   
\end{itemize}

\end{document}